\documentclass[fleqn,usenatbib]{mnras}

\usepackage{newtxtext,newtxmath}
\usepackage[T1]{fontenc}

\newcommand{\swift}{\textit{Swift}\xspace}
\newcommand{\fermi}{\textit{Fermi}\xspace}
\newcommand{\svom}{\textit{SVOM}\xspace}
\newcommand{\ep}{\textit{Einstein Probe}\xspace}
\newcommand{\bat}{BAT\xspace}
\newcommand{\xrt}{XRT\xspace}
\newcommand{\uvot}{UVOT\xspace}

\newcommand{\gbm}{GBM\xspace}
\newcommand{\lat}{LAT\xspace}

\newcommand{\eclairs}{ECLAIRs\xspace}
\newcommand{\grm}{GRM\xspace}
\newcommand{\mxt}{MXT\xspace}
\newcommand{\vt}{VT\xspace}

\newcommand{\swiftxrt}{\swift/\xrt}
\newcommand{\swiftbat}{\swift/\bat}
\newcommand{\swiftuvot}{\swift/\uvot}
\newcommand{\fermigbm}{\fermi/\gbm}
\newcommand{\fermilat}{\fermi/\lat}
\newcommand{\svomeclairs}{\svom/\eclairs}

\newcommand{\svomvt}{\svom/\vt}
\newcommand{\svomgrm}{\svom/\grm}

\newcommand{\grb}[1]{GRB~#1}

\DeclareRobustCommand{\VAN}[3]{#2}
\let\VANthebibliography\thebibliography
\def\thebibliography{\DeclareRobustCommand{\VAN}[3]{##3}\VANthebibliography}

\usepackage{graphicx}	% Including figure files
\usepackage{amsmath}	% Advanced maths commands
\usepackage{xspace}
\usepackage{longtable}
\usepackage[dvipsnames]{xcolor}
\usepackage{caption}
\title[GRB~250818B: Discovery \& Modelling]{GOTO identification and broadband modelling of the counterpart to the SVOM GRB~250818B}

\author[S. Belkin et al.]{
S.~Belkin,$^{1}$\thanks{E-mail: sergey.belkin@monash.edu (SB)}
G.~P.~Lamb,$^{2}$
K.~Ackley,$^{3}$
M.~E.~Wortley,$^{4,5}$
S.~McGee,$^{4,5}$
G.~Schroeder,$^{6}$
M.~Shrestha,$^{1}$
\newauthor
B.~P.~Gompertz,$^{4,5}$
D.~K.~Galloway,$^{1}$
R.~Starling,$^{7}$
W.-f.~Fong,$^{8,9}$
T.~Laskar,$^{8}$
C.~Liu,$^{8,9}$
A.~C.~Gordon,$^{8,9}$
\newauthor
N.~Pankov,$^{10,11}$
A.~E.~Volvach,$^{12}$
L.~N.~Volvach,$^{12}$
A.~Shein,$^{13}$
A.~Pozanenko,$^{10,11}$
M.~J.~Dyer,$^{14,15}$
\newauthor
J.~Lyman,$^{3}$
K.~Ulaczyk,$^{3}$
D.~Steeghs,$^{3}$
V.~S.~Dhillon,$^{14,16}$
P.~O'Brien,$^{7}$
G.~Ramsay,$^{17}$
K.~Noysena,$^{18}$
\newauthor
R.~Kotak,$^{19}$
R.~P.~Breton,$^{20}$
L.~K.~Nuttall,$^{21}$
D.~Pollacco,$^{3}$
S.~Awiphan,$^{18}$
J.~Casares,$^{16,22}$
P.~Chote,$^{3}$
\newauthor
A.~Chrimes,$^{23}$
R.~Eyles-Ferris,$^{7}$
B.~Godson,$^{3}$
P.~Irawati,$^{18}$
D.~Jarvis,$^{14}$
Y.~Julakanti,$^{7}$
L.~Kelsey,$^{24}$
\newauthor
M.~R.~Kennedy,$^{25}$
T.~Killestein,$^{3}$
A.~Kumar,$^{3}$
A.~Levan,$^{23}$
S.~Littlefair,$^{14}$
M.~Magee,$^{3}$
S.~Mandhai,$^{20}$
\newauthor
D.~Mata~S\'anchez,$^{16,22}$
S.~Mattila,$^{19,26}$
J.~McCormac,$^{3}$
D.~Mkrtichian,$^{18}$
S.~Moran,$^{7}$
J.~Mullaney,$^{14}$
\newauthor
D.~O'Neill,$^{3}$
M.~Patel,$^{7}$
K.~Pu,$^{1}$
M.~Pursiainen,$^{3}$
A.~Sahu,$^{3}$
U.~Sawangwit,$^{18}$
E.~Stanway,$^{3}$
Y.~Sun,$^{7}$
\newauthor
B.~Warwick,$^{3}$
and K.~Wiersema$^{27}$
\\
\textit{Affiliations are listed at the end of the paper}
}

\date{Accepted XXX. Received YYY; in original form ZZZ}

\pubyear{\the\year{}}

\begin{document}
\label{firstpage}
\pagerange{\pageref{firstpage}--\pageref{lastpage}}
\maketitle

\begin{abstract}
Rapid localisation and follow-up of gamma-ray bursts (GRBs) increasingly rely on low-latency triggers from new missions coupled to wide-field robotic optical facilities. We present the discovery and multi-wavelength follow-up of \grb{250818B}, detected by the \textit{Space Variable Objects Monitor} (\svom) and localised optically by the Gravitational-wave Optical Transient Observer (GOTO). We compile and homogenise X-ray, optical/NIR, and radio data to build broadband light curves and spectral energy distributions. The afterglow is unusually luminous for a nominal short GRB, lying on the bright end of the short-GRB population in X-rays and optical and among the most luminous high-redshift short-GRB afterglows in the radio. MeerKAT detects the source at 3.1~GHz, while the Atacama Large Millimeter/submillimeter Array (ALMA) provides deep higher-frequency limits. Keck/LRIS spectroscopy shows continuum and metal absorption (Fe\,{\sc ii}, Mg\,{\sc ii}, Mg\,{\sc i}), giving \( z=1.216 \). Synchrotron forward-shock modelling favours a constant-density medium and strongly prefers refreshed (energy-injection) emission, well described by a two-component jet with \( E_{\rm K,iso}\sim4\times10^{52} \)~erg, \( n_0\sim3.6\,\mathrm{cm^{-3}} \), \( \theta_j\simeq0.10 \)~rad (\( \sim5.7^\circ \)), and \( p\simeq1.64 \). The host association is ambiguous: the nearest LS~DR10 galaxy candidate (\( r_{\rm AB}\sim24.7 \)) is offset by \( \sim4\arcsec \) (\( \sim34 \)~kpc) with chance-alignment probability \( P_{\rm cc}\sim0.2 \), and current imaging does not exclude a fainter, near-coincident host. SED fitting of the candidate host suggests a low-mass galaxy. \grb{250818B} highlights the power of rapid wide-field counterpart identification in the \svom\ era, while host-association uncertainty can still limit offset-based interpretation.
\end{abstract}


%%%%%%%%%%%%%%%%%%%%%%%%%%%%%%%%%%%%%%%%%%%%%%%%%%

%%%%%%%%%%%%%%%%% BODY OF PAPER %%%%%%%%%%%%%%%%%%

\section{Introduction}
% \sergey{General intro to GRBs. If we plan to expand on the origin of GRBs, we can shed some light on the classification picture (mergers for short GRBs; collapsars for long GRBs). SVOM + GOTO context. Why \grb{250818b} is interesting (redshift, colour/reddening, radio hints, if any).
% \begin{itemize}
%     \item Gamma-ray bursts (GRBs)
%     \item Classification: long, short, ambiguous
%     \item Detection landscape: from \fermi and \swift \( \rightarrow \) \svom and \ep
%     \item GOTO as a wide-field facility for large localisations (briefly; not the paper’s focus)
%     \item \grb{250818B}: discovery and follow-up
%     \item This work: goals, data, and analysis outline
% \end{itemize}
% }

Gamma-ray bursts (GRBs) were discovered serendipitously in the late 1960s as brief flashes of high-energy photons on the sky~\citep{Klebesadel1973}. Subsequent observations have established GRBs as among the most luminous electromagnetic explosions in the Universe, with isotropic-equivalent \( \gamma \)-ray energy releases spanning \( \sim10^{51} \)--\( 10^{55} \)~erg and durations from milliseconds to minutes~\citep[e.g.][]{Piran2004,Kumar2015,Wang2015,Atteia2017,OConnor2023,LHAASO2023,Peer2024}. A key step in organising this diversity was the recognition that the distribution of prompt-emission durations is bimodal, allowing GRBs to be divided into two broad classes, ``short'' and ``long'', typically separated at \( T_{90} \simeq 2 \)~s\footnote{\(T_{90}\) is the time interval containing 90\% of the prompt \(\gamma\)-ray fluence.}~\citep{Kouveliotou1993}. Long-duration GRBs are most commonly associated with the core collapse of massive, stripped-envelope stars, as demonstrated by numerous cases with accompanying broad-lined Type~Ic supernovae~\citep[e.g.][]{Galama1998,Hjorth2003,Woosley2006,Cano2017,Kumar2024mag}, whereas short-duration GRBs are predominantly associated with compact-object mergers involving neutron stars and/or black holes~\citep[e.g.][]{Eichler1989,Narayan1992,Tanaka2016}, supported by the detection of kilonovae and, most dramatically, by the joint GW170817/\grb{170817A} event~\citep[e.g.][]{Abbott2017a,Goldstein2017,Savchenko2017,Pian2017,Troja2017,Valenti2017,Wang2017,Tanvir2017}. At the same time, recent discoveries have revealed events that blur this simple mapping between duration and progenitor: the short (rest-frame \( T_{90}\sim 0.5 \)~s) \grb{200826A} showed a collapsar-like supernova~\citep{Zhang2021,Rossi2022}, while long-duration bursts such as \grb{211211A}~\citep{Rastinejad2022,Troja2022,Yang2022,Gompertz2023} and \grb{230307A}~\citep{Dai2024,Levan2024,Sun2025} exhibit kilonova signatures indicative of compact-binary mergers. A further notable example is \grb{191019A}, which has been interpreted as a merger produced by dynamical capture~\citep{Levan2023,Stratta2025}, perhaps within the accretion disc of an active galactic nucleus~\citep{Lazzati2023}. Taken together, these ``hybrid'' cases demonstrate that \( T_{90} \) and spectral hardness alone are imperfect discriminants of progenitor type, and that broadband afterglow and host-galaxy information are essential for interpreting the diversity of the GRB population and placing individual events in their physical context.

Broadband afterglow emission is commonly modelled as synchrotron radiation from a relativistic outflow decelerating in the circumburst medium, and encodes the kinetic energy, ambient density profile, microphysical parameters, and outflow geometry~\citep[e.g.][]{Sari1998,Granot2002,Panaitescu2002,Miceli2022}. Multi-band light-curve and SED evolution can therefore test the simplest forward-shock scenario and reveal additional ingredients such as jet breaks and angular structure~\citep[e.g.][]{Rhoads1999,Sari1999}, refreshed shocks/energy injection~\citep[e.g.][]{Rees1998,Zhang2002,Nousek2006}, and reverse-shock or other extra components~\citep[e.g.][]{Kobayashi2000,Zhang2003}. Host-galaxy context provides complementary constraints on the environment and potential progenitor channel through offsets and association statistics~\citep[e.g.][]{Bloom2002,Fong2013a,Berger2014,Blanchard2016,Heintz2020,Fong2022,2024ApJ...962....5N,Castrejon2025}, but faint hosts near survey limits can complicate interpretation, particularly at \( z\gtrsim1 \) where even moderate-luminosity galaxies begin to approach detectability limits in wide-field imaging. These considerations are especially relevant for events whose prompt-emission classification is uncertain or whose prompt constraints are incomplete, such that the afterglow and host environment carry a disproportionate share of the interpretive weight~\citep[e.g.][]{Zhang2009,Bromberg2012,Bromberg2013}.

The detection and characterisation of GRBs have been transformed by the advent of dedicated space-based observatories.
% I've commented this out and moved a bit about Swift in Section 2.1 seems it looked a bit reduntant to talk that much about these instruments in this paper
%The \swift observatory combines the wide-field Burst Alert Telescope~\citep[\bat; 15--150~keV][]{Barthelmy2005} with two narrow-field instruments, the X-Ray Telescope~\citep[\xrt, 0.3--10~keV;][]{Burrows2005} and the Ultraviolet/Optical Telescope~\citep[\uvot, 170--650~nm;][]{Roming2005}, which can rapidly slew to BAT triggers and localise afterglows to arcsecond precision, enabling systematic X-ray and UV/optical follow-up. The \fermi satellite, and in particular its Gamma-ray Burst Monitor~\citep[\gbm, 8~keV--40~MeV;][]{Meegan2009,vonKienlin2020}, dominates the GRB detection rate, discovering of order one burst per day but typically with degree-scale localisation uncertainties. The \fermi Large Area Telescope~\citep[\lat, \(\sim\)20~MeV--\(>300\)~GeV][]{Atwood2009} extends the energy coverage to GeV energies and detects only a subset of the brightest events, at a rate of \( \sim 10 \) GRBs per year~\citep[e.g.][]{Kumar2015}. Because \fermi carries no narrow-field X-ray or optical instruments, GRBs detected solely by GBM rely on external facilities for afterglow identification and characterisation.
The \swift mission detects \( \sim 90 \) GRBs per year~\citep{Lien2016} with the Burst Alert Telescope~\citep[\bat;][]{Barthelmy2005} and provides rapid follow-up and arcsecond localisations of GRB afterglows with its narrow-field instruments, enabling systematic X-ray and UV/optical afterglow studies~\citep{Barthelmy2005,Burrows2005,Roming2005}. Complementing this, the \fermi Gamma-ray Burst Monitor~\citep[\gbm;][]{Meegan2009,vonKienlin2020} discovers \( \sim 250 \) GRBs per year but typically with degree-scale localisation uncertainties~\citep[e.g.][]{Connaughton2015,Lopez2024}, so bursts detected solely by \gbm often rely on external facilities for afterglow identification and characterisation~\citep[e.g.][]{Mong2021,Ahumada2022,Kumar2025}. More recently, new missions such as the \ep~\citep[\textit{EP};][]{Yuan2015} and the \textit{Space Variable Objects Monitor}~\citep[\svom;][]{Wei2016,Atteia2022}, both launched in 2024, provide complementary coverage and low-latency triggering: \ep combines the Wide-field X-ray Telescope (WXT; 0.5--4~keV) with the Follow-up X-ray Telescope (FXT; 0.5--10~keV), while \svom detects GRBs with \eclairs (4--150~keV) and \grm (15~keV--5~MeV) and refines afterglows with \swiftxrt-like follow-up via \mxt (0.2--10~keV) and the visible telescope \vt (450--950~nm).

For GRBs discovered by wide-field \( \gamma \)-ray monitors, initial localisations can range from arcminutes (e.g. \svomeclairs; typically \(\lesssim10\arcmin\), down to \(\sim3\arcmin\) for bright events) to tens of degrees (e.g. \fermigbm; localisation uncertainties can be \(\sim10{-}15^\circ \);~\citep{Connaughton2015}). Note that improved localisations can be obtained with more advanced methods~\citep[e.g.][]{Burgess2018,Berlato2019}. In the degree-scale regime in particular, efficiently identifying the optical afterglow requires ground-based facilities capable of tiling tens to hundreds of square degrees on timescales of minutes to a few hours. Wide-field, robotic optical surveys can fill this gap by rapidly covering the localisation region and performing real-time transient discovery.

The Gravitational-wave Optical Transient Observer~\citep[GOTO;][]{Steeghs2022,Dyer2024} has been designed as an array of wide-field optical telescopes with sites at Roque de los Muchachos Observatory on La Palma and at Siding Spring Observatory in Australia, together providing near-continuous coverage of both hemispheres and an instantaneous field of view (FoV) of \( \gtrsim 80 \)~deg\(^{2}\) per site (two mounts). GOTO’s fast-response scheduling and wide FoV allow it to begin tiling degree-scale \fermigbm localisations and arcminute-scale \svomeclairs error regions within minutes of a trigger, offering a complementary discovery channel to narrow-field facilities and helping to mitigate localisation-driven selection effects~\citep[e.g.][]{Singer2015,Turpin2016}. Previous campaigns have already demonstrated that GOTO can effectively search large \fermigbm error regions and identify GRB afterglows~\citep[e.g.][]{Mong2021,Belkin2024,Kumar2025}, illustrating its potential as a discovery engine for GRBs with large high-energy localisations and motivating its use in conjunction with \svom triggers such as \grb{250818B}.

\grb{250818B} provides a timely case study in the \svom era: rapid triggering enabled extensive multi-wavelength follow-up. Throughout this work, we adopt the nominal short-GRB classification reported in discovery notices~\citep{Wang2025gcn}; our physical interpretation is driven primarily by the broadband afterglow behaviour and the host-galaxy context. Keck/LRIS afterglow spectroscopy established an absorption redshift of \(z=1.216\)~\citep{Fong2025gcn41419}, enabling rest-frame modelling of the afterglow and host properties.

At the same time, the host association is non-trivial: the nearest catalogued LS~DR10 galaxy candidate is faint and offset by several arcseconds, while the imaging depth does not exclude a fainter, near-coincident host below the survey threshold. Because these scenarios imply very different physical offsets (and therefore limit the use of offset-based arguments), we place particular emphasis on constraints that arise directly from the broadband afterglow data. Specifically, we use the multi-band evolution to infer the circumburst environment and outflow dynamics, and to test whether a standard single-component forward-shock model is adequate or whether additional ingredients (e.g. energy injection and/or multiple components) are required. Where relevant, we comment on which inferences are insensitive to the uncertain physical offset and which depend on assumptions about the host association.

In this paper, we present the multi-wavelength follow-up of \grb{250818B} and the identification of its optical afterglow by GOTO, and we use the resulting dataset to model the broadband afterglow and assess the host-galaxy association. In Section~\ref{sec:obs} we describe the available observations across X-ray, optical/NIR, and radio bands, together with the afterglow spectroscopy used to establish the absorption redshift. In Section~\ref{sec:results} we construct broadband light curves and SEDs, perform afterglow modelling with \texttt{afterglowpy} (via \texttt{redback}, including tests for energy injection and/or additional components), and analyse the putative host galaxy through association metrics (Bayesian and chance-coincidence statistics) and host-galaxy SED fitting with \texttt{Prospector}; we also construct a single-epoch optical afterglow SED to constrain line-of-sight extinction. In Section~\ref{sec:discussion} we discuss the implications for the burst energetics, environment, and progenitor interpretation, including the host-association ambiguity and its impact on offset-based arguments. Our conclusions are summarised in Section~\ref{sec:conclusions}.

Throughout this paper, we assume a flat \( \Lambda \)CDM cosmology with \( H_{0} = 70~\mathrm{km~s^{-1}~Mpc^{-1}} \), \( \Omega_{\rm m} = 0.3 \) and \( \Omega_{\Lambda} = 0.7 \), and quote magnitudes in the AB system unless stated otherwise.

\section{Observations}\label{sec:obs}
At 2025-08-18T03:29:09~UTC (hereafter \(T_{0}\)), \svomeclairs triggered on \grb{250818B} and distributed via the General Coordinates Network (GCN) an on-board localisation with a \(90\%\) confidence radius of \(7\farcm44\)~\citep{Wang2025gcn}. The burst also triggered \svomgrm at \(T_{0}\) using a 1~s integration timescale~\citep{Wang2025gcn}.

We inspected the publicly available \fermigbm time-tagged event (TTE) data around \(T_{0}\) and found that the spacecraft appears to have been traversing the South Atlantic Anomaly at that time; accordingly, no on-board or ground trigger was reported. A targeted search of the available TTE data also showed no significant excess.\footnote{GBM burst data products are publicly available via HEASARC: \url{https://heasarc.gsfc.nasa.gov/FTP/fermi/data/gbm/bursts/}.} We therefore do not report prompt-emission parameters from \fermigbm for this event, and no useful high-energy constraints from \fermilat are available from public burst products.\footnote{LAT burst data products are publicly available via Fermi/SSC: \url{https://fermi.gsfc.nasa.gov/cgi-bin/ssc/LAT/LATDataQuery.cgi}.}

Wide-field optical imaging reported a fading counterpart within the \svom localisation region, providing a precise position and enabling rapid coordination of follow-up (Sec.~\ref{subsec:opt_nir};~\citet{Kumar2025gcn}). Subsequent \swiftxrt observations detected an uncatalogued X-ray source consistent with the \svom localisation and coincident with the optical counterpart position~\citep{Ferro2025gcn}. The \svom Visible Telescope (VT) also reported the same optical transient at a consistent position, and the \ep Follow-up X-ray Telescope (FXT) detected an X-ray source consistent with the counterpart~\citep{Li2025gcn}. These detections established the broadband afterglow and triggered an extensive multi-wavelength campaign from optical to radio. In the following subsections, we describe the high-energy, X-ray, optical/near-infrared, and radio datasets in more detail and outline the processing steps used to construct the broadband light curves and spectra employed in our analysis.

%Follow-up X-ray observations by \swiftxrt detected an uncatalogued source consistent with the \svom localisation~\citep{Ferro2025gcn}, and the \ep Follow-up X-ray Telescope (FXT) also detected the source at a consistent position~\citep{Li2025gcn}. These observations established the presence of an X-ray afterglow and triggered a multi-wavelength campaign, from optical to radio. In the following subsections, we describe the high-energy, X-ray, optical/near-infrared, and radio datasets in more detail and outline the processing steps used to construct the broadband light curves and spectra employed in our analysis.

% \subsection{High-energy Observations}
% \subsubsection{\fermigbm}
% \sergey{There was no \textit{Fermi} GCN for this GRB. However, based on \href{https://heasarc.gsfc.nasa.gov/db-perl/W3Browse/w3table.pl?tablehead=name\%3Dfermigdays\&Action=More+Options}{the Fermi GBM Daily Data}, there appears to be coverage on 2025-08-18 between 03:00 and 04:00~UTC. 14nov2025: nope, it's just SAA and nothing to analyse and present here, so delete this part. So we'll merge it with the above preamble part.}

% \subsubsection{\svomeclairs{}}
% \sergey{It detected the GRB, but the data are not publicly available, so I assume we don't have this either. 14nov2025: do we want to add our rough  estimations of the \( T_{90}\) from the figure \svom shared in their detection GCN? Because it's the only thing can be done here. RLCS: I sent a formal request to the SVOM team on this, and they responded that they will be writing their own paper and do not wish to provide svom data to us or, if it were offered, to include our contribution.\sergey{Yup, we will exclude this section with the high-energy one}}

\subsection{X-rays}\label{subsec:xrt}
The \swift X-Ray Telescope~\citep[\xrt, 0.3--10~keV;][]{Burrows2005} began follow-up observations of \grb{250818B} at \( T_0+1.7\)~ks after the \svom trigger. We use the \xrt data to perform a time-resolved spectral analysis of the afterglow. Using the UK Swift Science Data Centre \textsc{swifttools} API~\citep{Evans2009}\footnote{https://www.swift.ac.uk/API/}, we downloaded and binned the data into bins with a signal-to-noise threshold of S/N~=~10, which we adopt for spectral extraction and fitting in Section~\ref{subsec:xray_spec}.

\subsection{Optical and NIR}\label{subsec:opt_nir}
%\sergey{Facilities (GOTO, LT/NOT/\ldots; a subsubsection for each if we have data), pipelines, calibration to AB, Galactic extinction. If there is no data, just indicate what was fetched from GCNs. Yeah, probably we don't expect too much data aside from our: most of it is coming from the GCNs.}

GOTO responded to the \svom alert and began observations of the \svomeclairs localisation at \(T_{0}+0.54\)~h, obtaining a sequence of \(4\times90\)~s exposures in the wide \(L\) band (400--700~nm) between \(T_{0}+0.54\) and \(T_{0}+1.67\)~h. Images were processed with the standard GOTO pipeline, including difference imaging against deeper templates and automated candidate filtering, followed by human vetting. These observations revealed a fading optical source, GOTO25fzq/AT~2025ukm~\citep{Kumar2025tns}, within the \eclairs \(90\%\) localisation region at R.A. = \(03^{\rm h}04^{\rm m}13\fs52\), Dec. = \(-03^{\circ}07^\prime30\farcs82\) (J2000), with \(L = 18.71\pm0.14\)~mag at \(T_{0}+0.54\)~h and \(L = 19.49\pm0.18\)~mag (AB) at \(T_{0}+1.67\)~h. No source was detected at this position in pre-trigger GOTO images obtained 9.23~h before the burst, down to a \(3\sigma\) limit of \(L>20.3\)~mag, supporting the association with \grb{250818B}~\citep{Kumar2025gcn}.

The optical counterpart was subsequently detected by \svomvt in the VT\_B band at \(T_{0}+198.5\) and \(T_{0}+3963.5\)~s~\citep{Yao2025gcn41409}, by \swiftuvot in the \(u\) band~\citep{Siegel2025gcn41435}, and by numerous ground-based facilities including KAIT~\citep{Zheng2025gcn41417}, NOT~\citep{BroeBendtsen2025gcn41426}, SAO~RAS~\citep{Moskvitin2025gcn41428}, TRT~\citep{An2025gcn41430}, Lesedi~\citep{Kumar2025gcnLesedi}, LT~\citep{Dimple2025gcn41442}, the GRANDMA/Kilonova-Catcher network~\citep{Hellot2025gcn41444}, and the Wendelstein FTW telescope~\citep{Busmann2025gcn41445}, providing multi-epoch photometry from minutes to several days after the burst.

%Unless stated otherwise, all times are reported relative to the \svomeclairs trigger time \(T_0\). Optical/NIR photometry from the literature and GCN Circulars was converted to the AB system where necessary and corrected for Galactic extinction using the line-of-sight \(E(B-V)\) from \citet{Schlafly2011} and the \citet{Fitzpatrick1999} law with \(R_V=3.1\). Upper limits are quoted as \(3\sigma\) limits based on the local background RMS.

The full optical/NIR dataset compiled in this work is given in Table~\ref{tab:phot_250818B}. Where necessary, we converted literature photometry to the AB system using the offsets in Table~\ref{tab:vega_ab_offsets} and corrected for Galactic extinction using the coefficients in Table~\ref{tab:mw_extinction_coeffs}. We also list the catalogue photometry of the candidate host galaxy in Table~\ref{tab:host_phot_legacy}.

\subsection{Radio and millimeter/submillimeter}\label{subsec:radio}
We observed \grb{250818B} with the MeerKAT radio telescope (in the Karoo desert, South Africa) through Director's Discretionary Time (DDT-20250822-GS-01 PI Schroeder) at mean frequencies of 1.3~GHz (0.856~GHz bandwidth) and 3.1~GHz (0.875~GHz bandwidth) for four epochs in each band, spanning mid-times of 5.0--32.1\,days. J0323+0534 was used as a complex gain calibrator, and J0408-6545 was used as a flux calibrator for all observations and bands. We downloaded the SARAO Science Data Processor pipeline images for all observations and used the \texttt{pwkit/imtool} program~\citep{Williams2017} to measure the flux density and image RMS at the GOTO position. The radio afterglow of \grb{250818B} is detected in the first 3.1~GHz observations, as reported in \citet{Schroeder2025gcn41516}, and rises and fades over the course of the observations. The radio afterglow is only detected in the 1.3~GHz epoch at \( \sim 15 \)~days.

We observed GRB~250818B with the Atacama Large Millimeter/Submillimeter Array (ALMA; Program 2024.1.01131.T, PI: W. Fong) in standard continuum mode with 4\,GHz bandwidth at a mean frequency of 97.5~GHz for two epochs at mid-times of 10.14\,days and 17.19\,days, respectively. Both observations utilised J0312+0133 as a complex gain calibrator and J0327+0044 as a check source, while J0238+1636 and J0423-0120 were employed as the bandpass and flux density calibrators for the first and second epochs, respectively. We downloaded the final image products from the ALMA science archive. In the Quality Assurance Level 2 (QA2) images, no significant emission is seen in a \( 0\farcs5 \) circle at the GOTO position. The image root-mean-square (rms) at the expected position of the target in the observations is 13.2\,\(\mu\)Jy and 10.7\,\(\mu\)Jy, respectively.

The 36.8~GHz observations were carried out with the 22-m RT-22 radio telescope located in Simeiz (Crimea). The antenna has a half-power beam width of \( \sim 100\arcsec \). Observations were performed in a beam-switching mode: the telescope was alternately pointed at the source with each of the two beam lobes produced by diagram modulation and having mutually orthogonal polarisations.

The antenna temperature of the source was determined from the difference between the radiometer outputs, each averaged over 30~s at the two antenna positions. For each epoch, we obtained a series of 200--250 such measurements, from which we computed the mean signal and its rms error. The measured antenna temperatures, corrected for atmospheric attenuation, were converted to flux densities by comparison with observations of standard calibration sources~\citep{Volvach2025}. The resulting 36.8~GHz flux densities are listed in Table~\ref{tab:radio_250818B} and plotted in Fig.~\ref{fig:lc_multiband_mjy}.

In addition, we include the 10~GHz flux density reported in GCN~41455~\citep{Ricci2025gcn41455}. As no uncertainty was provided, the point is shown as indicative in Fig.~\ref{fig:lc_multiband_mjy}; for modelling, we assign an assumed uncertainty to prevent the fit from being dominated by this measurement.

\subsection{Spectroscopy}\label{subsec:spectroscopy}
\begin{figure*}
    \centering
    \includegraphics[width=\textwidth]{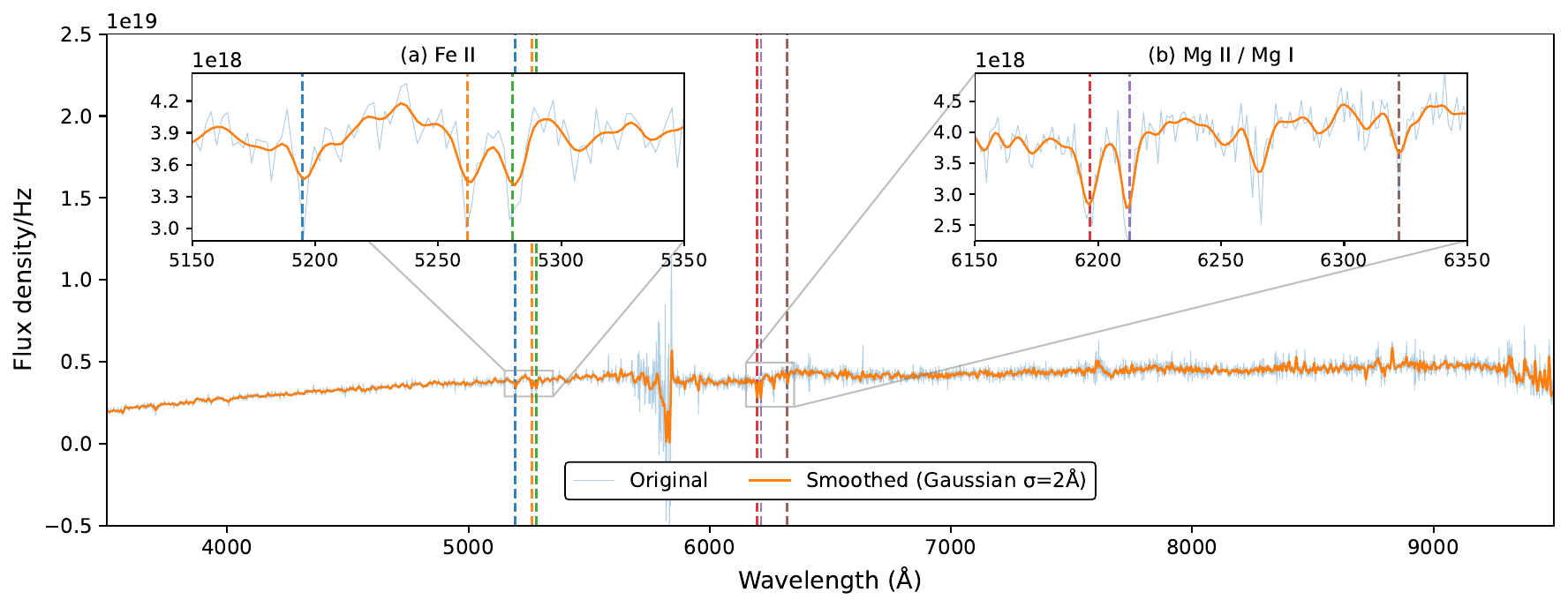}
    \caption{
    Keck~I/LRIS flux-calibrated, Galactic-extinction-corrected spectrum of the optical afterglow of \grb{250818B} obtained at \(\approx 10.7\)~hr post-burst; wavelengths are shown in the observer frame. The light-blue curve shows the original spectrum and the orange curve shows a Gaussian-smoothed version (\(\sigma = 2\)~\AA) for display purposes. Vertical dashed lines mark the expected observed-frame wavelengths of the Fe\,{\sc ii}~\(\lambda\lambda2344, 2374, 2382\), Mg\,{\sc ii}~\(\lambda\lambda2796, 2803\), and Mg\,{\sc i}~\(\lambda2852\) absorption transitions at the best-fit redshift \(z = 1.216\). Insets highlight the Fe\,{\sc ii} and Mg\,{\sc ii}/Mg\,{\sc i} regions used to refine the redshift estimate.}
    \label{fig:keck_lris_spec}
\end{figure*}

We obtained optical spectroscopy of the \grb{250818B} afterglow with the Low-Resolution Imaging Spectrograph~\citep[LRIS;][]{Oke1995} mounted on the Keck~I telescope (PI: C. Liu; Program O397). The observations comprised three 300~s exposures using the 400/3400 grism on the blue arm and the 400/8500 grating on the red arm, starting at 2025 August 18 14:10~UT (\( \approx 10.7 \)~hr post-burst), under clear conditions with \( \sim0.9\arcsec \) seeing at airmass \( \sim1.2 \). The combined spectrum covers \( \approx 3500 \) -- \( 9500 \)~\AA\ in the observer frame (Fig.~\ref{fig:keck_lris_spec}).

We reduce and co-add the data using the Python Spectroscopic Data Reduction Pipeline ({\tt PypeIt\footnote{\url{https://pypeit.readthedocs.io/en/latest/}}}; \citealt{Prochaska2020joss,Prochaska2020zonedo}). \texttt{PypeIt} performs bias-subtraction, flat-fielding, cosmic ray masking, and wavelength calibrations on the individual frames. We perform 1D extraction and then co-add the spectrum. We then perform flux calibration and telluric correction, and corrected for foreground Milky Way extinction along the line of sight to the GRB, adopting \(E(B-V)_{\rm MW}=0.0633\,\mathrm{mag}\) from the Schlafly \& Finkbeiner recalibration of the Schlegel et al.\ dust map \citep{Schlegel1998,Schlafly2011}, and an \(R_V=3.1\) Milky Way extinction law \citep{FitzpatrickMassa2007}. The continuum is detected across most of the wavelength range and exhibits several prominent metal absorption features.

We determine the redshift by identifying a consistent set of metal absorption lines, including Fe\,{\sc ii}~\(\lambda\lambda2344, 2374, 2382\), the Mg\,{\sc ii} doublet~\(\lambda\lambda2796, 2803\), and Mg\,{\sc i}~\(\lambda2852\). We refine the redshift by maximising the alignment of these transitions with absorption troughs in the spectrum, yielding \( z = 1.216 \). We adopt this value as the redshift of \grb{250818B} throughout the paper.

\section{Analysis and Results}\label{sec:results}
\subsection{Broadband Light-Curve Construction}
For our analysis, we compiled all publicly available photometric measurements reported in these GCN circulars and in our own follow-up before constructing the optical/near-infrared light curves. All optical and near-infrared measurements were homogenised onto the AB magnitude system~\citep{Oke1983}. For data reported in the Vega system, we applied fixed offsets of the form \( m_{\rm AB} = m_{\rm Vega} + \Delta m \) appropriate for each filter. For Johnson--Cousins \( UBVRI \) and near-infrared \( JHK_s \) filters, as well as Sloan-like \( griz \), we adopted the standard Vega-to-AB offsets from~\citet{BlantonRoweis2007}. For the \swift/UVOT \( u \), \( b \) and \( v \) bands we used the AB--Vega corrections from the UVOT photometric calibration~\citep[e.g.][]{Poole2008,Breeveld2011}. The values of \( \Delta m\) used in this work are listed in Table~\ref{tab:vega_ab_offsets}. Measurements already reported in AB magnitudes were left unchanged.

We corrected all photometry for Milky Way extinction using a single colour excess \(E(B-V)_{\rm MW}=0.0633\)~mag along the GRB sightline (consistent with Section~\ref{subsec:spectroscopy}), adopting the Schlafly \& Finkbeiner recalibration of the Schlegel et al.\ dust map~\citep{Schlegel1998,Schlafly2011}. We assumed a standard \(R_V=3.1\) Milky Way extinction curve. For Sloan-like \(ugriz\) filters we used the tabulated coefficients \(A_\lambda/E(B\!-\!V)\) from \citet{Schlafly2011}; for Johnson--Cousins \(UBVRI\) and near-infrared bands we used coefficients appropriate for an \(R_V=3.1\) Milky Way law~\citep[e.g.][]{Fitzpatrick1999}. For the \swiftuvot\ filters we employed the passband-integrated \(A_\lambda/E(B\!-\!V)\) values from \citet{Yi2023}, rescaled to be consistent with our adopted \(E(B\!-\!V)\). For non-standard filters, we adopted approximate coefficients: for the GOTO \(L\) band, we used \(A_L \simeq 0.997\,A_V\), and for the \svomvt\_B band we set \(A_{VT_B}\) to the average of the SDSS \(g\) and \(r\) coefficients. The resulting \(A_\lambda\) values for the filters used in this work are given in Table~\ref{tab:mw_extinction_coeffs}.

\begin{figure*}
    \centering
    \includegraphics[width=0.75\textwidth]{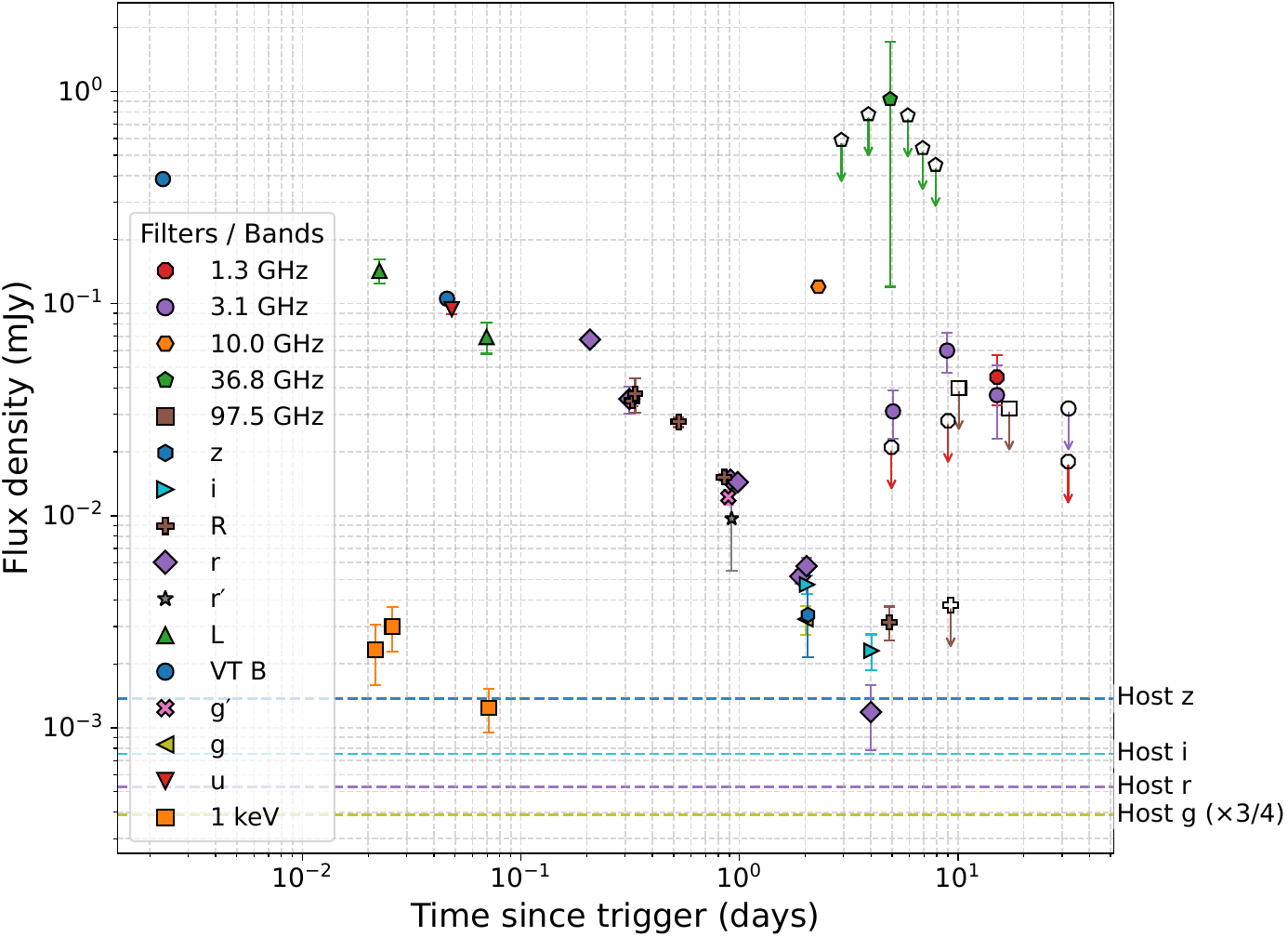}
    \caption{
    Multi-wavelength light curve of \grb{250818B}, showing X-ray, optical/near-infrared, and radio flux densities as a function of time since the \svomeclairs trigger. Optical points are corrected for Galactic extinction and converted to AB flux densities. Triangles with arrows indicate \( 3\sigma \) upper limits. The host-galaxy fluxes shown in \(g\), \(r\), \(i\), and \(z\) correspond to the LS~DR10 candidate host galaxy (Tractor \texttt{objid}~5790) discussed in Section~\ref{sec:host_association}. For visual clarity, the plotted \( g \)-band host level is vertically offset by a factor of \( 3/4 \) to avoid overlap with the \( r \)-band host level; all analysis uses the unscaled host flux.
    }
    \label{fig:lc_multiband_mjy}
\end{figure*}

After correcting for Galactic extinction, we converted all magnitudes to monochromatic flux densities \(F_\nu\) following the definition of the AB system \citep{Oke1983}, adopting a zeropoint \(F_{\nu,0} = 3631~\mathrm{Jy}\). For non-detections, we converted the limiting magnitudes to \(3\sigma\) upper limits on \(F_\nu\). The resulting extinction-corrected optical/near-infrared flux densities, together with the \swiftxrt and radio measurements (Sec.~\ref{subsec:xrt},~\ref{subsec:radio}), are shown in Fig.~\ref{fig:lc_multiband_mjy} and form the basis for the afterglow analysis in the following subsections.

\subsection{X-ray Spectral Analysis}\label{subsec:xray_spec}
\begin{figure}
    \centering
    \includegraphics[width=\columnwidth]{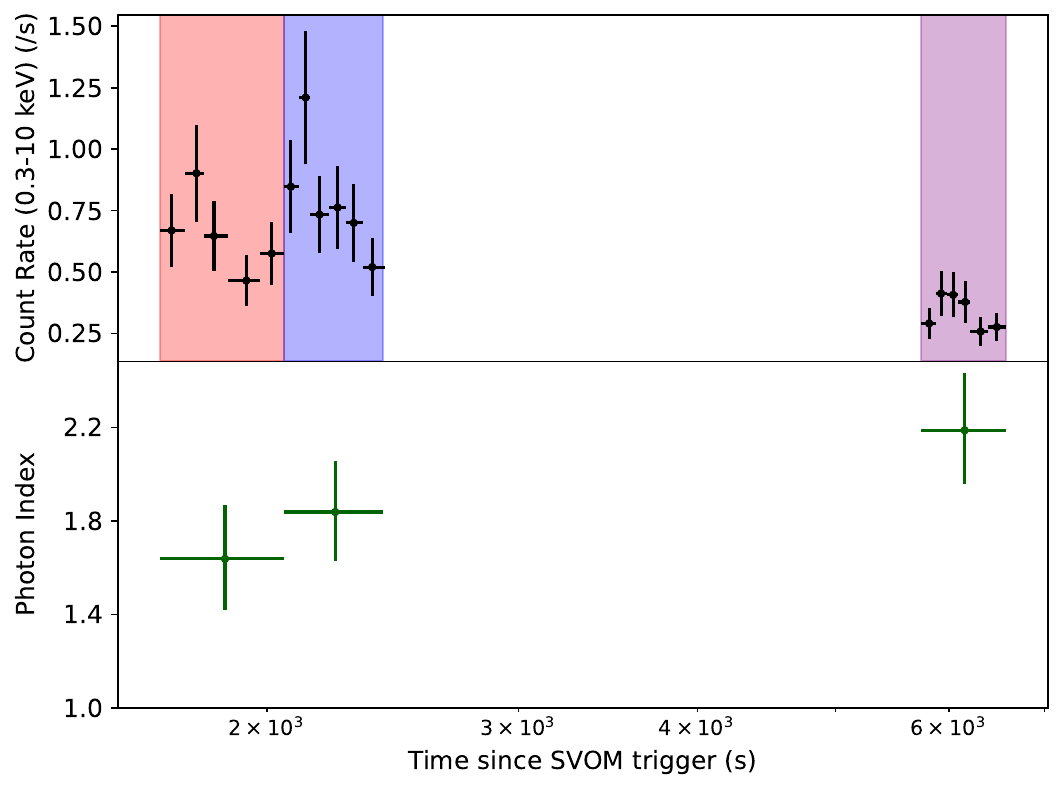}
    \caption{
    Top: XRT count rate light curve of the three flaring episodes, where the highlighted sections indicate the three SNR = 10 spectral bins. The late-time \xrt data at \( \sim 3.35 \)~d does not meet the SNR threshold and so is not included in the spectral analysis. Bottom: evolution of the best-fit photon index with time.
    }
    \label{fig:xrtlc_gamma_snr10}
\end{figure}

There were a total of three intervals found from the SNR = 10 binning, from which spectra were extracted and fit simultaneously. We fit the \swiftxrt spectra to a power-law model using \textsc{PyXspec} version 2.1.4 (\textsc{xspec} version 12.15.0) using Cash statistic~\citep{Cash1979} and a Wilms solar abundance~\citep{Wilms2000}. The absorption from the neutral hydrogen column along the line of sight is accounted for using a model components of the form \textsc{tbabs*ztbabs*cflux*powerlaw}. \textsc{tbabs} accounts for the neutral hydrogen in the Milky Way, fixed to \( N_{\text{H, Gal}} = 6.49 \times 10^{20} \)~cm\(^{-2}\)~\citep{Willingale2013}, \textsc{ztbabs} parameterises the absorption contribution from the host (\( N_{\text{H, Intr}} \)) with redshift fixed to the spectroscopic value of 1.216, and \textsc{cflux} calculates the flux of the model components. Two sets of fitting were performed: one calculating the unabsorbed flux (with the model format as above), and another calculating the absorbed flux (\textsc{cflux*tbabs*ztbabs*powerlaw}). \( N_{\text{H, Intr}} \) was tied between each spectrum as it is not expected to evolve with time, while the photon index \( \Gamma \) was left free to vary between spectra, as was the flux (calculated between 0.3 -- 10 keV). The absorbed power-law model provides an adequate description of the data (CSTAT/DoF = 0.24), with no systematic residuals; given the limited photon statistics, we therefore do not explore more complex spectral models. The time intervals for the extracted spectra and the evolution of \( \Gamma \) are shown in Fig.~\ref{fig:xrtlc_gamma_snr10}. % For the full best-fit parameters from this analysis, see Appendix~\ref{app:xray}, Table~\ref{tab:xrt_250818B_snr10}.
We further investigated spectral variability using finer spectral bins with SNR = 5, but found no evidence for additional spectral evolution due to the larger measurement uncertainties resulting from the lower-significance bins. 
% See Table~\ref{tab:xrt_250818B_snr5} in Appendix~\ref{app:xray} and Figure~\ref{fig:xrtlc_gamma_snr5} in Appendix~\ref{app:extra} for details.

%\subsection{Spectral Energy Distribution}

\subsection{Afterglow}

\subsubsection{X-ray afterglow brightness in context}
\label{subsubsec:ag_swift_context}
\begin{figure}
    \centering
    \includegraphics[width=\columnwidth]{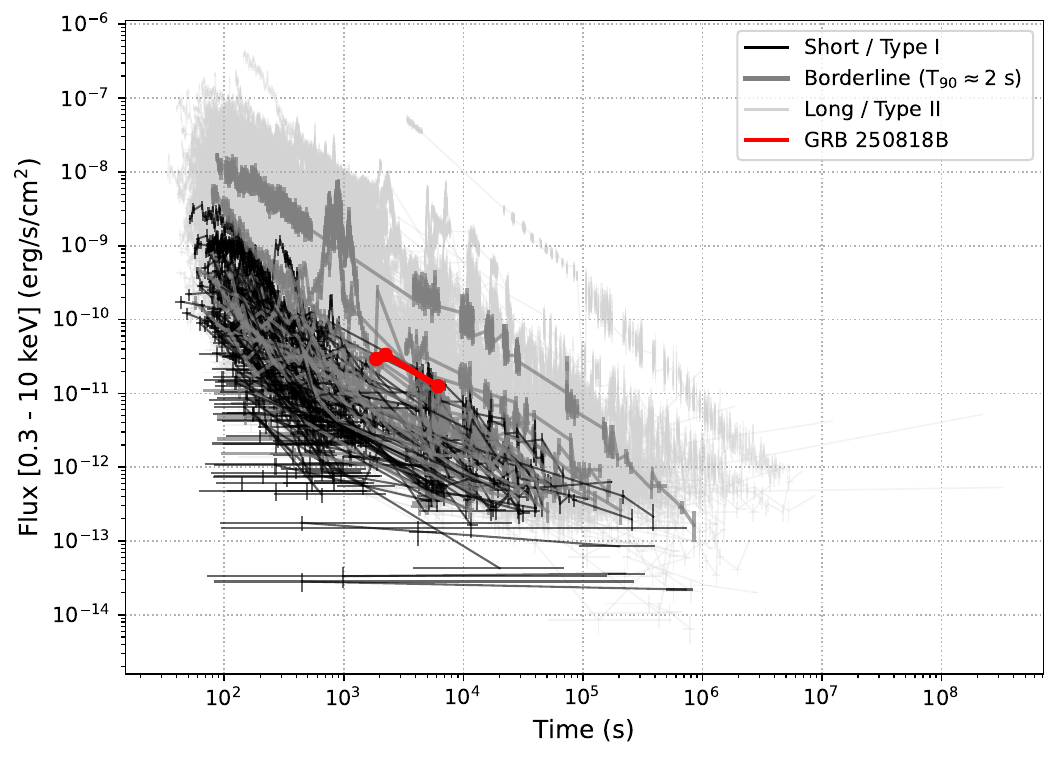}
    \caption{
    \swiftxrt observer-frame 0.3--10~keV afterglow comparison for \grb{250818B}. Grey curves show 0.3--10~keV \xrt light curves of \swift-detected GRBs downloaded from the UKSSDC repository. Short-duration bursts are highlighted in darker grey, while long-duration events are shown in light grey, using a Swift/BAT \(T_{90}\)-based duration classification from the \swiftbat GRB summary catalogue (excluding events with unidentified \(T_{90}\)). Bursts whose \bat \(T_{90}\pm\sigma\) interval straddles 2~s (``borderline'' between short and long) are plotted in medium grey. Red points show the unabsorbed 0.3--10~keV fluxes of \grb{250818B} from our \xrt spectral analysis with SNR=10 binning (Table~\ref{tab:xrt_250818B_snr10}); the red line connects the measurements to guide the eye. Times are measured relative to the prompt trigger in the observer frame (\bat trigger for the \swift comparison sample; \svomeclairs \(T_{0}\) for \grb{250818B}).}
    \label{fig:ag_xrt_population}
\end{figure}

To place the early-time X-ray behaviour of \grb{250818B} in context, we compared its \swiftxrt 0.3--10~keV afterglow to the population of \swift-detected events using UKSSDC \xrt light curves (Fig.~\ref{fig:ag_xrt_population}). Short-duration and long-duration bursts are shown in darker and lighter grey, respectively, using a \(T_{90}\)-based duration classification from the \swiftbat GRB summary catalogue. Bursts whose \bat \(T_{90}\pm\sigma\) interval straddles 2~s are flagged as ``borderline'' between the two classes and plotted in an intermediate grey shade; they are not included when computing the short-GRB medians quoted below. In this view, \grb{250818B} sits on the bright end of the short-GRB distribution at early times. At the three epochs corresponding to our SNR=10 \xrt spectral bins (1869, 2232, and 6153~s; Table~\ref{tab:xrt_250818B_snr10}), its unabsorbed flux is \(\sim 12.7\), \(\sim 17.1\), and \(\sim 9.5\) times higher than the median 0.3--10~keV flux of the \swift short-duration sample with coverage at these times (\(N_{\rm short}=46, 47,\) and \(35\), respectively). While this comparison is made in the observer frame and does not account for redshift or sample heterogeneity, it supports the qualitative impression that the early afterglow of \grb{250818B} is unusually X-ray bright relative to typical \swift short-duration GRBs.

The elevated early-time X-ray flux of \grb{250818B} could reflect a combination of higher-than-average afterglow energetics, and/or a favourable viewing geometry (e.g., a more on-axis sightline or a narrow/structured jet component; e.g.~\citealt{Sari1998,Granot2002,Panaitescu2002}). In the standard forward-shock synchrotron framework, if the X-ray band lies above the cooling break (\( \nu_X > \nu_c \); as suggested by the optical--X-ray spectral slopes), the X-ray flux depends only weakly on the external density, while density variations primarily affect the location and evolution of \( \nu_c \) and hence the optical/radio behaviour~\citep[e.g.][]{Sari1998,Granot2002}. If instead \( \nu_c \) lies above the X-ray band (\( \nu_c > \nu_X \)), the X-ray flux would have a stronger dependence on the circumburst environment and the expected spectral/temporal indices would differ, but this appears less consistent with the observed broadband slopes. A contribution from early-time energy injection is also plausible~\citep[e.g.][]{Rees1998,Zhang2002,Nousek2006}; however, the X-ray comparison is not diagnostic, and distinguishing between these possibilities requires dedicated broadband modelling (and, where possible, rest-frame comparisons).

\subsubsection{Optical afterglow brightness in context}
\label{subsubsec:ag_kann_context}
\begin{figure}
    \centering
    \includegraphics[width=\columnwidth]{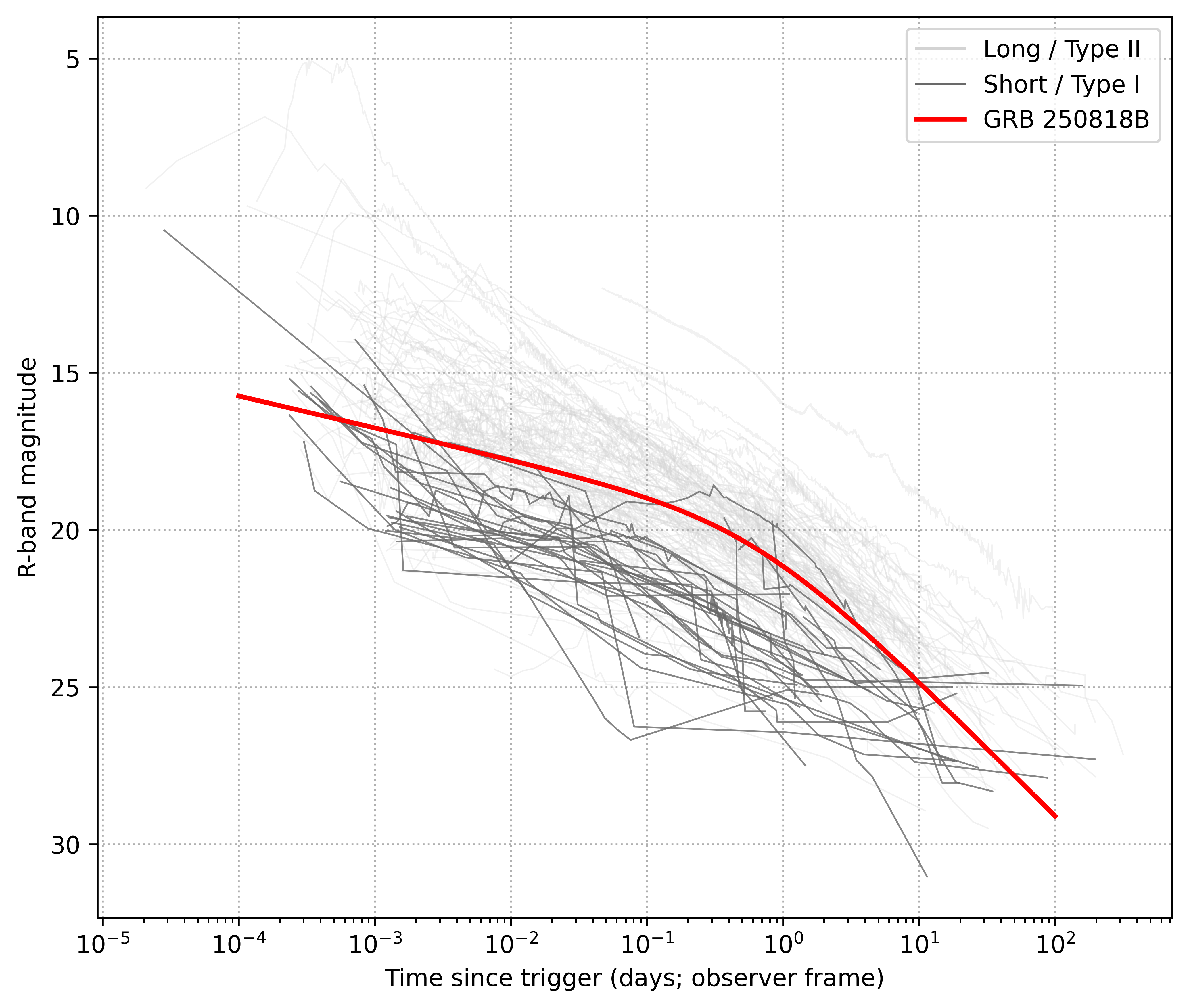}
    \caption{
    Observer-frame \( R \)-band afterglow comparison for GRB~250818B. Grey curves show literature \( R \)-band afterglow light curves compiled from \citet{Kann2006,Kann2010,Kann2011,NicuesaGuelbenzu2012}. The short/Type~I subset (as classified in \citealt{Kann2011}) is highlighted in darker grey, while long/Type~II events are shown in light grey. The red curve shows the \( R \)-band model of GRB~250818B derived from our simultaneous multiband SBPL fit. All magnitudes are corrected for Galactic extinction.
    }
    \label{fig:ag_kann_R}
\end{figure}

To place the optical afterglow of \grb{250818B} in the broader GRB population, we compare its observer-frame \( R \)-band evolution with the literature compilation of \citet{Kann2006,Kann2010,Kann2011} and the sGRB sample of \citet{NicuesaGuelbenzu2012}. For the \citet{Kann2011} comparison sample, we adopt the Type~I/Type~II classifications reported therein (a progenitor-motivated scheme that is not strictly equivalent to a \( T_{90}\)-based short/long division). Given the limited number of direct \( R \)-band measurements for \grb{250818B}, we show the \( R \)-band projection of our simultaneous multiband SBPL fit (Fig.~\ref{fig:ag_kann_R}). All magnitudes are corrected for Galactic extinction.

At the epochs corresponding to our \( R \)-band observations (0.32, 0.33, 0.53 and 0.85~d), we interpolate the comparison light curves in \( \log t \)--magnitude space, using only events with data bracketing each epoch (yielding \( N_{\rm all}\simeq136 \) -- \(142 \) and \( N_{\rm short}\simeq 29 \) -- \( 31 \)). The full sample spans \( R \approx 14.2 \) -- \( 26.7 \)~mag with medians of \( R \approx 20.4 \) -- \( 21.3 \)~mag, while the short/Type~I subset has fainter medians of \( R \approx 22.6 \) -- \( 23.9 \)~mag. \grb{250818B} has \( R \approx 19.9 \) -- \( 20.9 \)~mag at these times, placing it only \( \sim 0.3 \) -- \( 0.6 \)~mag brighter than the median of the overall population, but \( \sim 2.7 \) -- \( 3.0 \)~mag brighter than the median of the short/Type~I subset. For reference, relative to the long/Type~II subset alone, \grb{250818B} is within \( \lesssim 0.2 \)~mag of the median at these epochs. Thus, although \grb{250818B} is identified as a short GRB, its optical afterglow lies on the bright end of the sGRB distribution at \( \sim 0.3 \) -- \( 0.9 \)~d, while remaining broadly consistent with the full comparison sample (dominated by long GRBs) in the same band.

Having a securely defined redshift (\( z=1.216 \)) for \grb{250818B}, we can also express its optical afterglow brightness in the luminosity-space framework commonly used for short-GRB comparisons. Using our Galactic-extinction-corrected observer-frame \( R \)-band SBPL model and converting to \( \nu L_\nu \) at the corresponding rest-frame frequency, we find \( \log_{10}(\nu L_\nu)\approx 45.26 \) at a rest-frame time \( \delta t_{\rm RF}=3 \)~hr (i.e.\ the observer-frame time divided by \(1+z\); \( \nu L_\nu \simeq 1.8\times10^{45} \)~erg~s\(^{-1}\)), declining to \( \log_{10}(\nu L_\nu)\approx 44.24 \) at \( \delta t_{\rm RF}=1 \)~day. These values place \grb{250818B} toward the bright end of the short-GRB optical afterglow distribution at a common rest-frame epoch~\citep{Castrejon2025}, consistent with its location on the bright tail in the observer-frame \( R \)-band comparison shown in Fig.~\ref{fig:ag_kann_R}.

\subsubsection{Optical SED and line-of-sight extinction}\label{subsubsec:ag_sed_extinction}

To constrain the line-of-sight extinction in the host galaxy, we constructed a single-epoch optical SED at \(t_{\rm ref}=2.035\)~d after the trigger using nearly contemporaneous LT \(gri\) photometry and late-time DECam imaging to subtract the host contribution (Sec.~\ref{subsubsec:host_phot}). Throughout this subsection, we adopt the standard convention \(F_\nu \propto \nu^{-\beta}\) for the flux-density spectral index \(\beta\). For reference, the X-ray photon index \(\Gamma\) (defined by the photon spectrum \( N_E \equiv \mathrm{d}N/\mathrm{d}E \propto E^{-\Gamma}\)) corresponds to \(\beta_X=\Gamma_X-1\).

\begin{figure}
    \centering
    \includegraphics[width=\columnwidth]{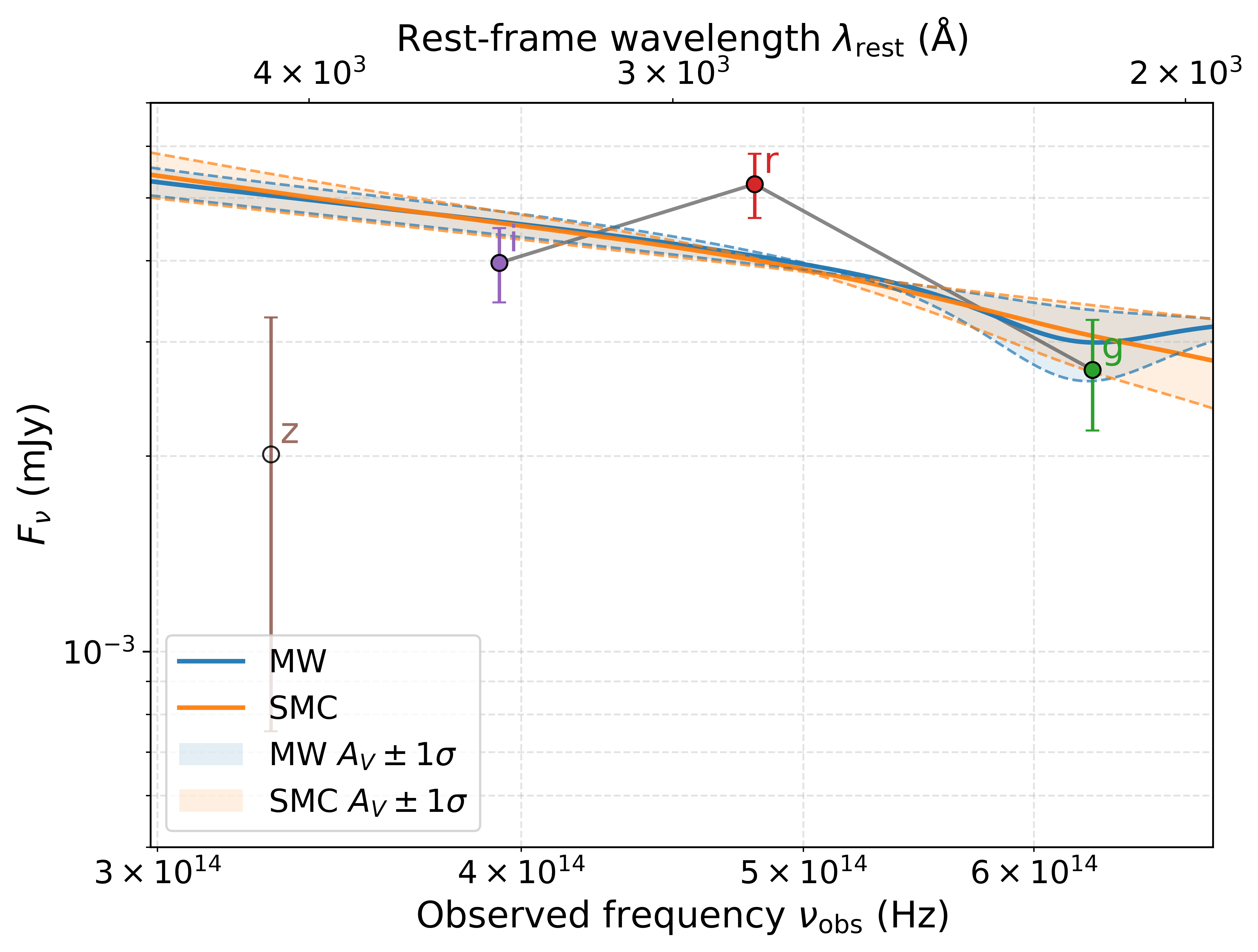}
    \caption{
    Host-subtracted optical SED of \grb{250818B} at \(t\simeq2.0\)~d. Filled points show the host-subtracted \(gri\) flux densities; the open \(z\) point is shown for comparison but excluded from the fit. Solid curves show the best-fitting dust-attenuated power-law models for a Milky Way extinction law (F99; \(R_V=3.1\)) and an SMC Bar law (G03; \(R_V=2.74\)) at \(z=1.216\). Dashed curves and shaded bands illustrate the effect of varying the fitted \(A_V\) by \(\pm1\sigma\) (with the power-law normalisation and slope refit at fixed \(A_V\)). The two laws are nearly indistinguishable across the observed optical bands, and the host-frame extinction is weakly constrained (\(A_V \sim 0.2\)).
    }
    \label{fig:ag_sed}
\end{figure}

All measurements were corrected for Galactic foreground extinction~\citep{Schlafly2011}. We subtracted the DECam host flux in each band from the corresponding LT flux, propagated the statistical errors in quadrature, and added a \(5\%\) fractional uncertainty in flux to account for cross-calibration and host-subtraction systematics. The resulting host-subtracted flux densities are shown in Fig.~\ref{fig:ag_sed} as a function of observed frequency (bottom axis; the top axis gives the corresponding rest-frame wavelength at \(z=1.216\)). The \(z\)-band point is plotted for reference but excluded from the fit because its uncertainty is dominated by low S/N and host-subtraction systematics.

We model the SED as a power law modified by dust extinction in the host galaxy,
\begin{equation}
    F_\nu(\nu_{\rm obs}) = C
    \left(\frac{\nu_{\rm obs}}{\nu_0}\right)^{-\beta}
    10^{-0.4\,A_V\,k(\lambda_{\rm rest})},
    \label{eq:dust_pl}
\end{equation}
where \(C\) is a normalisation at reference frequency \(\nu_0\), \(\beta\) is the intrinsic optical spectral index, \(A_V\) is the host-frame \(V\)-band extinction, and \(k(\lambda_{\rm rest})\equiv A_\lambda/A_V\) is the extinction curve evaluated at \(\lambda_{\rm rest}=c/[\nu_{\rm obs}(1+z)]\). We consider a Milky Way law \citep[][F99]{Fitzpatrick1999} with fixed \(R_V=3.1\) and an SMC Bar law \citep[][G03]{Gordon2003} with \(R_V=2.74\), as implemented in \texttt{dust\_extinction}~\citep{Gordon2024}. We do not consider the M14/LMC family further because it is undefined below \(\lambda_{\rm rest}\sim3000\)~\AA, whereas at \(z=1.216\) the observed \(g\) and \(r\) bands probe \(\lambda_{\rm rest}\lesssim3000\)~\AA.

To reduce the strong \(\beta\)--\(A_V\) degeneracy inherent to three-band photometry, we incorporate an X-ray-informed prior on \(\beta\). Our time-resolved \swiftxrt\ spectroscopy yields photon indices \(\Gamma_X\simeq1.6{-}2.2\), with a weighted mean \(\langle\Gamma_X\rangle\approx1.9\), i.e. \(\beta_X\approx0.9\) (Sec.~\ref{subsec:xrt}). In the slow-cooling synchrotron model the spectrum steepens by \(\Delta\beta=0.5\) across the cooling break, so a physically plausible configuration is \(\nu_{\rm opt}<\nu_c<\nu_X\), which would imply \(\beta_{\rm opt}\approx\beta_X-0.5\simeq0.4\). We therefore adopt a broad Gaussian prior \(\beta \sim \mathcal{N}(0.40,\,0.30)\), which also allows the possibility that optical and X-rays lie on the same spectral segment. We fit Eq.~\ref{eq:dust_pl} in flux space using \(\chi^2\) minimisation (\texttt{scipy.optimize.least\_squares}; \citealt{Virtanen2020}), with the prior implemented as an additional term in the residual vector. The XRT flux itself is not used in the optical extinction fit.

Fitting the host-subtracted \(gri\) SED yields consistent results for both extinction laws (Fig.~\ref{fig:ag_sed}). For the MW law we obtain \(A_V^{\rm host}=0.22\pm0.18\)~mag and \(\beta_{\rm opt}=0.28\pm0.30\) (\(\chi^2/{\rm dof}\approx5.67/1\)); for the SMC law we find \(A_V^{\rm host}=0.20\pm0.19\)~mag and \(\beta_{\rm opt}=0.31\pm0.30\) (\(\chi^2/{\rm dof}\approx6.14/1\)). Given that the fit is constrained by only three photometric points (one degree of freedom) and is sensitive to host-subtraction and cross-calibration systematics, these numerical values should be regarded as indicative rather than definitive. The relatively large \( \chi^2 \) values and the sparse three-point SED emphasise that \( A_V^{\rm host} \) is only weakly constrained and sensitive to systematics (Fig.~\ref{fig:ag_sed}). Nonetheless, both laws point to low-to-moderate extinction and remain compatible with \(A_V^{\rm host}\approx0\) within \(\sim1\sigma\).

We also investigated joint optical+X-ray SED fits at the same epoch imposing \(\beta_X=\beta_{\rm opt}+0.5\), as expected for a standard slow-cooling synchrotron spectrum when the cooling break lies between the optical and X-ray bands~\citep[e.g.][]{Sari1998,Granot2002}. However, at \(t\simeq2\)~d the contemporaneous 1~keV flux is only a marginal detection and does not meet the quality cuts adopted for our X-ray analysis. Including this point does not materially change the best-fitting \(A_V\), but it introduces additional assumptions (e.g. the location of \( \nu_c \), the validity of the fixed \( \Delta \beta=0.5 \) relation, and potential X-ray absorption/cross-calibration systematics) that can dominate the inference when the X-ray constraint is weak. We therefore adopt the optical-only fit with the XRT-informed \(\beta\) prior as our baseline.

\subsubsection{Radio afterglow brightness in context}
\label{subsubsec:ag_radio_context}

\begin{figure}
    \centering
    \includegraphics[width=\columnwidth]{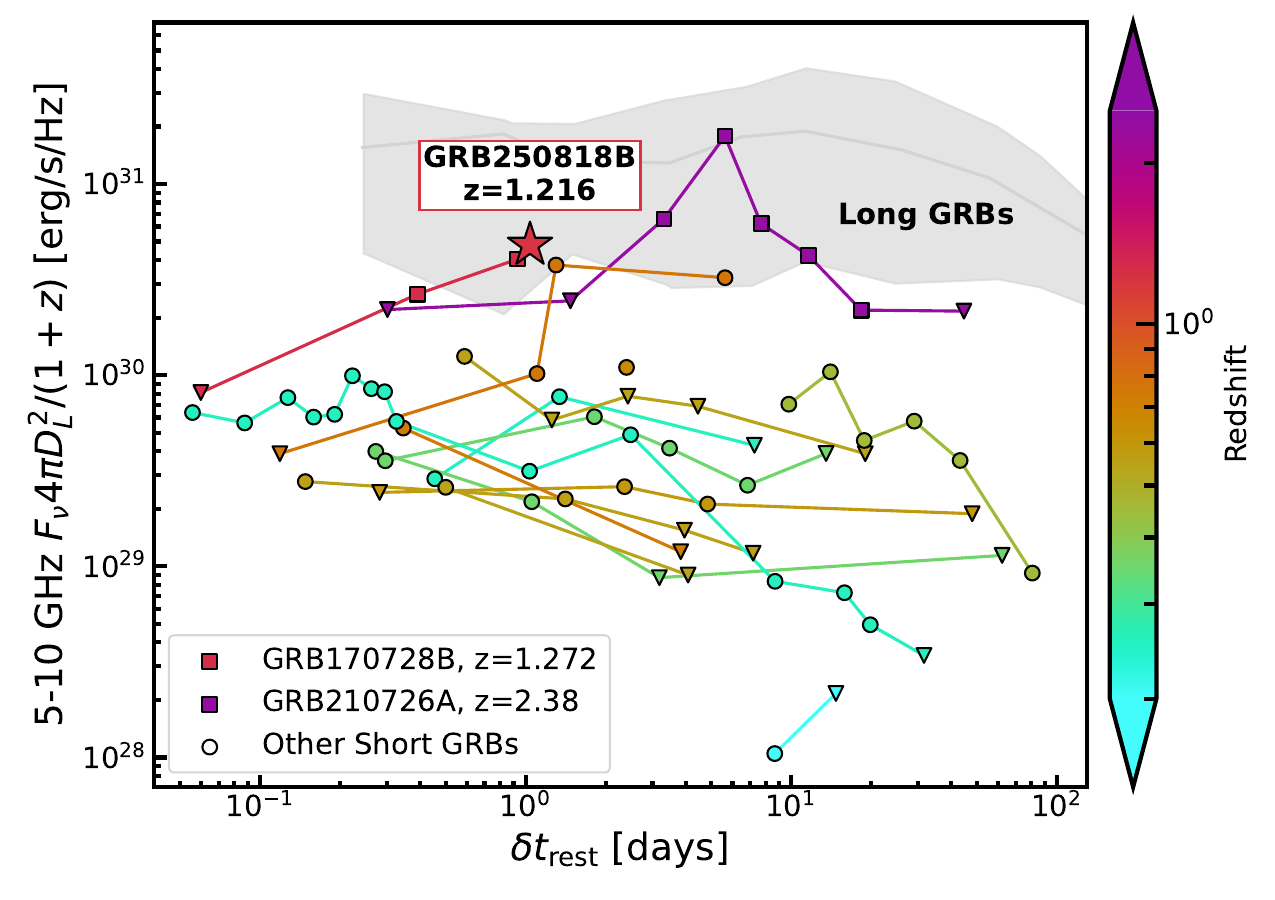}
    \caption{5--10~GHz (observer frame) afterglow luminosity vs rest frame time of radio detected short GRBs \citep{2005Natur.438..988B, 2006ApJ...650..261S, 2006MNRAS.367L..42P, 2014ApJ...780..118F, 2015ApJ...815..102F, 2019ApJ...883...48L, 2021ApJ...906..127F, 2022ApJ...935L..11L, 2024ApJ...970..139S, 2025ApJ...982...42S, 2025ApJ...994....5A, 2025arXiv251019132A, 2025GCN.39501....1G, 2025GCN.41046....1R}, with points colored by redshift \citep[redshifts from ][]{2022ApJ...940...56F, 2024ApJ...970..139S, 2024ApJ...962....5N, 2025arXiv251019132A, 2025ApJ...982...42S}. Triangles represent \( 3\sigma \) upper limits. The 10~GHz afterglow of GRB\,250818B \citep{Ricci2025gcn41455} is represented as a star, and the two other radio detected short GRBs at \( z > 1 \) (\grb{170728B} at \( z = 1.272 \) and \grb{210726A} at \( z = 2.38 \)) are represented as squares. Also shown is the typical radio afterglow range for long GRBs \citep[shaded grey,][]{2012ApJ...746..156C}.
    }
    \label{fig:radio_comparison}
\end{figure}

\grb{250818B} is only the third short GRB to have a detected radio afterglow at \( z > 1 \), with the other two short GRBs being \grb{170728B} \citep[with a spectroscopic redshift of \( z = 1.272 \pm 0.002 \);][]{2022ApJ...940...56F} and \grb{210726A} \citep[with a photometric redshift of \( z=2.38^{+0.39}_{-0.75} \);][]{2024ApJ...970..139S}. As a result, at a spectroscopic redshift of \( z = 1.216 \), the radio afterglow of \grb{250818B} is among the most luminous in the short GRB population, with a 10~GHz luminosity of \( \sim 5 \times 10^{30}~{\rm erg}~{\rm s}^{-1}~{\rm Hz}^{-1} \) at a rest-frame time of \( \delta t_{\rm RF} \approx 1\)~days, based on the 10~GHz measurement reported in GCN~41455 (no formal uncertainty was provided)~\citep{Ricci2025gcn41455}. In comparison, most radio detected short GRBs have 10~GHz afterglow luminosities of \( \sim 2 \) -- \( 8\times 10^{29}~{\rm erg}~{\rm s}^{-1}~{\rm Hz}^{-1} \) at similar rest-frame times (Fig.~\ref{fig:radio_comparison}). \grb{170728B} and \grb{250221A} \citep[spectroscopic redshift of \( z = 0.768 \);][]{2025arXiv251019132A} have comparable radio afterglow luminosities to \grb{250818B} at \( \delta t \approx 1\)~day, whereas \grb{210726A} peaked at \( \sim 2 \times 10^{31}~{\rm erg}~{\rm s}^{-1}~{\rm Hz}^{-1} \) at a rest frame time of \( \delta t_{\rm RF} \approx 6\)~days (Fig.~\ref{fig:radio_comparison}). Several of the most radio-luminous short GRBs have been modelled with additional components beyond a single forward shock (e.g. energy injection and/or reverse-shock contributions; see Sec.~\ref{sec:discussion} for further discussion).
%\textcolor{red}{Comment from Genevieve: It might be worth noting that 210726A and 250221A both had energy injection/reverse shocks invoked to explain their radio behavior, but so have most radio detected short GRBs, even at lower redshifts. See \citep{2025ApJ...982...42S, 2024ApJ...970..139S, 2025ApJ...994....5A} for some discussions on the invocation of RS/energ injections in radio detected short GRBs.}
% Notably, both \grb{210726A} and \grb{250221A} required additional components beyond a single forward shock to explain their radio behaviour, commonly attributed to energy injection and/or a reverse-shock contribution \citep[e.g.][]{2025ApJ...982...42S,2024ApJ...970..139S,2025ApJ...994....5A}. More broadly, energy injection (refreshed shocks) and reverse-shock emission have been invoked for a substantial fraction of radio-detected short GRBs, suggesting that the radio-bright subset may preferentially include events with prolonged central-engine activity and/or significant ejecta stratification.
The MeerKAT detection of \grb{250818B} adds to the growing population of short GRBs securely detected at \( \lesssim 5\)~GHz, now totaling 7 \citep[GRBs\,140903A, 210726A, 230217A, 231117A, 250520A, 250704B, and 250818B;][]{2014GCN.16815....1N, 2024ApJ...970..139S, 2024MNRAS.532.2820C, 2024ApJ...975L..13A, 2025ApJ...982...42S, 2025GCN.40546....1S, 2025GCN.41060....1S, 2025ApJ...994....5A}. A tentative very-low-frequency radio counterpart has also been reported for \grb{201006A}~\citep{Rowlinson2024}. The late-time rise of the 1.3~GHz afterglow is reminiscent of \grb{210726A} \citep{2024ApJ...970..139S}, and highlights the utility of low-frequency (\( \lesssim 5~\)~GHz) observations extended to later times (\( \gtrsim 10\)~days) to capture the afterglow peak as the synchrotron spectrum evolves to lower frequencies~\citep[e.g.][]{Sari1998,Granot2002}. In particular, when prompt-emission constraints are weak or unavailable, the afterglow becomes the primary handle on the explosion energetics and environment. Locating the low-frequency peak and its time evolution anchors the synchrotron spectrum (i.e. the characteristic frequency and flux scale), helping to constrain the kinetic energy and circumburst density and to reduce degeneracies among the microphysical parameters in broadband modelling.

Measuring the peak time and flux at these frequencies provides direct leverage on the afterglow energetics and circumburst environment (and helps break degeneracies in broadband modelling), which is particularly valuable when prompt-emission constraints are weak or unavailable.

\subsubsection{\texttt{afterglowpy} Baseline Modelling}
\begin{figure}
    \centering
    \includegraphics[width=\columnwidth]{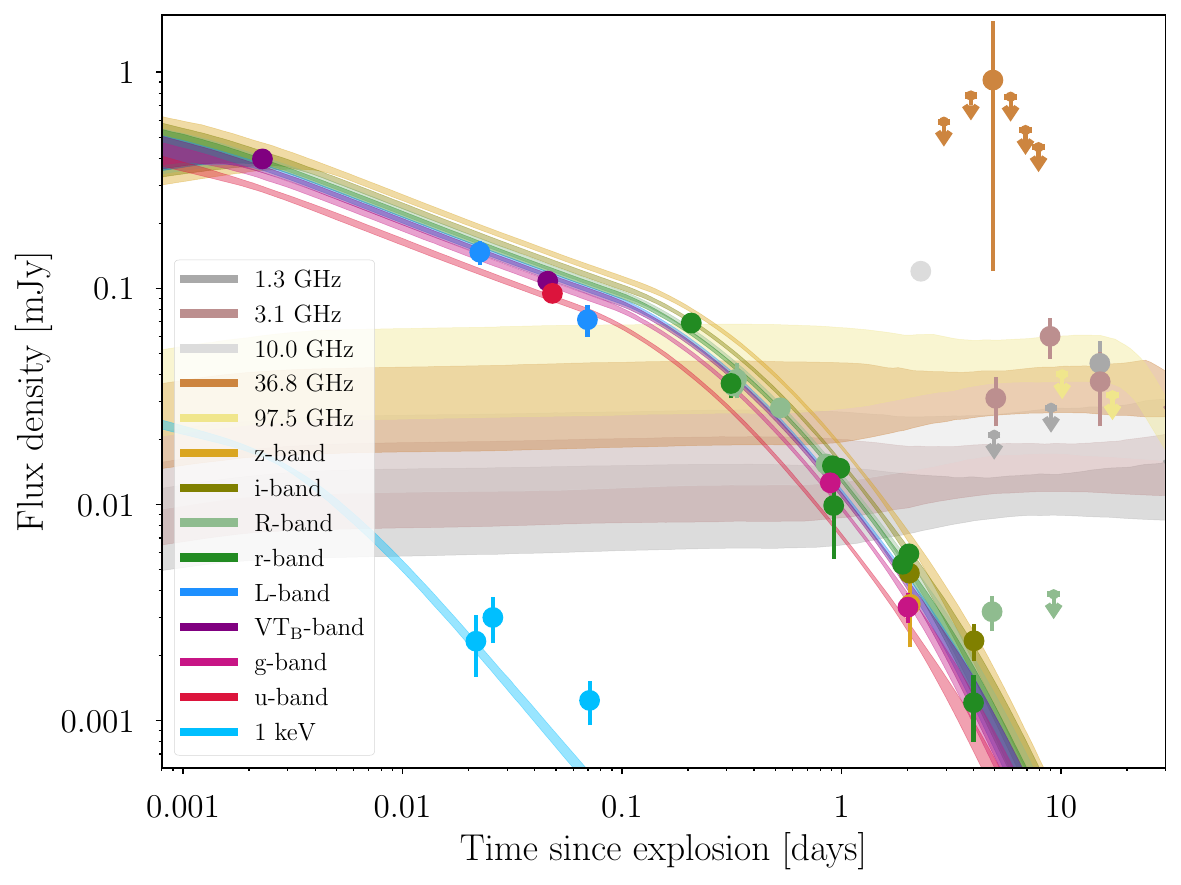}
    \caption{The 1$\sigma$ confidence interval for light curves drawn from the posterior sample for a simple top-hat afterglow model from \texttt{afterglowpy}.}
    \label{fig:lc_afterglowpy}
\end{figure}

A rapid afterglow modelling tool is provided by \texttt{afterglowpy} \citep{ryan2020}, and in combination with \texttt{redback} \citep{sarin2024} this enables end-to-end modelling and inference. Using the available detections and upper limits, we fit a simple uniform (top-hat) jet model to the data. We sample the posterior with the nested sampler \texttt{nessai} \citep{williams2023}, adopting the priors listed in Table~\ref{tab:pri_post}. The resulting \(1\sigma\) credible intervals for each observing band are shown in Fig.~\ref{fig:lc_afterglowpy}.

\begin{table*}
    \centering
    \caption{The model parameter prior distributions and constraints for the \texttt{afterglowpy} and the \texttt{twocomponent\_refreshed} model fit to data via \texttt{redback} using the \texttt{nessai} sampler. The model parameters are: \( \theta_{\rm obs} \), the observer angle from the jet central axis; \( E_{\rm iso} \), the initial isotropic equivalent jet kinetic energy for a point at the jet axis; \( \theta_c \), the jet core half-opening angle; \( \theta_j \), the maximum angular extent of the jet; \( n_0 \), the ambient particle number density; \( p \), the accelerated electron distribution index within the afterglow shock; \( \varepsilon_e \), the fraction of energy in shocked electrons; \( \varepsilon_B \), the fraction of energy in the shock-induced magnetic field; \( \xi_N \), the shocked electron, synchrotron participation fraction; \( \Gamma_0 \), the initial bulk Lorentz factor of the jet core; \( \gamma_{\rm col} \), the Lorentz factor of the impulsive shell when energy injection (refreshed) commences; \( e_t \), the fractional increase in the energy of the core due to energy injection; \( \Gamma_{0,2C} \), the initial bulk Lorentz factor of the second, sheath component; \( e_s \), the fraction of the core isotropic kinetic energy in the sheath.}
    \label{tab:pri_post}
    \begin{tabular}{c|c|ccc|c|ccc}
        & & & \texttt{Afterglowpy} & & & & \texttt{twocomponent\_refreshed}  & \\
       Parameter  & \hspace{1cm} & Prior & Condition & Posterior & \hspace{1cm} &  Prior & Condition & Posterior \\
       \hline
       $\theta_{\rm obs}$ (rad) & & $0, \pi/2$  & $\theta_{\rm obs}/\theta_j \leq 3/2$ & $0.07^{+0.03}_{-0.02}$ && 0 & fixed & 0 \\
       $\log E_{\rm iso}$ (erg) & & $44, 54$ & - & $52.48^{+0.23}_{-0.22}$ && $44, 54$ & - & $52.60^{+0.12}_{-0.13}$ \\
       $\theta_c$ (rad) & &  - & - & - && $0, 0.1$ & $\theta_c < \theta_j$ & $0.07^{+0.01}_{-0.01}$ \\
       $\theta_j$ (rad) & & $0, 0.3$ & - & $0.07^{+0.03}_{-0.02}$ && $0.02, 0.12$ & $\theta_j > \theta_c$ & $0.10^{+0.01}_{-0.02}$ \\
       $\log n_0$ (cm$^{-3}$) & & $-5,2$ & - & $-0.21^{+0.93}_{-0.67}$ && $-5,2$ & - & $0.56^{+0.07}_{-0.07}$ \\
       $p$ & & $2, 3$ & - & $2.05^{+0.03}_{-0.02}$ && $1.4, 3.1$ & - & $1.64^{+0.05}_{-0.05}$\\
       $\log \varepsilon_e$  & & $-5, 0$ & $\varepsilon_e + \varepsilon_B \leq 0.9$ & $-0.26^{+0.13}_{-0.19}$ && $-1$ & fixed & $-1$ \\
       $\log \varepsilon_B$ & & $-5, 0$ &  $\varepsilon_e - \varepsilon_B > 0.0$ & $-1.12^{+0.43}_{-0.62}$ && $-2$ & fixed & $-2$ \\
       $\xi_N$ & & $0.1, 1$ & - & $0.16^{+0.08}_{-0.04}$ && $1$ & fixed & $1$\\
       $\Gamma_0$ & & $100, 2000$ & - & $1055^{+634}_{-634}$ && $40,150$ & - & $130.9^{+9.6}_{-9.7}$\\
       $\gamma_{\rm col}$ & & - & - & - && $2,50$ & $ \gamma_{\rm col} < \Gamma_0/2$ & $41.9^{+5.6}_{-7.7}$ \\
       $e_t$ & & -  & - & - && $1,50$ & - & $25.81^{+16.01}_{-16.07}$ \\
       $\Gamma_{0,2C}$ & & - & - & - && $3,20$ & - & $6.0^{+1.9}_{-1.2}$ \\
       $e_s$ & & - & - & - && $0.1,10$ & - & $1.03^{+1.72}_{-0.74}$ \\
       \hline
        & & Information & & $\ln$ Bayes & & Information & & $\ln$ Bayes \\
        & & 25.82 & & $-42.537$ & & 21.98 & & $42.537$ \\
       \hline
    \end{tabular}
    
\end{table*}

Although these fits return parameters that are consistent with expectations for GRB afterglows, it should be noted that the posterior distribution for \( p \) is pushing up against the imposed lower boundary of \( p=2 \). This suggests that, within this simple top-hat forward-shock model family, the data would prefer a harder electron spectrum (formally \( p<2\)); we therefore interpret this as an indication of model tension rather than a precise physical constraint on \(p\). The baseline \texttt{afterglowpy} top-hat model does not reproduce the full X-ray evolution, with noticeable mismatch after the first data point (see Fig.~\ref{fig:lc_afterglowpy}).
The optical data from \( \sim 3 \)~days are additionally in excess of the model light curve, suggesting either an additional component to the emission at late times or limitations of the simple top-hat forward-shock description.
The radio data and limits sit well above the model radio posterior distribution, which does not violate the limits used; however, the model underpredicts by \( > 1\sigma \) uncertainty the 36.8~GHz detection at \( \sim 4 \) -- \( 5 \)~days, noting that this detection has a large uncertainty.

\subsubsection{\texttt{redback} energy-injection test and model comparison}
%\sergey{Add energy injection, compare to the baseline via the Bayes factor; interpret the suspected break. Or we can merge it with the previous section under the title of "Afterglow modelling". Or separate it here, but discuss together in a Discussion section. 20nov2025: I can start this section with where the idea of energy injection came from (SBPL fit)}

%\gavin{This is in-progress, however, is taking far longer than anticipated. I have run a afterglowpy injection model, however the injection scenario only allows an end time and not a start time, assuming injection commences immediately. This does not let the refreshed shock model to be tested, else a magnetar. The fit is analogous to the tophat fit.}

\begin{figure}
    \centering\includegraphics[width=\columnwidth]{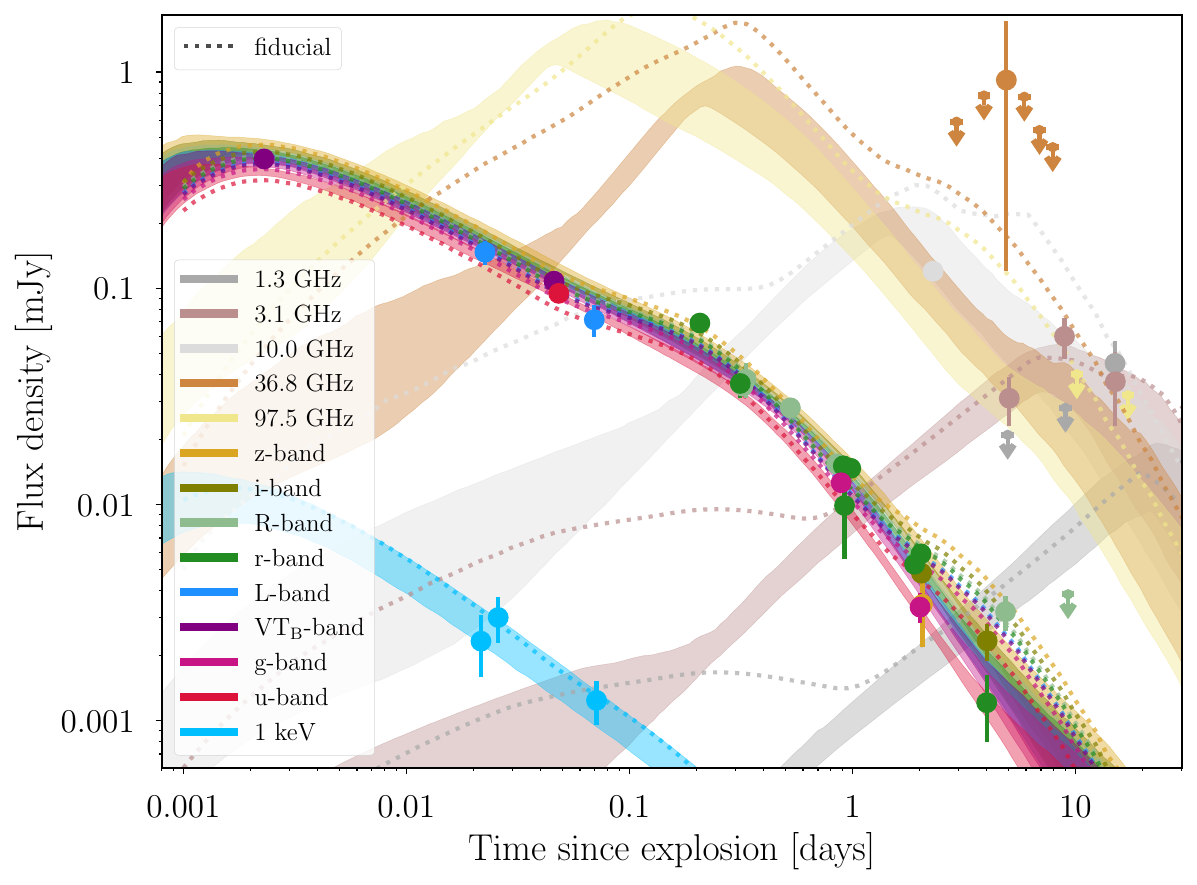}
    \caption{A refreshed shock and two-component structured jet where the second component is a narrow, energetic and baryon-loaded (lower Lorentz factor) sheath that surrounds the rapid core. Shaded regions indicate the \( 1\sigma \) confidence intervals inferred from 250 randomly drawn light curves from the posterior distribution. The x-axis shows the observer time in days and the y-axis the flux density in units of milli-Jansky. The fiducial proof-of-principle model, with parameters listed in the text, is shown as a dotted line. The jet has a refreshed shock episode early in the deceleration, characterised by a collision at \( \sim 0.05 \)~days in the observer frame. The sheath begins to dominate emission at optical and radio from \( \gtrsim 1 \) -- \( 2 \)~days. Sheath emission dominating at later times in this structured jet CSE is due to the larger isotropic equivalent energy within the second component than the jet core.}
    \label{fig:2C_injection}
\end{figure}

The variability within the afterglow indicates that there may have been a period of energy injection.
Additionally, the late data appears to indicate either a further period of injection or an energetic wider component coming into view~\cite[i.e.,][a sheath surrounding a hollow/lower-energy jet]{takahashi2021}.
To demonstrate the viability of this model, we pick fiducial model parameters informed via the \texttt{afterglowpy} fit, however, setting \( p = 1.8 \) (i.e., \( p<2 \)), isotropic kinetic energy, \( E_{\rm iso} = 3.15 \times 10^{52}\)~erg, an ambient particle number density, \( n_0 = 1 \)~cm\( ^{-3} \), and fixing microphysical parameters \( \varepsilon_e = 0.1 \), \( \varepsilon_B = 0.01 \), \( \xi_N = 1 \), and the bulk Lorentz factor, \( \Gamma_0 = 140 \), for the fiducial illustrative model shown as the dotted curve in Fig.~\ref{fig:2C_injection}. We adopt \( \varepsilon_B = 0.01 \) as a conventional fiducial choice often used when the microphysics are not directly constrained~\citep[e.g.][]{2015ApJ...815..102F}; however, published estimates of \( \varepsilon_B \) span several orders of magnitude~\citep[e.g.][]{Hennessy2025}. The normalisation of this illustrative curve (and the implied energy/density scalings) should therefore be regarded as conditional on the assumed microphysical parameters.

We use a refreshed shock, two-component model~\citep[see e.g.,][]{lamb2017, lamb2018, lamb2020} from \texttt{redback}, the \texttt{twocomponent\_redback\_refreshed} from their model library, and set additional parameters as:
the Lorentz factor of the impulsive shell at collision, \( \gamma_{\rm col} = 28 \), the total factor by which energy increases, \( e_t = 3.5 \), assuming an instantaneous injection, and a second component with a bulk Lorentz factor, \( \Gamma_{0,{\rm 2C}}=12 \), with a factor, \( e_s = 2.5 \), of the isotropic equivalent energy within the core region.
In this two-component configuration, a relativistic core is surrounded by a slower, baryon-loaded sheath extending to \( \theta_j \); for this fit, we assume an on-axis viewing geometry (\( \theta_{\rm obs} = 0 \)).
For the fiducial proof-of-principle model (dotted curve in Fig.~\ref{fig:2C_injection}), we adopted narrow opening angles (\( \theta_c \sim 2\fdg2 \), \(\theta_j \sim 2\fdg5 \)) to reproduce the apparent break in the light curve.
For the posterior fit, the sampler prefers broader angles; we therefore quote the fitted values from Table~\ref{tab:pri_post} (median \( \theta_c \approx 0.07 \)~rad and \( \theta_j \approx 0.10 \)~rad).
%The core is $\theta_c\sim 2\fdg 2$ and the out component is a narrow sheath that extends to $\theta_j \sim 2\fdg 5$, while the model is assumed to be viewed on-axis.
This model is shown against the data in Fig.~\ref{fig:2C_injection}; additionally plotted are the confidence intervals inferred via 250 random draws from the posterior distribution fit for this model -- the prior was chosen based on our fiducial parameters and is listed, along with posterior ranges, in Table~\ref{tab:pri_post}.

The sampler, \texttt{nessai}, provides a log Evidence and an information score~\citep[see,][for details]{williams2021}; despite the increased number of parameters for the \texttt{twocomponent\_redback\_refreshed} model, this is the preferred model.
The respective information scores and the log Bayes factor for model preference between the two are shown -- where the preferred model will have the lowest value and the most positive difference in log evidence, respectively (see Table~\ref{tab:pri_post}).

\subsection{Host Galaxy}
\subsubsection{Association}\label{sec:host_association}
\begin{figure}
    \centering
    \includegraphics[width=\columnwidth]{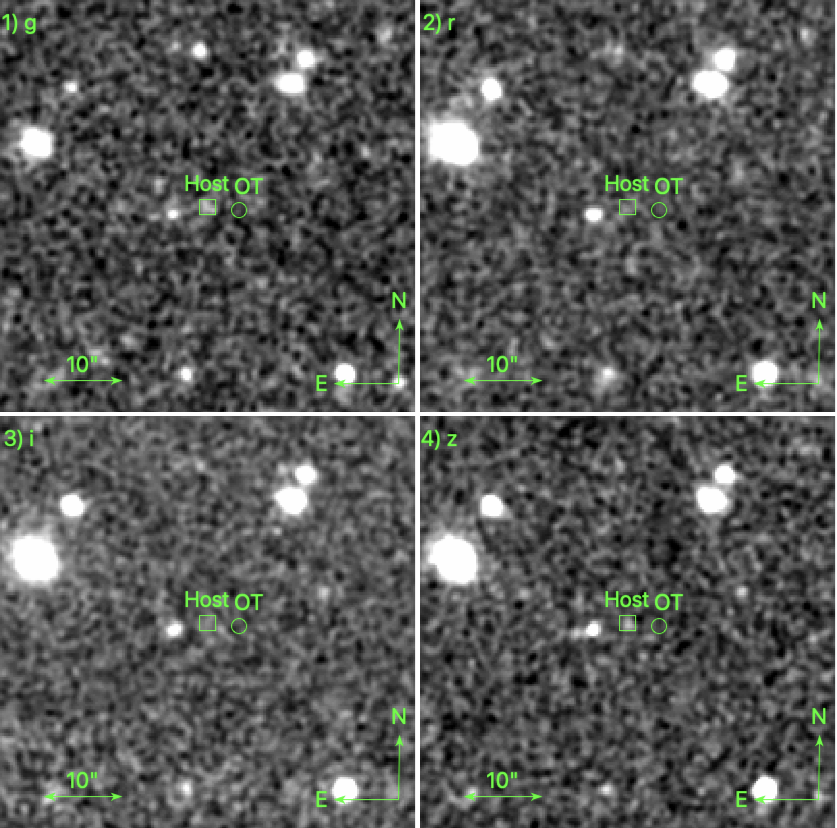}
    \caption{
    Legacy Survey DR10 image cutouts of the \grb{250818B} field in the optical bands: (1) \(g\), (2) \(r\), (3) \(i\), and (4) \(z\). The green circle marks the position of the optical transient (OT), while the green square marks the position of the LS~DR10 candidate host galaxy (objid~5790). A faint flux excess is visible at the candidate-host location, most clearly in \(g\), and is less significant in \(r\) and \(i\), consistent with the catalogue photometric uncertainties. North and east are indicated, and the scale bar corresponds to \(10\arcsec\).
    }
    \label{fig:host_lsdr10_griz}
\end{figure}

We searched the Legacy Survey (LS~DR10; \citealt{Dey2019}) catalogue around the position of \grb{250818B} and identified a faint galaxy at \( \alpha = 03^{\rm h}04^{\rm m}13\fs79 \), \( \delta = -03\degr07\arcmin30\farcs4 \) (J2000), classified by \textit{The Tractor}~\citep{Lang2016} as a round exponential source (\texttt{REX}, \texttt{objid}~5790). The location of this candidate host is shown in Fig.~\ref{fig:host_lsdr10_griz}. The LS~DR10 \(r\)-band catalogue magnitude is \(r_{\rm AB}=24.74\pm0.36\), corresponding to a marginal detection (\(\mathrm{S/N}\approx3\)). Its angular separation from the GOTO afterglow position (\( \alpha = 03^{\rm h}04^{\rm m}13\fs52 \), \( \delta = -03\degr07\arcmin30\farcs8 \)) is \( R_{\rm off} = 4\farcs03 \). Adopting \( z = 1.216 \) (from afterglow absorption), this corresponds to a projected physical offset of
\( R_{\rm off} = 33.51~{\rm kpc} \)\footnote{We assume a flat \(\Lambda\)CDM cosmology with \(H_{0}=70~\mathrm{km~s^{-1}~Mpc^{-1}}\), \(\Omega_{\rm m}=0.3\), and \(\Omega_{\Lambda}=0.7\), and quote magnitudes in the AB system unless stated otherwise.}.

To quantify the probability of the host galaxy association to this event, we follow the Bayesian association method of \citet{Ackley2025}. In this framework, for a set of $N$ candidate hosts $G = \{ G_j\}_{j=1}^N $ within a search region, the probability that $G_j$ is the true host is
\begin{equation}
p(G_j\,|\, d) = \frac{p(d\,|\,G_j)\,\pi(G_j)}{\sum_{k=1}^N p(d|G_k)\pi(G_k) + p(d|\varnothing)\,\pi(\varnothing)},
\label{eq:host}
\end{equation}
where $d$ is the GRB afterglow position and redshift, $p(d\,|\,G_j)$ is the likelihood, $\pi(G_j)$ is the prior on each galaxy, and $p(d|\varnothing)\,\pi(\varnothing)$ accounts for the possibility that the true host is not represented by any catalogued object, either because it is outside of the search region or below detection thresholds. We adopt a uniform prior for the $N$ candidates in the search cone, $\pi(G_j) = 1/N$.

The host-association likelihood is factorized into an offset term and a redshift-consistency term,
\begin{equation}
    p(d\,|\,G_j) = p(r_{\rm DLR} \,|\,G_j)\,\times\, p(z\,|\,G_j) \equiv p_{\mathrm{offset}}(G_j) \times p_{z}(G_j),
\end{equation}
where $r_{\rm DLR} \equiv \Delta\theta/\mathrm{DLR}$ is the dimensionless host-normalised separation, $\Delta\theta$ is the angular separation between the afterglow and the galaxy centroid, and DLR is the directional light radius of the galaxy in the direction of the afterglow, computed from the galaxy morphology parameters calculated by \textit{the Tractor}~\cite{Lang2016}.

For the offset likelihood, we marginalise over the unknown true host-normalised offset $r_0$ as
\begin{equation}
    p_{\rm{offset}}(G_j) = \sum_{c}\pi(c)\,\int^{\infty}_{0} p(r_{\rm DLR} \,|\,r_0, \sigma_{r_0})\,\pi(r_0\,|\,G_j, c) \,dr_0,
\label{eq:host:offset}
\end{equation}
where $c\in\{\mathrm{short, long}\}$ is the class condition.
Assuming isotropic 2D Gaussian astrometric uncertainties, the radial likelihood $p(r_{\rm DLR} \,|\,r_0, \sigma_{r_0})$ is a Rice distribution with non-centrality $r_0$ and scale $\sigma_{r_0} \equiv \sigma_{\rm ast}/\mathrm{DLR}$. The astrometric error $\sigma_{\rm ast}$ includes both centroid and registration terms, each conservatively estimated to be on the order of $\sim0\farcs2$. We note that for the host candidates considered here $\sigma_{r_0} \ll \Delta\theta$, and thus the association is dominated by the offset prior.

We adopt a class-conditional Gamma-family prior to the host-normalised offset
\begin{equation}
\pi (r_0\,|\, c) = \mathrm{Gamma}(r_0; \lambda_c,\kappa_c) \propto r_0^{\lambda_c-1}\,\exp(-r_0/\kappa_c)
\label{eq:host:offsetprior}
\end{equation}
with hyperparameters chosen to reproduce the observed host-normalised offset distributions of short and long GRBs from \citep{Fong2022,Blanchard2016}. In our analysis we fix $\lambda_c=2$ and allow $\kappa_c$ to vary with class $c$. To avoid implicitly encoding a progenitor class assumption in our host selection, we search for sources within a large angular radius ($R_{\max} = 0.1^{\circ}$), for which truncation of the offset prior is negligible for both classes.

For the redshift term, the host galaxy candidates have photometric redshift estimates, while the GRB afterglow has a precise spectroscopic redshift. Therefore, the redshift likelihood $p(z\,|\,G_j)$ treats the GRB afterglow redshift as a delta function and the host redshift as a Gaussian with uncertainty given by the photometric redshift uncertainty. In practice, this reduces to evaluating the host redshift probability distribution function at the GRB afterglow redshift, $p_{z}(G_j) \propto p_j(z)$.

In addition to the set of catalogue candidates, the normalisation term of Eq.~\ref{eq:host} includes an uncatalouged/missing host term that accounts for the possibility that the true host is not present in the LS~DR10 source list. For \grb{250818B}, we model this contribution using a limiting-magnitude completeness prescription as implemented in \citet{Ackley2025}. We define a catalogue completeness fraction $C(z)$ as the probability that a typical host at $z=1.216$ would be brighter than the local LS~DR10 depth at the burst position. Using the local $r$-band limiting magnitude $m_{\rm lim}$ (Sec.~\ref{subsubsec:host_limit_coincident}), we convert to an absolute threshold $M_{\rm lim} \approx m_{\rm lim} -{\rm DM}(z)$. Given a Schechter luminosity function, we compute $C(z)$ by integrating the galaxy luminosity function above $M_{\rm lim}$, where
\begin{equation}
    C(z) = \frac{\int^{M_{\rm lim}}_{-\infty} \phi(M) dM}{\int^{M_{\rm max}}_{-\infty} \phi(M) dM}.
\end{equation}
The faint/missed host probability is then 
\begin{equation}
    P_{\rm faint} = 1 - C(z).
\end{equation}
We note that we keep $C(z)$ as a term independent of the GRB source type and do not take into consideration any weights to the luminosity function.
We also include an outside search cone term that accounts for the fraction of the offset prior that lies beyond the maximum search radius, $R_{\max}$,
\begin{equation}
    P_{\rm out} = \int_{R_{\max}}^{\infty} p(R)\, dR.
\end{equation}
The resulting uncatalogued host term is then approximated as $P_{\varnothing} \approx P_{\rm faint} + P_{\rm out}$.

We identify our candidate hosts by querying the LS~DR10 catalogue within a \( 0.1\degr \) cone centred on the GRB with the following filters. We select galaxy-like sources with \texttt{type}~ \(\in\{\texttt{REX,EXP,DEV,SER}\} \), \texttt{brick\_primary = 1}, \texttt{allmask\_g = allmask\_r = allmask\_z = 0}, and significant $r$-band flux (\texttt{flux\_r > 0}, \texttt{flux\_ivar\_r > 0}). For each candidate returned, we calculate the host-association probability, as shown in Table~\ref{tab:galaxies}.

Given the limiting magnitude of the catalogue, we find that none of the host galaxy candidates in our selection is strongly favoured as the host galaxy. The LS~DR10 source (\texttt{objid}~5790) has the highest association probability, $P_{\mathrm{assoc}}=0.025$, which is primarily driven by the offset likelihood term under a short-GRB offset prior assumption with moderate probabilistic support at the reported photometric redshift of the galaxy. Thus, within the LS~DR10 catalogue and in this framework, we cannot definitely identify that \texttt{objid}~5790 is the true host and instead find support for the host being outside of the galaxy catalogue, $P_{\varnothing} \approx 0.882$, where we find that $P_{\rm out} \approx 0.084$ and $P_{\rm faint} \approx 0.829$.

\begin{table*}
    \centering
\caption{Bayesian host-association probability for LS~DR10 sources within a \( 0.1\degr \) cone centered on \grb{250818B}. The columns list the sky position and morphological\texttt{type}, the host-normalised offset $r_{\rm DLR}\equiv \Delta\theta/\mathrm{DLR}$, the angular separation from \grb{250818B} in units of arcseconds, the projected physical separation at the redshift of \grb{250818B} ($z=1.216$), and the reported redshift of the host with uncertainty ($z_{\rm gal} \pm \sigma_{z_{\rm gal}}$). The probability terms represent the offset likelihood $p_{\mathrm{offset}}$, the redshift likelihood $p_{z}$, the probability that the host is missing from the catalogue $P_{\varnothing}$, and the total probability for the host association $P_{\mathrm{assoc}}$ (Eq.~\ref{eq:host}).}
    \label{tab:galaxies}
\begin{tabular}{ccccccccccccc}
RA (J2000) & Dec (J2000) & Type & $r_{\rm DLR}$ & Separation (\(\Delta\theta\), \(\arcsec\)) & Separation (kpc) & $z_{\rm gal} \pm \sigma_{z_{\rm gal}}$ & $P_{\mathrm{offset}}$ & $P_{z}$ & $P_{\varnothing}$ & $P_{\mathrm{assoc}}$ \\
\hline\hline
$03^{\rm h}04^{\rm m}13\fs79$ & $-03^{\circ}07^\prime30\farcs40$
 & REX & 15.99 & 4.03 & 33.50 & $1.10 \pm 0.32$ & 1.000 & 0.752 & 0.882 & 0.025 \\
$03^{\rm h}04^{\rm m}14\fs08$ & $-03^{\circ}07^\prime31\farcs20$ & PSF & 14.07 & 8.35 & 69.40 & $0.67 \pm 0.41$ & 0.013 & 0.538 & 0.882 & 0. \\
$03^{\rm h}04^{\rm m}12\fs82$ & $-03^{\circ}07^\prime26\farcs40$ & PSF & 19.08 & 11.32 & 94.06 & $0.75 \pm 0.21$ & 0. & 0.323 & 0.882 & 0. \\
$03^{\rm h}04^{\rm m}13\fs08$ & $-03^{\circ}07^\prime14\farcs81$ & EXP & 20.53 & 17.30 & 143.77 & $0.48 \pm 0.07$ & 0. & 0.023 & 0.882 & 0. \\
    \end{tabular}
    
\end{table*}

We compare our host association probability results against the chance-coincidence formalism of \citet{Bloom2002}. The chance-coincidence probability that an unrelated field galaxy of apparent magnitude \( \le m_r \) lies within an angular distance \( R_{\rm off} \) of the GRB is
\begin{equation}\label{eq:bloom_prob}
    P_{\rm cc} \;=\; 1 - \exp\!\left[-\pi R_{\rm off}^2 \,
    \sigma(<m_r)\right],
\end{equation}
where \( \sigma(<m_r) \) is the surface density of galaxies brighter than \( m_r \), in units of arcsec\(^{-2}\). Rather than adopting a generic number--counts model, we measure \( \sigma(< m_r) \) directly from the LS~DR10 field around \grb{250818B} (within \(0.1^\circ\) of the burst position, applying the same catalogue quality and morphology cuts used throughout this work). The resulting cumulative counts \( N(<m_r) \), shown in Fig.~\ref{fig:lsdr10_counts}, follow an approximately log--linear relation with slope \( {\rm d}\log_{10} N / {\rm d}m_r \simeq 0.32 \) over \( 22.5 \lesssim m_r \lesssim 24.5 \), consistent with standard faint galaxy number counts in the \( R \) band \citep[e.g.][]{Hogg1997}, but with an overall normalisation lower by a factor of \( \sim 2.4 \) relative to that model, likely due to our strict source selection and the limited field size.

At the magnitude of the candidate host (\(m_r=24.74\)), the empirical counts imply a surface density of \(\sigma(<m_r)\simeq 4.3\times10^{-3}~{\rm arcsec}^{-2}\) (see Fig.~\ref{fig:lsdr10_counts}), which yields \(P_{\rm cc}\approx0.195\) (range \(\approx0.18\)--\(0.20\) when propagating the \(r\)-band uncertainty). This relatively high \(P_{\rm cc}\) reflects the large angular offset and the high surface density of faint galaxies, and implies that the association is tentative rather than definitive. For comparison, using a simple Hogg-like number--counts model normalised to deep \(R\)-band surveys would increase \(P_{\rm cc}\) to \(\sim0.4\); we adopt the local empirical counts as our baseline, noting that this estimate depends on the chosen field size and the quality/morphology cuts applied to the catalogue.

We also compute \( P_{\rm cc} \) for all LS~DR10 galaxies in the same \( 0.1^\circ \) cone and rank them by increasing chance probability. The LS~DR10 object~5790 is among the lowest-\( P_{\rm cc} \) galaxies within \( \sim 40\arcsec \) of the GRB position. The only source with a formally comparable \( P_{\rm cc} \) lies at an angular distance of \( \simeq 27\arcsec \), corresponding to a projected offset of \( \simeq 225~{\rm kpc} \) at \( z=1.216 \), which is implausibly large for a GRB host~\citep[e.g.][]{Mandhai2022}. On this basis, we consider LS~DR10~\texttt{objid}~5790 to be a possible host galaxy of \grb{250818B}. We caution that the current data do not exclude an even fainter, near-coincident host galaxy candidate below the LS~DR10 detection threshold, which would reduce the inferred offset and therefore weaken host-offset-based arguments for a short-GRB classification. We note that \( P_{\rm cc} \) quantifies the probability of a chance alignment given only local number counts, whereas the Bayesian framework additionally incorporates host-transient offset priors (via DLR) and redshift consistency; this increases the relative weight of objid~5790 compared to other nearby sources.

To further constrain the redshift of the probable host galaxy, we compare its photometry to the galaxy population in the COSMOS-Web DR1 catalogue \citep{2026A&A...704A.339S}. COSMOS-Web DR1 provides photometric redshifts for over 700,000 galaxies in the Cosmic Evolution Survey (COSMOS) field, derived from deep JWST near-infrared imaging combined with extensive HST and ground-based data. In total, 37 photometric bands spanning 0.3--8 $\mu$m are used, yielding some of the most tightly constrained photometric redshifts available for any extragalactic field. 

We select COSMOS-Web galaxies whose photometric measurements are consistent with those of the probable host galaxy, requiring agreement within \( 1\sigma \) of the detected fluxes or fainter than the corresponding upper limits. This selection identifies galaxies with photometry compatible with the host, but whose redshifts are constrained by substantially richer multi-wavelength data. The resulting photometric-redshift distribution therefore provides an empirical estimate of the host-galaxy redshift. Using this approach, we obtain a median redshift of \( z = 1.118 \pm 0.202 \). This result is robust to reasonable variations in the matching criteria (e.g. adopting \( 3\sigma \) rather than \( 1\sigma \) thresholds). Notably, this is close to the measured spectroscopic afterglow redshift. 

\subsubsection{Limit on an underlying coincident host galaxy candidate}
\label{subsubsec:host_limit_coincident}
The LS~DR10 candidate host galaxy (objid~5790) is offset by \(R_{\rm off}=4\farcs03\) from the afterglow position and is detected only marginally in the LS~DR10 catalogue (\(r_{\rm AB}=24.74\pm0.36\), \(\mathrm{S/N}\approx3\)). Motivated by the relatively large offset and the proximity of the candidate to the survey detection threshold, and noting that the Keck/LRIS spectrum of the afterglow shows metal absorption at the burst redshift (imprinted on the afterglow continuum; Sec.~\ref{subsec:spectroscopy}), we assess whether the current imaging could plausibly miss a fainter, near-coincident host at (or very close to) the afterglow position.
%Motivated by the relatively large offset and the proximity of the candidate to the survey detection threshold, we assess whether the current imaging could plausibly miss a fainter, near-coincident host at (or very close to) the afterglow position. Such a host, if present, would reduce the inferred physical offset and therefore weaken host-offset-based arguments.

We use the LS~DR10 Tractor depth quantities, which are designed to quantify the limiting flux that would have been detected at a given sky position. The LS~DR10 catalogues are produced with the \textsc{Tractor} image-modelling code \citep{Lang2016}. The \texttt{psfdepth\_<band>} and \texttt{galdepth\_<band>} columns encode inverse-variance depth in flux units (\(\mathrm{nMgy}^{-2}\)), appropriate for point-source and small-galaxy detection, respectively\footnote{\url{https://www.legacysurvey.org/dr10/catalogs/}}. For an \(N\sigma\) detection threshold, the corresponding limiting flux in nanomaggies is \(f_{N\sigma}=N/\sqrt{\mathrm{depth}}\), which converts to an AB magnitude via \(m_{\rm AB}=22.5-2.5\log_{10}(f_{\rm nMgy})\).
We query the LS~DR10 Tractor catalogue within a \(1'\) radius of the afterglow position and adopt the median depth values in this region as representative of the local detection threshold.
For the \(r\) band, this yields median depths of \(\texttt{psfdepth}_r\simeq 666\) and \(\texttt{galdepth}_r\simeq 439~\mathrm{nMgy}^{-2}\), corresponding to
\(r^{\rm psf}_{3\sigma}\approx 24.84\) and \(r^{\rm gal}_{3\sigma}\approx 24.61\)
(\(r^{\rm psf}_{5\sigma}\approx 24.28\), \(r^{\rm gal}_{5\sigma}\approx 24.06\)).
We treat \(m_{\rm lim}\equiv r^{\rm gal}_{3\sigma}\) as a conservative limit for an underlying host galaxy candidate, and quote \(r^{\rm psf}_{3\sigma}\) as an optimistic (compact-source) case.

As in Section~\ref{sec:host_association} (Fig.~\ref{fig:lsdr10_counts}), we measure the cumulative surface density \(\sigma(<m_r)\) directly from LS~DR10 sources in a \(0.1^\circ\) cone around the GRB, selecting galaxy-like morphologies and applying strict quality cuts (\texttt{brick\_primary}, zero \texttt{allmask}, and positive \texttt{flux\_r} with \texttt{flux\_ivar\_r}). We then evaluate Eq.~\ref{eq:bloom_prob} for a hypothetical host located essentially at the afterglow position by substituting \(R_{\rm off}\rightarrow R_0\) and \(m_r\rightarrow m_{\rm lim}\), where \(R_0\) is a small ``coincident'' radius.
For \(R_0=0\farcs5\) (with \(0\farcs3\) in parentheses) and \(m_{\rm lim}=r^{\rm gal}_{3\sigma}=24.61\), we obtain
\begin{equation}
    P_{\rm cc}(m_r=m_{\rm lim}, R_0)
    \;\approx\;
    4.2\times 10^{-3}
    \quad
    (1.5\times 10^{-3}).
\end{equation}
Adopting the point-source limit \(r^{\rm psf}_{3\sigma}=24.84\) changes these values only slightly (\(\approx 4.3\times10^{-3}\) and \(1.6\times10^{-3}\) for \(R_0=0\farcs5\) and \(0\farcs3\), respectively). Thus, if a host galaxy were present at (or very near) the afterglow position and were bright enough to be detected at the LS~DR10 threshold, it would have a very small chance-coincidence probability and would constitute a much stronger association than the offset candidate host.

Conversely, we estimate how faint a near-coincident host galaxy would need to be for its chance-coincidence probability to be comparable to that of the offset candidate (\(P_{\rm cc}\approx0.24\); Section~\ref{sec:host_association}). Solving Eq.~\ref{eq:bloom_prob} for the required surface density gives
\begin{equation}
    \sigma_{\rm req} \;=\; \frac{-\ln\!\left(1-P_{\rm cc}\right)}{\pi R_0^2}.
\end{equation}
For \(P_{\rm cc}=0.24\), we obtain \(\sigma_{\rm req}\approx 3.5\times10^{-1}~\mathrm{arcsec^{-2}}\) at \(R_0=0\farcs5\) (\(\sigma_{\rm req}\approx 9.8\times10^{-1}~\mathrm{arcsec^{-2}}\) at \(0\farcs3\)). In contrast, our empirical LS~DR10 counts in this field reach a maximum of only \(\sigma_{\max}\approx 5.7\times10^{-3}~\mathrm{arcsec^{-2}}\) by the faintest magnitudes sampled (\(r\simeq 27.3\)). Therefore, the magnitude at which a near-coincident source would attain \(P_{\rm cc}\approx0.24\) lies far beyond the regime probed by our catalogue (and would require much deeper data), implying that any genuinely near-coincident host would have \(P_{\rm cc}\ll 0.24\).

No galaxy is detected at the afterglow position down to the local LS~DR10 limits (\(m_{\rm lim}=r^{\rm gal}_{3\sigma}\approx24.61\)), so the current data do not exclude a fainter, near-coincident host below the DR10 detection threshold. If present, such a host would reduce the inferred physical offset relative to objid~5790 and weaken offset-based arguments in favour of a short-GRB origin; we discuss the implications further in Section~\ref{sec:discussion_host}.

\subsubsection{Photometry}\label{subsubsec:host_phot}
We obtained host-galaxy photometry from the DESI Legacy Imaging Surveys DR10 Tractor catalogue~\citep{Dey2019} at the position of the candidate host (Sec.~\ref{sec:host_association}), using DECam \( griz \) and unWISE/WISE \( W1 \)-\( W4 \) bands~\citep{Flaugher2015,Dey2019}. Catalogue fluxes and inverse variances were converted to AB magnitudes and \(1\sigma\) uncertainties following the standard nanomaggy formalism. In the optical, the host galaxy candidate is detected in DECam \( g \), \( r \), and \( z \), while the \( i \)-band measurement has low signal-to-noise and is treated as an upper limit in our SED fitting. The mid-infrared WISE bands are also of low significance and are used as upper limits only. The LS~DR10 Tractor photometry is based on forward modelling of the imaging with PSF convolution, returning band-by-band model fluxes for a common source morphology; we therefore treat the DECam \( grz \) fluxes as internally consistent total-flux estimates across those bands. The SkyMapper and \textit{WISE} measurements are non-detections and are used only as upper limits; since these limits are substantially shallower than the LS~DR10 detections, the SED constraints are dominated by the DECam \( grz \) points and are insensitive to modest cross-survey aperture/PSF differences. All magnitudes listed in Table~\ref{tab:host_phot_legacy} are on the AB system and are not corrected for Galactic extinction; for SED modelling we apply the Milky Way extinction corrections described in Sec.~\ref{subsec:opt_nir}. Given the faintness of the source (\( g \simeq 24.8 \)~mag), substantially deeper host imaging and spectroscopy would require large-aperture facilities and lie beyond the scope of this work.

\subsubsection{\texttt{Prospector} SED Modelling}\label{sec:host_sed}
%\sergey{We have data from the Legacy Survey that might be used for SED modelling with \texttt{Prospector}. I’m setting it up now to see what can be done.}
%\sergey{11feb2026: I've reran it using smaller amount of parameters. I'll update the text with the figures soon.}

\begin{figure}
    \centering
    \includegraphics[width=\columnwidth]{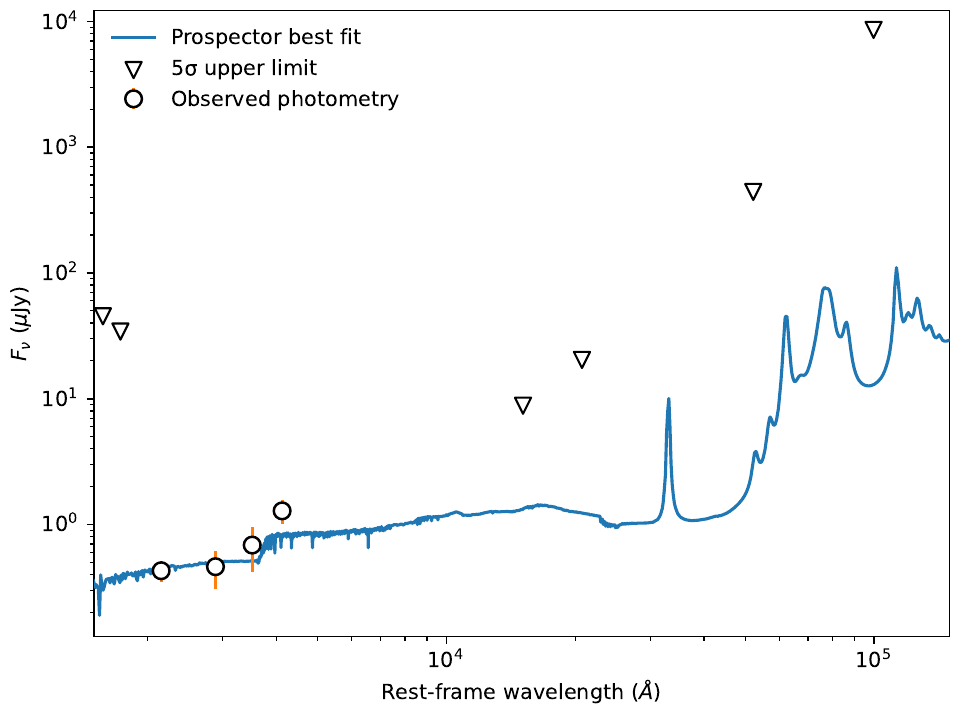}
    \caption{
    Rest-frame SED of the \grb{250818B} host with the \textsc{Prospector} best-fit spectrum (solid line). Points show DECam \(griz\) detections (AB, corrected for Galactic extinction); triangles indicate \(5\sigma\) upper limits (SkyMapper \(u,v\) and \textit{WISE} bands). Limited blue/UV and NIR leverage leaves age--dust--metallicity partially degenerate, but the overall SED is well reproduced. The structured emission at \(\gtrsim 3~\mu\)m arises from the model dust/PAH emission component in FSPS/\textsc{Prospector}; with only upper limits in the mid-IR, this part of the spectrum is effectively unconstrained and is shown for completeness only.
    }
    \label{fig:host_sed}
\end{figure}

We modelled the host galaxy of \grb{250818B} with \textsc{Prospector}~\citep{Leja2017,Johnson2021}, using the FSPS stellar-population models~\citep{Conroy2009,Conroy2010}, a Chabrier IMF~\citep{Chabrier2003}, and a Calzetti foreground dust law~\citep{Calzetti2000}. We fix the redshift to \(z=1.216\) from the afterglow absorption spectrum~\citep{Fong2025gcn41419}; therefore, the inferred host properties below are conditional on the assumption that the LS~DR10 candidate galaxy is at this redshift (see Sec.~\ref{sec:discussion_host}). The star-formation history was modelled with a parametric delayed-\(\tau\) form. As photometric constraints we used DECam \(griz\) detections from the DESI Legacy Surveys DR10 Tractor catalogue (Table~\ref{tab:host_phot_legacy}), together with \textit{WISE} \(W1\)--\(W4\) and SkyMapper \(u,v\) non-detections as \(5\sigma\) upper limits.\footnote{All measurements are AB and were corrected for Galactic extinction using \(E(B\!-\!V)_{\rm MW}=0.0633\)~\citep{Schlegel1998,Schlafly2011} and the band coefficients in Table~\ref{tab:mw_extinction_coeffs}. The upper limits are included in the fit but, given their depth relative to the DECam detections, do not materially impact the inferred posteriors.} As a check, repeating the fit using only the DECam \(griz\) detections (i.e. omitting the upper-limit bands) yields consistent posteriors for \(M_{\star,\mathrm{formed}}\) and \(A_V\). The rest-frame SED fit is shown in Fig.~\ref{fig:host_sed}; wavelengths are rest-frame unless stated otherwise.

\begin{figure}
    \centering
    \includegraphics[width=\columnwidth]{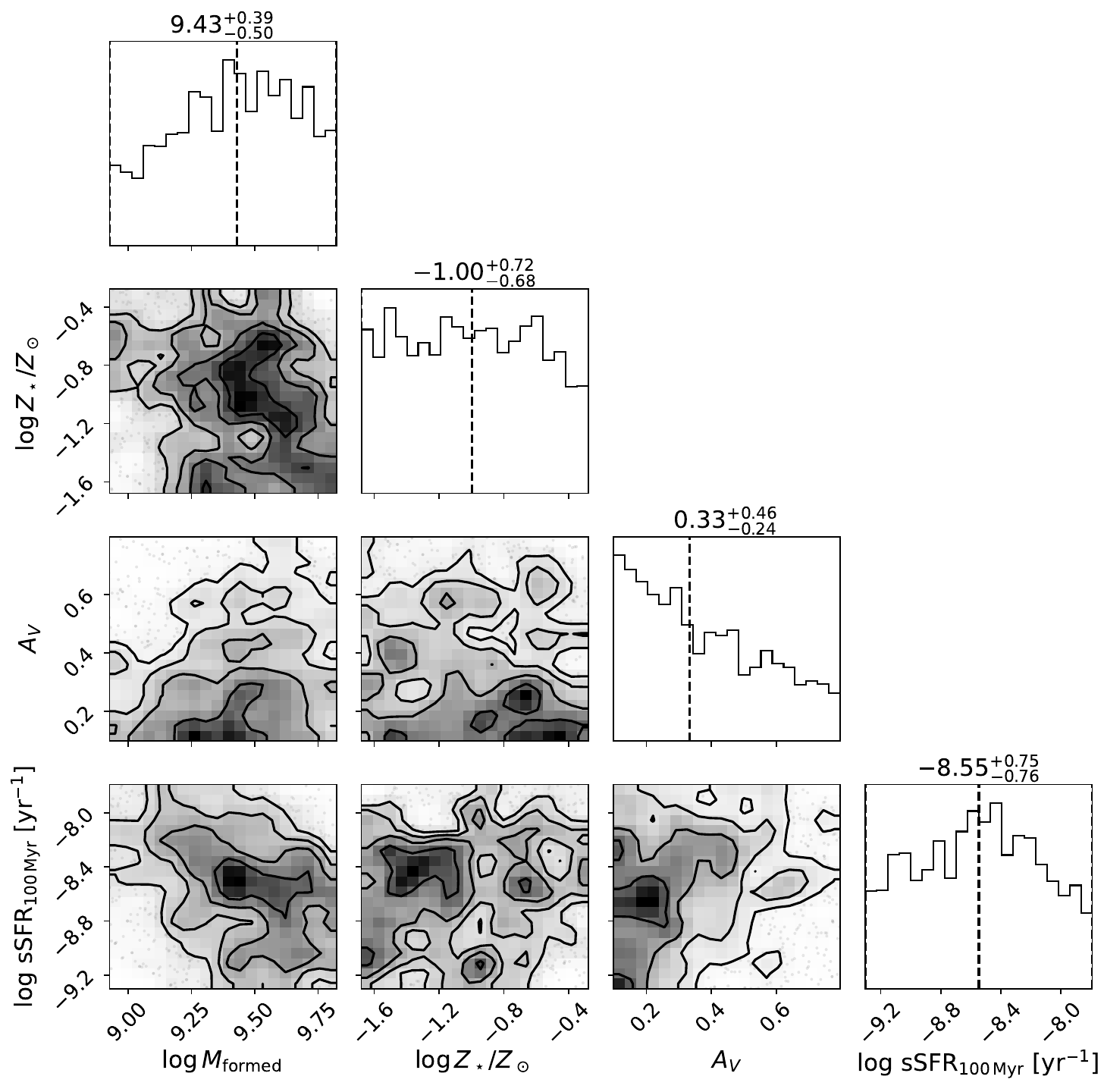}
    \caption{
    Posterior constraints on derived host properties from the \textsc{Prospector} fit (Age\(_\mathrm{MW}\) panel omitted). Medians and 68\% credible intervals:
    \( \log(M_{\rm formed}/M_\odot)=9.77^{+0.31}_{-0.37} \),
    \( \log(Z_\star/Z_\odot)=-0.19^{+0.25}_{-0.22} \),
    \( A_V=0.39^{+0.34}_{-0.27} \)~mag,
    \( \log{\rm sSFR}_{100\,{\rm Myr}}\,[{\rm yr}^{-1}]=-9.28^{+0.47}_{-0.42} \).
    The host is consistent with modest attenuation and near-solar metallicity; recent (\(\lesssim 100\)\,Myr) activity remains weakly constrained. For completeness, the full \textsc{Prospector} parameter corner plot is shown in Fig.~\ref{fig:host_corner_full}.
    }
    \label{fig:host_corner_phys}
\end{figure}

From the posterior, we quote the following physical parameters (medians with 16--84\% credible intervals; Fig.~\ref{fig:host_corner_phys}; see Fig.~\ref{fig:host_corner_full} for the full \textsc{Prospector} parameter corner plot):
\( \log(M_{\star,\mathrm{formed}}/M_\odot)=9.77^{+0.31}_{-0.37} \),
\( \log(Z_\star/Z_\odot)=-0.19^{+0.25}_{-0.22} \),
\( A_V=0.39^{+0.34}_{-0.27}\,\mathrm{mag} \), and
\( \log\mathrm{sSFR}_{100\,\mathrm{Myr}}\,[\mathrm{yr}^{-1}]=-9.28^{+0.47}_{-0.42} \).

The posterior favours a moderate-mass host with modest internal attenuation. In particular, \( A_V=0.39^{+0.34}_{-0.27} \)\,mag is weakly constrained: the 68\% interval spans \( \sim 0.12 \)--\( 0.73 \)~mag, so values consistent with very low attenuation are not excluded, while heavier dust is disfavoured by the blue SED. Given the limited UV/blue and NIR leverage (no deep \(u/v\) detection and only mid-IR limits), the age--dust--metallicity degeneracy remains significant and propagates into the posterior uncertainties in \( \log Z_\star \) and \( \mathrm{sSFR}_{100\,\mathrm{Myr}} \). We therefore treat the global host attenuation as small to moderate and avoid using it to draw strong conclusions about the recent (\( \lesssim 100 \)\,Myr) star-formation activity. These inferences are driven primarily by the DECam \(griz\) detections; deeper UV and a redder/NIR detection (e.g. \(YJH\)) would substantially tighten the constraints. For completeness, the full \textsc{Prospector} parameter corner and the SFH summary are shown in Fig.~\ref{fig:host_corner_full} and Fig.~\ref{fig:host_sfh}, respectively.

\subsubsection{COSMOS-Web host constraints}\label{sec:cosmos_web}
%\sergey{We used the photometry and COSMOS-Web to determine our host galaxy candidate's photo-z. It's close to the spectroscopic one. So, we could try to extract some other physical parameters about our candidate from COSMOS-Web as well.}
%As described in section \ref{sec:host_association}, we used the photometry and COSMOS-Web to determine our host galaxy candidate’s photo-z. We can use the same method to extract other physical parameters from COSMOS-Web, including stellar masses and star formation rates that we created with CIGALE. Using the same matching methods, we find a stellar mass of 9.351 $\pm$ 0.22 and sfr = 1.02 $\pm$ 3.78. This is the equivalent of a median sSFR of -9.34. Both the stellar mass and sSFR are consistent with the \textsc{Prospector} fit to within approximately 1$\sigma$. This complementary approach furthers the case for a low-mass, star forming host.
As described in Section~\ref{sec:host_association}, we used COSMOS-Web DR1 to place empirical constraints on the redshift of the candidate host by selecting COSMOS-Web galaxies with photometry consistent with the LS~DR10 measurements (within the adopted tolerances). We extend the same matching approach to additional physical parameters available for the COSMOS-Web comparison sample. Using the matched galaxies, we infer a typical stellar mass of \(\log_{10}(M_\star/M_\odot)=9.35\pm0.22\) and a typical star-formation rate of \(\mathrm{SFR}\approx 1.0\pm3.8~M_\odot~\mathrm{yr}^{-1}\), corresponding to a median \(\log_{10}(\mathrm{sSFR}/\mathrm{yr}^{-1})\approx -9.34\). These values are consistent with the \textsc{Prospector} constraints (noting that \textsc{Prospector} reports a formed stellar mass) at the \(\sim 1\sigma\) level. While this COSMOS-Web comparison does not constitute a direct measurement for the specific LS~DR10 source, it provides an independent sanity check that the candidate host is plausibly a low-mass galaxy with at most modest ongoing star formation.

\section{Discussion}\label{sec:discussion}
\subsection{Afterglow brightness and broadband interpretation}
\grb{250818B} displays an unusually bright afterglow for a short/Type~I event across X-ray, optical, and radio bands. In the observer-frame \swiftxrt 0.3--10~keV light curve, it lies on the bright end of the short-GRB population at early times (Sec.~\ref{subsubsec:ag_swift_context}). In the optical, its R-band brightness at \( \sim 0.3 \)--\( 0.9 \)~d is comparable to the median of the overall GRB population, but \( \sim 2.8 \)--\( 3.2 \)~mag brighter than the median of the short/Type~I subset (Sec.~\ref{subsubsec:ag_kann_context}). In the radio, \grb{250818B} is among the most luminous short-GRB afterglows at (\( z > 1\)) (Sec.~\ref{subsubsec:ag_radio_context}). Together, these comparisons indicate that the afterglow emission is bright relative to typical short GRBs, motivating dedicated broadband modelling.

The optical SED at \( t_{\rm ref} = 2.035 \)~d after the trigger suggests low-to-moderate host-frame dust extinction along the line of sight, \(A_V^{\rm host} \sim 0.2\)~mag, with no strong preference between MW- and SMC-like extinction curves (Sec.~\ref{subsubsec:ag_sed_extinction}). The inferred optical spectral index is consistent with the X-ray spectral slope and with a standard slow-cooling synchrotron spectrum, potentially with the cooling break between the optical and X-ray bands \citep[e.g.][]{Sari1998,Granot2002}. Thus, the observed optical brightness is unlikely to be dominated by dust obscuration, and the broadband behaviour can be interpreted within the synchrotron afterglow framework.

We modelled the multi-wavelength afterglow using \texttt{afterglowpy} (via \texttt{redback}) as a baseline forward-shock description. A single-component top-hat jet provides a reasonable overall fit but does not reproduce the full evolution: it exhibits a noticeable mismatch to the X-ray light curve from \( \sim 0.02 \)~d, underpredicts the optical emission at \( \gtrsim 3 \)~d, and underpredicts the 36.8~GHz detection at \( \sim 4 \) -- \( 5 \)~d (noting the large uncertainty on this point). These discrepancies suggest additional structure beyond a simple top-hat forward shock.

Motivated by this behaviour, we explore an energy-injection scenario with a refreshed shock, and a two-component jet model (inner core surrounded by a wider, sheath-like component). In this picture, an early injection episode and a slower, baryon-loaded sheath component surrounding the relativistic core can naturally produce a more complex multi-band evolution, with the sheath contribution becoming important at optical and radio wavelengths from \( \gtrsim 1 \) -- \( 2 \)~d. Bayesian model comparison prefers the refreshed two-component model over the single-component top-hat fit (Table~\ref{tab:pri_post}), supporting the interpretation that additional dynamical structure (energy injection and/or angular stratification) is a required feature of this GRB afterglow.

Physically, such energy injection can be produced either by slower ejecta catching up with the decelerating blast wave (refreshed shocks; \citealt{Rees1998,Sari2000}) or by sustained power from a long-lived central engine (e.g., magnetar spin-down; \citealt{Dai1998,Zhang2001,Metzger2011,Rowlinson2013}), while reverse-shock contributions can further enhance very early optical/radio emission at the afterglow onset (deceleration time, \citep[e.g.][]{Sari1999,Kobayashi2000,Zhang2005}; as the excess emission is seen at times later than the afterglow peak, we can rule out any reverse shock component for the afterglow variability. %distinguishing between these scenarios would require denser early-time X-ray coverage and/or earlier multi-frequency radio constraints.

Overall, the broadband brightness of \grb{250818B} can be understood as the outcome of one (or more) of the standard levers that control synchrotron afterglow luminosity: higher kinetic energy, a denser circumburst medium, and/or a favourable viewing geometry (e.g. a more on-axis line of sight or a narrow/structured energetic component), potentially augmented by energy injection \citep[e.g.][]{Nakar2007,Berger2014}. In the baseline fit, relatively high inferred microphysical parameters could also contribute to the flux normalisation; however, these parameters are strongly degenerate with energetics, density, and the location of spectral breaks. In the refreshed two-component model, the microphysics are fixed, and the preference for this model points instead to energy injection and angular structure as a natural explanation for the combination of early-time variability and late-time optical/radio excesses. The total isotropic kinetic energy for the refreshed two-component model is $\sim 10^{54}$\,erg (after energy injection), with a jet-corrected energy of $E_{j}\sim5\times10^{51}$ erg -- such energetics are approximately two orders of magnitude more energetic than typical short-duration GRBs, and fits more naturally within the long GRB population.

Similar modelling ingredients (energy injection and/or reverse-shock-related components) are commonly invoked to explain the radio behaviour of the most radio-luminous short GRBs, including high-redshift events such as \grb{210726A} and \grb{250221A} (Sec.~\ref{subsubsec:ag_radio_context}; \citealt{2024ApJ...970..139S,2025ApJ...982...42S,2025ApJ...994....5A}). However, energy injection and/or a two-component jet structure are commonly invoked for long-duration GRBs \citep[e.g.,][]{granot2003,sato2023}. In the context of short-duration bursts, \grb{250818B} strengthens the emerging picture that the brightest radio afterglows in the short-GRB population may preferentially require departures from the simplest single-component forward-shock description. And in the context of collapsar GRBs, \grb{250818B} reinforces the growing evidence for late-time energy injection via refreshed shocks and an energetic wide-angle second component of a structured jet.

\grb{250818B} additionally requires an accelerated electron index, $p<2$, where a value of $p>2$ typically supports Fermi acceleration at a relativistic shock.
A preference for $p<2$ is often seen, when parameter priors include it or the analysis method does not exclude the $p<2$ range \citep[e.g.,][where the distribution is found to extend below $p=2$]{Wang2015b}, and may well indicate that multiple- or alternative particle acceleration mechanisms are at work within the forward shock.

\subsection{Host-galaxy association: two viable scenarios}
\label{sec:discussion_host}

The available imaging and catalogue information admit (at least) two plausible host-galaxy scenarios for \grb{250818B}. With the current data, we cannot unambiguously distinguish between them, and we therefore treat the host association as tentative and discuss both possibilities below. We stress that the redshift \( z=1.216 \) is derived from metal absorption features imprinted on the afterglow continuum, i.e. material along the GRB sightline, and therefore does not uniquely identify which galaxy (if any) in the LS~DR10 imaging is the host. If objid~5790 is the host, the absorption could plausibly arise in its extended circumgalactic medium at projected separations of tens of kpc~\citep[e.g.][]{Tumlinson2017,Nielsen2013,Nielsen2016}. Alternatively, it could originate in a fainter, near-coincident host below the LS~DR10 detection threshold, or in an unrelated intervening absorber along the line of sight~\citep[e.g.][]{Prochter2006}.

\textbf{Scenario A: the offset LS~DR10 galaxy (objid~5790) is the host.}
In LS~DR10 we identify a faint galaxy-like source (objid~5790; \( r_{\rm AB}=24.74\pm0.36 \), \( \mathrm{S/N}\approx3 \)) at an angular separation \( R_{\rm off}=4\farcs03 \) from the afterglow position (Sec.~\ref{sec:host_association}). At \( z=1.216 \), this corresponds to a projected physical offset of \( \approx 33.51 \)~kpc. Using the chance-coincidence formalism of \citet{Bloom2002} together with empirical galaxy number counts measured locally from the LS~DR10 field (Fig.~\ref{fig:lsdr10_counts}), we obtain a relatively high chance-alignment probability of \( P_{\rm cc}\approx0.20 \), implying that the association is \emph{plausible but not definitive} when judged by angular proximity alone. Nevertheless, we retain objid~5790 as a viable candidate because (i) it is among the lowest-\( P_{\rm cc} \) galaxy-like sources in the immediate field at reasonable offsets, and (ii) association information beyond angular proximity (e.g. DLR-based offset priors and redshift consistency) increases its relative weight compared to other nearby catalogued sources within the same framework.

A complementary Bayesian association analysis (Sec.~\ref{sec:host_association}; \citealt{Ackley2025}) ranks objid~5790 as the most probable host among nearby catalogued sources once additional information is incorporated (e.g. offset priors via DLR and redshift consistency). In addition, an empirical photometric-analogue test using COSMOS-Web DR1 yields a median photometric redshift \( z\approx 1.12\pm0.20 \) for galaxies matched in photometry to the candidate host, consistent with the afterglow absorption redshift within uncertainties (Sec.~\ref{sec:host_association}). We emphasise that this analogue test evaluates whether the \emph{candidate’s} photometry is consistent with galaxies at \( z\sim1 \); it does not by itself demonstrate that the absorber at \( z=1.216 \) is physically associated with objid~5790.
%Under this scenario, the large projected offset would be consistent with the broad offset distribution observed for short/Type~I GRBs~\citep[e.g.][]{Fong2010,Fong2013a,Berger2014}, but the present imaging does not provide a high-confidence association in the strict \( P_{\rm cc} \) sense.
We note that if Scenario~A is correct, the implied projected offset of \(R_{\rm off}\simeq 33.5\)~kpc can provide additional context for the progenitor interpretation; however, we defer any population-based comparison to Sec.~\ref{subsec:discussion_progenitor}, since offset-based arguments are conditional on the host association being correct.

\textbf{Scenario B: the true host is a fainter, near-coincident galaxy below the LS~DR10 detection threshold.}
Because objid~5790 lies close to the LS~DR10 detection threshold and is substantially offset, the current data do not exclude an undetected host at (or very near) the afterglow position (Sec.~\ref{subsubsec:host_limit_coincident}). At \( z\sim1.2 \), low-luminosity systems can readily fall below the LS~DR10 depth, particularly because the observed \( r \) band probes the rest-frame near-UV. Using LS~DR10 Tractor depth quantities at the GRB position, we estimate local \( 3\sigma \) limits of \( r^{\rm gal}_{3\sigma}\simeq 24.61 \) for a small galaxy and \( r^{\rm psf}_{3\sigma}\simeq 24.84 \) for a point source (with corresponding \( 5\sigma \) limits \( r^{\rm gal}_{5\sigma}\simeq 24.06 \) and \( r^{\rm psf}_{5\sigma}\simeq 24.28 \)). Combining these limits with the locally measured number counts, a hypothetical coincident host within \( R_0=0\farcs3{-}0\farcs5 \) would have a very small chance-coincidence probability if it were detectable at the LS~DR10 threshold, \( P_{\rm cc}(m_{\rm lim},R_0)\approx 1.5\times10^{-3} \) (for \( 0\farcs3 \)) to \( \approx 4.2\times10^{-3} \) (for \( 0\farcs5 \)). No such galaxy is detected at the afterglow position down to these limits, implying that any genuinely near-coincident host must be fainter than LS~DR10.

At \( z=1.216 \), the LS~DR10 \( r \) band (\(\lambda_{\rm eff}\sim6200~\)~\AA) samples the rest-frame near-UV, \(\lambda_{\rm rest}\approx \lambda_{\rm obs}/(1+z)\simeq 2800~\)~\AA, so translating an observed-frame \( r \)-band non-detection into a stellar-mass constraint is highly model dependent (recent star formation and dust can dominate the UV luminosity; e.g. \citealt{Conroy2013}). In the same manner, the local \( z \)-band depth gives \( z^{\rm gal}_{3\sigma}\simeq 23.90 \), which corresponds to rest-frame \(\lambda_{\rm rest}\sim 4100~\)~\AA\ and is therefore somewhat less sensitive to short-timescale UV/SFR variations than \( r \).

Adopting our fiducial cosmology, the distance modulus at \( z=1.216 \) is \( \mu\simeq 44.63 \) mag, so (ignoring the spectral part of the \( K \)-correction) these limits correspond to absolute-magnitude \emph{equivalents} at the corresponding rest wavelengths of \( M_{2800}\gtrsim r^{\rm gal}_{3\sigma}-\mu \simeq -20.0 \) and \( M_{4100}\gtrsim z^{\rm gal}_{3\sigma}-\mu \simeq -20.7 \). For plausible UV--optical spectral slopes, the additional spectral \( K \)-correction term is typically of order \(\lesssim 1\) mag \citep[e.g.][]{Hogg2002}, so these values should be regarded as order-of-magnitude. Taken at face value, they imply that the current LS~DR10 data are sensitive only to relatively luminous galaxies at \( z\sim1.2 \), and thus can at best begin to probe the \emph{upper} end of the dwarf-like stellar-mass regime (\( M_\star \sim 10^{7} - 10^{9}\,M_\odot \); e.g. \citealt{Tolstoy2009,McConnachie2012,Geha2012}), depending on the assumed SED and mass-to-light ratio. Substantially deeper imaging (e.g. \( r\sim 26 \) or deeper) would be required to meaningfully test a typical dwarf-mass coincident host at \( z\sim1.2 \).

\textbf{Summary and outlook.}
In summary, Scenario~A is the \emph{best-supported association among catalogued sources} (and is favoured by the Bayesian ranking when DLR priors and redshift consistency are included), but the relatively large \( P_{\rm cc} \) and the survey depth mean that Scenario~B cannot be ruled out. Accordingly, we do not treat the large projected offset of objid~5790 as decisive evidence on its own, given the wide offset distribution and host-luminosity diversity in the short-GRB population; if Scenario~B holds, the true offset could be small and offset-based classification arguments would be correspondingly weakened. Resolving this ambiguity will require deeper, higher-resolution imaging (to search for a faint coincident host and/or to confirm the morphology and colours of objid~5790) and, ideally, host spectroscopy.

\subsection{Progenitor interpretation: merger versus collapsar contamination}
\label{subsec:discussion_progenitor}
Although we adopt the nominal short-GRB classification reported in discovery notices, we note that prompt-duration-based classifications are not definitive, and a non-negligible fraction of apparently short events can arise from collapsars (``short'' GRBs drawn from the long/Type~II population; e.g. \citealt{Zhang2009,Bromberg2012,Bromberg2013}). We therefore focus on the afterglow and host-galaxy context, using the extensive multi-wavelength follow-up to assess whether the broadband behaviour is more naturally explained within a compact-object merger (Type~I) picture or whether a collapsar origin remains plausible.

Our afterglow modelling (constant-density environment and a preference for energy injection and/or angular structure) does not uniquely distinguish between a compact-object merger and a collapsar origin: similar forward-shock phenomenology and refreshed-shock/engine activity are invoked across both GRB classes, and the inferred circumburst density and energetics are not by themselves decisive (e.g. \citealt{Woosley2006,Berger2014}). Instead, the most discriminating information in our current dataset is the host association. If Scenario~A holds, the implied projected offset of \(R_{\rm off}\simeq 33.51~\)kpc would be unusual for a collapsar origin---which is expected to track star-forming regions closely---and would more naturally align with the broad offset distribution of short/Type~I GRBs \citep[e.g.][]{Fong2010,Fong2013b}. Conversely, if Scenario~B is correct and the burst is associated with a faint, near-coincident host below the LS~DR10 detection threshold, then a collapsar origin remains viable and offset-based arguments become correspondingly weaker.

\begin{figure}
    \centering
    \includegraphics[width=\columnwidth]{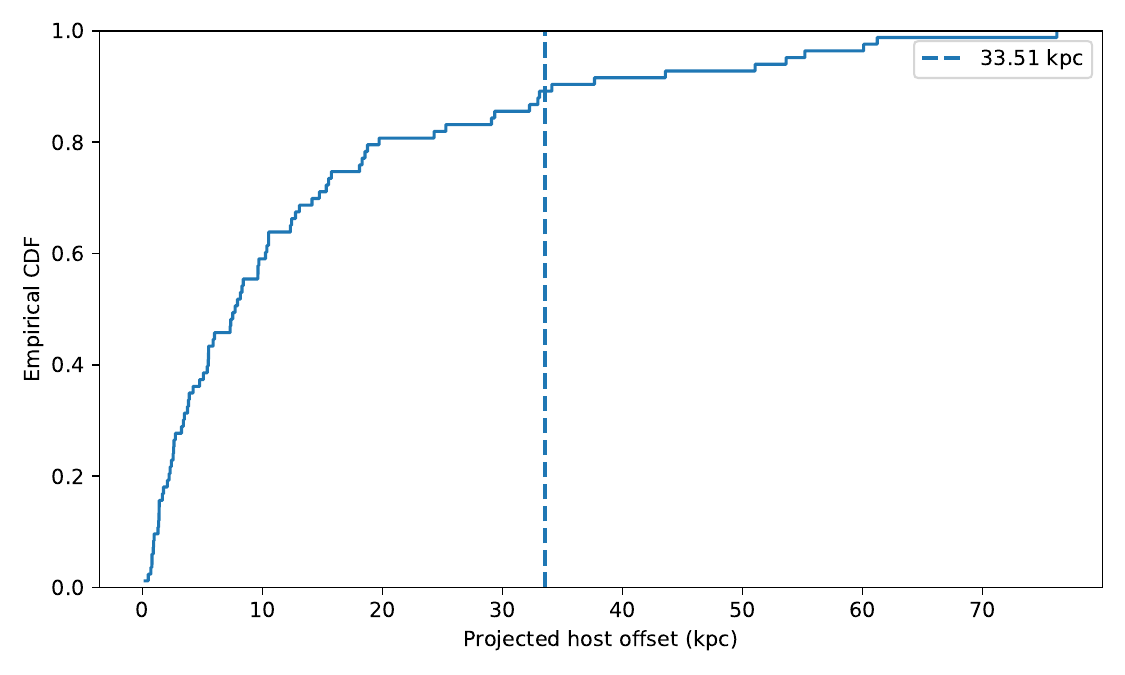}
    \caption{Empirical cumulative distribution of projected (2D) physical offsets for the BRIGHT short-GRB host sample \citep{Fong2022}, restricted to the cosmological subset with measured offsets (\( N=83 \); excluding \grb{170817A} as in \citealt{Fong2022}). The vertical line indicates the projected offset of \grb{250818B} under Scenario~A (\( R_{\rm off}=33.51 \)~kpc).}
    \label{fig:sgrb_offset_ecdf}
\end{figure}

To quantify how unusual the Scenario~A offset is in known GRB populations, we compare \(R_{\rm off}\) to published distributions of \emph{projected physical offsets} for short and long GRBs (noting that the underlying samples are not population-complete and have different selection functions). For the short-GRB population, we adopt the BRIGHT host catalogue~\citep{Fong2022} and restrict to the cosmological subset with measured offsets, excluding \grb{170817A} (included only as a local point of comparison in \citealt{Fong2022}), yielding \(N=83\) events. For our Scenario~A offset estimate \(R_{\rm off}=33.51\)~kpc, the empirical tail probability is \(f_{\rm short}=P(R\ge R_{\rm off})\simeq 0.108\) (equivalently \({\rm CDF}(R_{\rm off})\simeq 0.892\)). Bootstrap resampling of the \(N=83\) offsets yields a 16--84\% range of \(0.072\)--\(0.145\). Thus, an \(\approx 33.51\)~kpc offset lies in the upper tail of the short-GRB offset distribution but remains within the observed short-GRB range (Fig.~\ref{fig:sgrb_offset_ecdf}).

In contrast, long GRBs are expected to trace star-forming regions closely, and observed offsets are typically small. Using the \textit{HST}-based long-GRB host sample of \citet{Blanchard2016} and restricting to bursts with measured projected physical offsets (\(N=70\)), we find no events with \(R_{\rm off}\ge 33.51\)~kpc. For this sample, the corresponding one-sided 95\% Clopper--Pearson upper limit is \(f_{\rm long}=P(R\ge R_{\rm off})\lesssim 0.0419\). Therefore, if Scenario~A is correct, the large projected offset is atypical of the \citet{Blanchard2016} long-GRB host-offset distribution while remaining compatible with the observed short-GRB distribution; if Scenario~B holds, however, the true physical offset could be small, and this discriminator would not apply.

\begin{figure}
    \centering
    \includegraphics[width=\columnwidth]{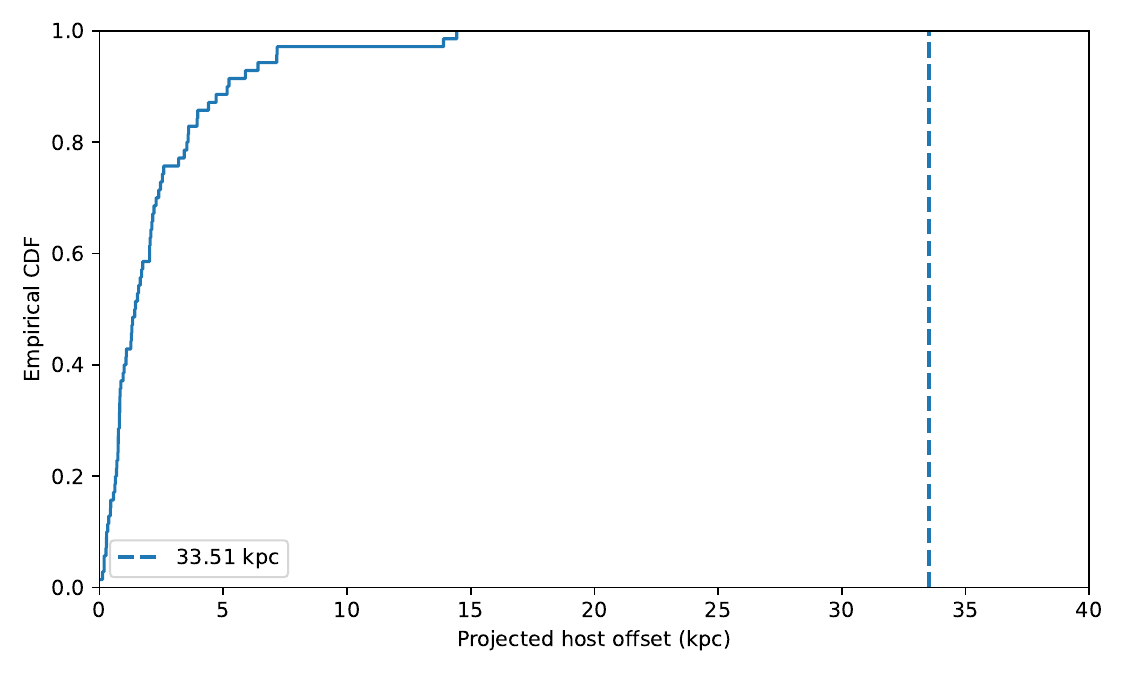}
    \caption{Empirical cumulative distribution of projected physical offsets for the long-GRB host sample of \citet{Blanchard2016}, restricted to bursts with measured projected physical offsets (\( N=70 \)). The vertical line indicates the projected offset of \grb{250818B} under Scenario~A (\( R_{\rm off}=33.51 \)~kpc).}
    \label{fig:lgrb_offset_ecdf}
\end{figure}

Additional, non-decisive clues come from the afterglow brightness in the optical and X-ray bands. As shown in Fig.~\ref{fig:ag_kann_R}, we compare the observer-frame \(R\)-band evolution of \grb{250818B} to the literature compilation of \citet{Kann2006,Kann2010,Kann2011} and the short-GRB sample of \citet{NicuesaGuelbenzu2012}. We show the \(R\)-band projection of our simultaneous multiband SBPL fit and place all measurements on a consistent photometric system. Interpolating the comparison light curves in \(\log t\)--magnitude space at the epochs of our \(R\)-band coverage (0.32, 0.33, 0.53, and 0.85~d), requiring the data to bracket each epoch, we find that \grb{250818B} is brighter than \(\approx 93\)--\(97\%\) of Type~I afterglows (\(N_{\rm I}=29\)--\(31\)). In contrast, it is brighter than \(\approx 58\)--\(62\%\) of Type~II afterglows (\(N_{\rm II}=105\)--\(113\)), and brighter than \(\approx 65\)--\(70\%\) of the combined sample. These percentiles are intended as indicative brightness context (the comparison samples are heterogeneous and not population-complete), but they show that while \grb{250818B} lies on the bright tail of the short/Type~I afterglow distribution at \(\sim 0.3\)--\(0.9\)~d, its optical brightness is broadly typical of the long/Type~II population.

A similar conclusion is obtained in X-rays. Building on the population comparison in Section~\ref{subsubsec:ag_swift_context} (Fig.~\ref{fig:ag_xrt_population}), we interpolate UKSSDC \swiftxrt\ 0.3--10~keV flux light curves in \(\log t\)--\(\log F\) space at the three early epochs used for our SNR=10 spectral bins (1869, 2232, and 6153~s), again requiring the data to bracket each epoch and excluding borderline-duration events (defined via \bat \(T_{90}\)). At these times, \grb{250818B} is brighter than \(\approx 96\)--\(99\%\) of the Swift short-duration sample (\(N_{\rm short}=35\)--\(47\)), but brighter than only \(\approx 67\)--\(73\%\) of the Swift long-duration sample (\(N_{\rm long}=994\)--\(1041\)). As for the optical comparison, these percentiles are meant as context rather than classification, but they reinforce that \grb{250818B} is an outlier among short-GRB afterglows while being comparatively typical of the long-GRB population at similar epochs.

Taken together, the afterglow-brightness context alone does not determine the progenitor class, but it is consistent with either an unusually luminous short-GRB afterglow or a collapsar-contaminated ``short'' event. Ultimately, the host association remains the strongest discriminator \citep[see][for similar conclusions regarding the nominally short \grb{241105A}]{dimple2025b}, and deeper, high-resolution imaging and spectroscopy of any putative host would provide the most direct test of the collapsar versus merger interpretations.

\section{Conclusions}\label{sec:conclusions}
We present the discovery and multi-wavelength follow-up of \grb{250818B}, a \svom-triggered burst whose optical afterglow was identified by the wide-field Gravitational-wave Optical Transient Observer (GOTO). We compile and homogenise X-ray, optical/NIR, and radio observations into broadband light curves and spectral energy distributions suitable for physical interpretation, and we investigate both the afterglow physics and the host-galaxy environment. Public prompt-emission constraints are limited, so we adopt the nominal short-GRB classification reported in discovery notices, while our interpretation is driven primarily by the broadband afterglow behaviour and the host-galaxy context.

Keck~I/LRIS spectroscopy obtained at \( \approx 10.7 \)~hr post-burst reveals a continuum detection with metal absorption features (Fe\,{\sc ii}, Mg\,{\sc ii}, Mg\,{\sc i}), yielding an absorption redshift of \( z=1.216 \) \citep{Fong2025gcn41419}, which we adopt throughout (Sec.~\ref{subsec:spectroscopy}; Fig.~\ref{fig:keck_lris_spec}). Using \swiftxrt\ data, we performed time-resolved spectral fitting across three independent intervals, and we find that the X-ray afterglow is well described by an absorbed power-law model with a single intrinsic neutral-hydrogen column density (tied between intervals) and a photon index that may vary with time; we find no compelling evidence that more complex spectral models are required.

We constrain dust extinction along the GRB sightline using a host-subtracted optical SED at \( t_{\rm ref}\simeq 2.0 \)~d, constructed from nearly contemporaneous LT \( gri \) photometry with late-time DECam imaging used to remove the host contribution (Fig.~\ref{fig:ag_sed}). Modelling the SED as a power law attenuated by host-frame dust yields only modest extinction, with \( A_V^{\rm host}=0.22\pm0.18 \)~mag (Milky Way law) and \( A_V^{\rm host}=0.20\pm0.20 \)~mag (SMC law), which are statistically indistinguishable and consistent with low extinction within the uncertainties. The inferred optical spectral index is consistent with the XRT-informed prior, supporting a standard synchrotron interpretation at this epoch.

Placing \grb{250818B} in context, we compare its afterglow brightness to literature samples and \swift GRB populations in the observer frame. The burst lies on the bright end of the \swift short-GRB distribution in X-rays at early times (Fig.~\ref{fig:ag_xrt_population}) and is similarly luminous in the optical relative to short/Type~I comparison samples at \( \sim 0.3 \)--\( 0.9 \)~d (Fig.~\ref{fig:ag_kann_R}). In the radio, MeerKAT detections at 3.1~GHz (and later 1.3~GHz), together with deep higher-frequency limits, indicate that \grb{250818B} is among the more radio-luminous short-GRB afterglows at \( z\gtrsim 1 \) in our comparison set.

We model the multi-wavelength afterglow using \texttt{afterglowpy} (via \texttt{redback}) within the synchrotron forward-shock framework. A single-component top-hat jet provides a reasonable overall description but does not reproduce the full broadband evolution, including mismatches in the X-ray light curve and late-time optical/radio behaviour. Bayesian model comparison favours a refreshed two-component jet scenario (interpretable as energy injection and/or angular structure), supporting the conclusion that departures from the simplest single-component forward shock are required to explain the data (Table~\ref{tab:pri_post}).

Finally, we investigate the host-galaxy environment and show that the association is non-trivial (Sec.~\ref{sec:discussion_host}). The nearest LS~DR10 galaxy-like source (Tractor \texttt{objid}~5790; \( r_{\rm AB}=24.74\pm0.36 \), \( {\rm S/N}\approx 3 \)) lies at an angular separation \( R_{\rm off}=4\farcs03 \) (\( \approx 33.51 \)~kpc at \( z=1.216 \)) and has a relatively high chance-alignment probability, \( P_{\rm cc}\approx 0.20 \), implying that the association is plausible but not definitive. A Bayesian host-association analysis following \citet{Ackley2025}, which incorporates DLR-normalised offsets and redshift consistency, likewise finds that no catalogued galaxy is strongly favoured: \texttt{objid}~5790 is ranked highest but with only \(P_{\mathrm{assoc}}=0.025\), while the probability that the host is missing from the LS~DR10 catalogue is \(P_{\varnothing}\approx 0.882\) (dominated by the faint-host term). At the same time, the imaging depth does not exclude a fainter, near-coincident host below the LS~DR10 detection threshold, which would reduce the true physical offset and weaken offset-based classification arguments. We emphasise that the redshift \( z=1.216 \) is measured from absorption along the GRB sightline and does not, by itself, uniquely identify the host in the available imaging. Conditional on \texttt{objid}~5790 being the host, a COSMOS-Web DR1 photometric-analogue test is also consistent with the candidate lying at \(z\sim 1\).

Assuming \texttt{objid}~5790 is the host, SED modelling with \textsc{Prospector} at fixed \( z=1.216 \) yields a moderately low-mass system (formed stellar mass \(\log(M_{\star,\rm formed}/M_\odot)\approx 9.77\)) with modest attenuation (\(A_V\approx 0.39\)~mag) and near-solar stellar metallicity (\(\log(Z_\star/Z_\odot)\approx -0.19\); Figs.~\ref{fig:host_sed} and \ref{fig:host_corner_phys}), albeit with remaining degeneracies due to limited UV/NIR leverage. Overall, \grb{250818B} exhibits an unusually bright afterglow for a nominal short GRB across X-ray, optical, and radio bands, enabled by rapid and sustained multi-wavelength follow-up. The homogenised dataset, afterglow modelling, and host-association analysis presented here provide the foundation for interpreting the burst energetics, circumburst environment, and viewing geometry (including possible additional energy supply). Deeper, higher-resolution imaging and host spectroscopy remain essential for resolving host-association ambiguities in high-redshift systems.

\section*{ACKNOWLEDGMENTS}
The Gravitational-wave Optical Transient Observer (GOTO) project acknowledges the support of the Monash-Warwick Alliance; University of Warwick; Monash University; University of Sheffield; University of Leicester; Armagh Observatory \& Planetarium; the National Astronomical Research Institute of Thailand (NARIT); Instituto de Astrofísica de Canarias (IAC); University of Portsmouth; University of Turku; University of Birmingham; and the UK Science and Technology Facilities Council (STFC, grant numbers ST/T007184/1, ST/T003103/1 and ST/Z000165/1). The National Radio Astronomy Observatory and Green Bank Observatory are facilities of the U.S. National Science Foundation operated under cooperative agreement by Associated Universities, Inc. This paper makes use of the following ALMA data: ADS/JAO.ALMA\#2024.1.01131.T. ALMA is a partnership of ESO (representing its member states), NSF (USA) and NINS (Japan), together with NRC (Canada), NSTC and ASIAA (Taiwan), and KASI (Republic of Korea), in cooperation with the Republic of Chile. The Joint ALMA Observatory is operated by ESO, AUI/NRAO and NAOJ. The MeerKAT telescope is operated by the South African Radio Astronomy Observatory, which is a facility of the National Research Foundation, an agency of the Department of Science and Innovation. This work made use of data supplied by the UK Swift Science Data Centre at the University of Leicester. This research has made use of data obtained through the High Energy Astrophysics Science Archive Research Center online service, provided by the NASA/Goddard Space Flight Center. W. M. Keck Observatory access was supported by Northwestern University and the Center for Interdisciplinary Exploration and Research in Astrophysics (CIERA). Some of the data presented herein were obtained at the W. M. Keck Observatory, which is operated as a scientific partnership among the California Institute of Technology, the University of California and the National Aeronautics and Space Administration. The Observatory was made possible by the generous financial support of the W. M. Keck Foundation. GPL acknowledges support from the Royal Society (grant Nos. DHF-R1-221175 and DHF-ERE-221005). ACG and the Fong Group at Northwestern acknowledge support by the National Science Foundation under grant Nos. AST-1909358, AST-2206494, AST-2308182 and CAREER grant No. AST-2047919. W. M. Keck Observatory access was supported by Northwestern University and the Center for Interdisciplinary Exploration and Research in Astrophysics (CIERA). MEW is supported by the Science and Technology Facilities Council (STFC). BPG acknowledges support from STFC grant No. ST/Y002253/1. BPG and DO acknowledge support from the Leverhulme Trust grant No. RPG-2024-117. RLCS and SM are supported by Leverhulme Trust grant RPG-2023-240. YS is supported by the Chinese Scholarship Council and the University of Leicester. DMS acknowledges support through the Ram\'on y Cajal grant RYC2023-044941, funded by MCIU/AEI/10.13039/501100011033 and FSE+. TLK acknowledges support from a Warwick Astrophysics prize post-doctoral fellowship made possible thanks to a generous philanthropic donation. SM acknowledges financial support from the Research Council of Finland project 350458. The authors express their gratitude to the Terskol Observatory Collective Use Center for organizing and making it possible to conduct observations using the Zeiss-2000 telescope of the INASAN observatory.

\section*{DATA AVAILABILITY}
All data sets supporting this study are provided in the paper. Additional data are available from the corresponding author upon reasonable request.

%%%%%%%%%%%%%%%%%%%%%%%%%%%%%%%%%%%%%%%%%%%%%%%%%%
%\section*{Data Availability}



\bibliographystyle{mnras}
\bibliography{example} % if your bibtex file is called example.bib

%\begin{thebibliography}{99}
%\end{thebibliography}

%%%%%%%%%%%%%%%%%%%%%%%%%%%%%%%%%%%%%%%%%%%%%%%%%%

%%%%%%%%%%%%%%%%% APPENDICES %%%%%%%%%%%%%%%%%%%%%
%\clearpage
\appendix

\section{Data tables}\label{app:data_tables}

\onecolumn
\begin{longtable}{|c|c|c|c|c|c|c|}
\caption{Optical and near-infrared afterglow observations of \grb{250818B} compiled in this work, including measurements reported in GCN Circulars. All magnitudes are in the AB system. Upper limits are \( 3\sigma \) and are indicated with a "\( > \)" sign. Times are measured relative to the \svomeclairs trigger at \( T_{0} = \,\) 2025-08-18T03:29:09~UTC.}
\label{tab:phot_250818B}\\
\hline
\( T-T_0 \) (h) & Instrument/Telescope & Exp. Time (s) & Filter & Mag & Mag err & Source \\
\hline
\endfirsthead
\multicolumn{7}{l}{\small \textit{Table \ref{tab:phot_250818B} continued from previous page.}}\\
\hline
\( T-T_0 \) (h) & Instrument/Telescope & Exp. Time & Filter & Mag & Mag err & Source \\
\hline
\endhead

\hline
\endfoot

\hline
\endlastfoot

% -------------------------
% Pre-burst imaging. If it makes sense.
% -------------------------
\( -9.23 \) & GOTO-S             & \( 4\times45 \)    & \( L \)      & \(>20.3\)   &          & This work \\
\hline
% -------------------------
% Afterglow photometry (AB, dereddened)
% -------------------------
\(0.06\)  & SVOM/VT             & \( 50 \)           & \(VT\_B\)    & \(17.40\)   & \(0.03\) & ~\citet{Yao2025gcn41409} \\
\(0.54\)   & GOTO                & \( 4\times90 \)    & \(L\)        & \(18.48\)   & \(0.14\) & This work \\
\(1.10\)  & SVOM/VT             & \( 50 \)           & \(VT\_B\)    & \(18.81\)   & \(0.03\) & ~\citet{Yao2025gcn41409} \\
\(1.16\)  & Swift/UVOT          & \( 1544 \)         & \(u\)        & \(18.96\)   & \(0.06\) & ~\citet{Siegel2025gcn41435} \\
\(1.67\)   & GOTO                & \( 4\times90 \)    & \(L\)        & \(19.26\)   & \(0.18\) & This work \\
\(4.97\)   & KNC/CDK17-AITP      & \( 15\times300 \)  & \(r\)        & \(19.30\)   & \(0.07\) & ~\citet{Hellot2025gcn41444} \\
\(7.53\)   & KNC/TEC160FL        & \( 11\times300 \)  & \(r\)        & \(20.00\)   & \(0.16\) & ~\citet{Hellot2025gcn41444} \\
\(7.75\)   & TRT/0.7m            & --                 & \(R\)        & \(20.03\)   & \(0.06\) & ~\citet{An2025gcn41430} \\
\(8.00\)    & KAIT                & \( 90\times60 \)   & \(R\)        & \(19.95\)   & \(0.20\) & ~\citet{Zheng2025gcn41417} \\
\(12.64\)  & TRT/0.7m            & --                 & \(R\)        & \(20.28\)   & \(0.06\) & ~\citet{An2025gcn41430} \\
\(20.42\)  & SAO~RAS/Zeiss-1000  & \( 13\times300 \)  & \(R\)        & \(20.93\)   & \(0.04\) & ~\citet{Moskvitin2025gcn41428} \\
\(21.37\)  & SAAO/1mLesedi       & \( 6\times450 \)   & \(g^\prime\) & \(21.15\)   & \(0.09\) & ~\citet{Kumar2025gcn41431} \\
\(21.84\)  & FTW/3KK             & \( 35\times180 \)  & \(r\)        & \(20.95\)   & \(0.02\) & ~\citet{Busmann2025gcn41445} \\
\(22.13\)  & SAAO/1mLesedi       & \( 6\times400 \)   & \(r^\prime\) & \(21.41\)   & \(0.47\) & ~\citet{Kumar2025gcn41431} \\
\(23.60\)  & NOT/ALFOSC          & \( 3\times300 \)   & \(r\)        & \(20.98\)   & \(0.04\) & ~\citet{BroeBendtsen2025gcn41426} \\
\(45.60\)  & FTW/3KK             & \( 30\times180 \)  & \(r\)        & \(22.09\)   & \(0.08\) & ~\citet{Busmann2025gcn41445} \\
\(48.24\)  & LT/IO:O             &                    & \(g\)        & \(22.59\)   & \(0.17\) & This work \\
\(48.72\)  & LT/IO:O             & \( 5\times240 \)   & \(r\)        & \(21.97\)   & \(0.10\) & ~\citet{Dimple2025gcn41442} \\
\(48.96\)  & LT/IO:O             &                    & \(i\)        & \(22.19\)   & \(0.11\) & This work \\
\(49.44\)  & LT/IO:O             &                    & \(z\)        & \(22.56\)   & \(0.40\) & This work \\
\(96.00\)   & LT/IO:O             &                    & \(r\)        & \(23.69\)   & \(0.37\) & This work \\
\(96.48\)  & LT/IO:O             &                    & \(i\)        & \(22.97\)   & \(0.21\) & This work \\
\(116.59\) & Terskol/Zeiss-2000  & \( 3420 \)         & \(R\)        & \(22.64\)   & \(0.20\) & This work \\
\(222.55\) & Terskol/Zeiss-2000  & \( 3720 \)         & \(R\)        & \(>22.44\)  &          & This work \\
\hline
\end{longtable}

\twocolumn
\begin{table}
    \centering
    \caption{
    Vega-to-AB magnitude offsets \( \Delta m\) adopted in this work, defined such that
    \( m_{\rm AB} = m_{\rm Vega} + \Delta m\).
    }
    \label{tab:vega_ab_offsets}
    \begin{tabular}{lcc}
        \hline
        Filter & Photometric system & \( \Delta m= m_{\rm AB} - m_{\rm Vega} \) \\
        \hline
        \( U \)      & Johnson--Cousins & 0.79  \\
        \( B \)      & Johnson--Cousins & \( -0.09 \) \\
        \( V \)      & Johnson--Cousins & 0.02  \\
        \( R \)      & Johnson--Cousins & 0.21  \\
        \( I \)      & Johnson--Cousins & 0.45  \\
        \( J \)      & NIR (2MASS-like) & 0.91  \\
        \( H \)      & NIR (2MASS-like) & 1.39  \\
        \( K_s \)    & NIR (2MASS-like) & 1.85  \\
        \( g \)      & Sloan-like       & \( -0.08 \) \\
        \( r \)      & Sloan-like       & 0.16  \\
        \( i \)      & Sloan-like       & 0.37  \\
        \( z \)      & Sloan-like       & 0.54  \\
        \( u \)      & \swiftuvot       & 1.02  \\
        \( b \)      & \swiftuvot       & \( -0.13 \) \\
        \( v \)      & \swiftuvot       & 0.00  \\
        \hline
    \end{tabular}
\end{table}

\begin{table}
    \centering
    \caption{Milky Way extinction coefficients \(A_\lambda/E(B\!-\!V)\) adopted in this work.}
    \label{tab:mw_extinction_coeffs}
    \begin{tabular}{lcc}
        \hline
        Filter & Origin & \(A_\lambda/E(B\!-\!V)\) \\
        \hline
        \( g \) (SDSS-like)    & \citet{Schlafly2011}     & 3.303 \\
        \( r \) (SDSS-like)    & \citet{Schlafly2011}     & 2.285 \\
        \( i \) (SDSS-like)    & \citet{Schlafly2011}     & 1.698 \\
        \( z \) (SDSS-like)    & \citet{Schlafly2011}     & 1.263 \\
        \( U \) (Landolt)      & \citet{Fitzpatrick1999}  & 4.334 \\
        \( B \) (Landolt)      & \citet{Fitzpatrick1999}  & 3.626 \\
        \( V \) (Landolt)      & \citet{Fitzpatrick1999}  & 2.742 \\
        \( R \) (Landolt)      & \citet{Fitzpatrick1999}  & 2.169 \\
        \( I \) (Landolt)      & \citet{Fitzpatrick1999}  & 1.505 \\
        \( u \) (UVOT)\(^{c}\) & \citet{Yi2023}           & 4.80 \\
        \( b \) (UVOT)\(^{c}\) & \citet{Yi2023}           & 3.97 \\
        \( v \) (UVOT)\(^{c}\) & \citet{Yi2023}           & 2.99 \\
        \( L \) (GOTO)         & this work\(^{a}\)        & \(0.997\,R_V \simeq 3.09\) \\
        \( \mathrm{VT_B} \) (\svom) & this work\(^{b}\)   & \((A_g + A_r)/2 \simeq 2.79\) \\
        \hline
    \end{tabular}
    \begin{flushleft}
    \(^{a}\) Approximated as \(A_L \simeq 0.997\,A_V\) with \(R_V=3.1\). \\
    \(^{b}\) Approximated as the mean of the SDSS \(g\) and \(r\) coefficients. \\
    \(^{c}\) UVOT coefficients are taken from \citet{Yi2023} and are tabulated per \(E(B\!-\!V)\) from the SFD98 map \citep{Schlegel1998}. Since we adopt the Schlafly \& Finkbeiner (2011) recalibration \(E(B\!-\!V)_{\rm SF11}\approx0.0633\,E(B\!-\!V)_{\rm SFD}\) \citep{Schlafly2011}, we rescale the UVOT coefficients by \(1/0.0633\) so that \(A_\lambda/E(B\!-\!V)\) is consistent with our adopted \(E(B\!-\!V)\).
    \end{flushleft}
\end{table}

\begin{table}
    \centering
    \caption{
    Host-galaxy photometry at the position of \grb{250818B}, from the DESI Legacy Imaging Surveys DR10 Tractor catalogue (DECam \( griz \)) and WISE (\( W1 \)--\( W4 \)).  All magnitudes are AB and are not corrected for Galactic extinction. For bands with low signal-to-noise, we quote \( 3\sigma \) upper limits.
    }
    \label{tab:host_phot_legacy}
    \begin{tabular}{lccc}
        \hline
        Band & Instrument & \( m_{\rm AB} \) & err \\
        \hline
        \( u \)   & SkyMapper & \( >19.76 \)  &      \\
        \( v \)   & SkyMapper & \( >20.07 \)  &      \\
        \( g \)   & DECam     & 24.82        & 0.20 \\
        \( r \)   & DECam     & 24.74        & 0.36 \\
        \( i \)   & DECam     & 24.31        & 0.43 \\
        \( z \)   & DECam     & 23.63        & 0.24 \\
        \( W1 \)  & WISE      & \( >21.54 \) &      \\
        \( W2 \)  & WISE      & \( >20.63 \) &      \\
        \( W3 \)  & WISE      & \( >17.29 \) &      \\
        \( W4 \)  & WISE      & \( >14.07 \) &      \\
        \hline
    \end{tabular}
\end{table}

\begin{table*}
    \centering
    \caption{Radio observations of \grb{250818B}. Upper limits are \( 3\sigma \).}
    \label{tab:radio_250818B}
    \begin{tabular}{l c c c c c}
        \hline
        Facility & Frequency (GHz) & \( T-T_{0} \) (d) & Flux density (\( \mu \)\,Jy) & Flux err (\( \mu \)\,Jy) & Source \\
        \hline
        VLA     & 10.0  & \(2.30\) & 120        &        & ~\citet{Ricci2025gcn41455} \\
        MeerKAT & 1.3  & \(4.97\) & <21         &        & This Work \\
        MeerKAT & 1.3  & \(9.02\) & <28         &        & This Work \\
        MeerKAT & 1.3  & \(15.05\) & 45         &   12   & This Work \\
        MeerKAT & 1.3  & \(32.00\) & <18        &        & This Work \\
        MeerKAT & 3.1  & \(5.05\) & 31          &   8    & This Work \\
        MeerKAT & 3.1  & \(8.93\) & 60          &   13   & This Work \\
        MeerKAT & 3.1  & \(15.1\) & 37          &   14   & This Work \\
        MeerKAT & 3.1  & \(32.1\) &  <32        &       & This Work \\
        CrAO    & 36.8  & \(2.93\) & \( <590 \) &        & This work \\
        CrAO    & 36.8  & \(3.89\) & \( <780 \) &        & This work \\
        CrAO    & 36.8  & \(4.90\) & 920        & 800    & This work \\
        CrAO    & 36.8  & \(5.90\) & \( <770 \) &        & This work \\
        CrAO    & 36.8  & \(6.90\) & \( <540 \) &        & This work \\
        CrAO    & 36.8  & \(7.91\) & \( <450 \) &        & This work \\
        ALMA    & 97.5  & \(10.14\) & \( <40 \) &        & This work \\
        ALMA    & 97.5  & \(17.19\) & \( <32 \) &        & This work \\
        \hline
    \end{tabular}
\end{table*}

\begin{table*}
    \centering
    \caption{
    \swiftxrt 0.3-10~keV fluxes of \grb{250818B} (SNR = 10 binning). The best-fit hydrogen absorption is N\(_{\text{H}}\) (intrinsic) $= \left(2.09_{-2.09}^{+2.51}\right) \times 10^{21}$cm\(^{-2}\), consistent with no excess above the galactic value of $6.49 \times 10^{20}$cm\(^{-2}\)~\citep{Willingale2013}. The reduced fit statistic is 0.24. All errors are to the $90\%$ tolerance level.}
    \label{tab:xrt_250818B_snr10}
    \begin{tabular}{c c c c c}
        \hline
        Bin Centroid (s) & Bin Width (s) & Absorbed Flux ($ 10^{-11}$erg\,cm\(^{-2}\)\,s\(^{-1}\)) & Unabsorbed Flux ($ 10^{-11}$erg\,cm\(^{-2}\)\,s\(^{-1}\)) & Photon Index \\
        \hline
        1869 & 371.1 & $2.58_{-0.45}^{+0.55}$ & $2.91_{-0.43}^{+0.51}$ & $1.64_{-0.22}^{+0.23}$ \\
        \hline
        2232 & 355.3 & $2.82_{-0.42}^{+0.51}$ & $3.32_{-0.42}^{+0.48}$ & $1.84_{-0.21}^{+0.22}$ \\
        \hline
        6153 & 837.4 & $0.96_{-0.14}^{+0.16}$ & $1.25_{-0.16}^{+0.18}$ & $2.19_{-0.23}^{+0.24}$ \\
        % fill in when ready
        % This is for the SNR = 10 results
        \hline
    \end{tabular}
\end{table*}

\begin{table*}
    \centering
    \caption{
    \swiftxrt 0.3-10~keV fluxes of \grb{250818B} (SNR = 5 binning). The best-fit hydrogen absorption is N\(_{\text{H}}\) (intrinsic) $= \left(0.79_{-0.79}^{+2.21}\right) \times 10^{21}$cm\(^{-2}\), consistent with no excess above the galactic value of $6.49 \times 10^{20}$cm\(^{-2}\)~\citep{Willingale2013}. The reduced fit statistic is 0.12. All errors are to the $90\%$ tolerance level.
    }
    \label{tab:xrt_250818B_snr5}
    \begin{tabular}{c c c c c}
        \hline
        Bin Centroid (s) & Bin Width (s) & Absorbed Flux ($ 10^{-11}$erg\,cm\(^{-2}\)\,s\(^{-1}\)) & Unabsorbed Flux ($ 10^{-11}$erg\,cm\(^{-2}\)\,s\(^{-1}\)) & Photon Index \\
        \hline
        1745 & 122.9 & $2.51_{-0.58}^{+0.82}$ & $2.97_{-0.65}^{+0.76}$ & $1.94_{-0.34}^{+0.35}$ \\
        \hline
        1892 & 170.5 & $2.05_{-0.38}^{+0.48}$ & $2.33_{-0.38}^{+0.46}$ & $1.79_{-0.25}^{+0.26}$ \\
        \hline
        2042 & 130.4 & $3.48_{-0.94}^{+1.28}$ & $3.68_{-0.92}^{+1.26}$ & $1.29_{-0.32}^{+0.33}$ \\
        \hline
        2159 & 102.8 & $2.96_{-0.74}^{+0.99}$ & $3.48_{-0.77}^{+0.96}$ & $1.96_{-0.35}^{+0.37}$ \\
        \hline
        2310 & 199.9 & $2.42_{-0.50}^{+0.64}$ & $2.79_{-0.51}^{+0.62}$ & $1.87_{-0.28}^{+0.29}$ \\
        \hline
        6153 & 837.4 & $0.99_{-0.14}^{+0.17}$ & $1.21_{-0.15}^{+0.17}$ & $2.11_{-0.21}^{+0.23}$ \\
        % fill in when ready
        % This is for the SNR = 5 results
        \hline
    \end{tabular}
\end{table*}

\clearpage
\twocolumn
\section{Supplementary figures}\label{app:extra}

\begin{figure}
    \centering
    \includegraphics[width=\columnwidth]{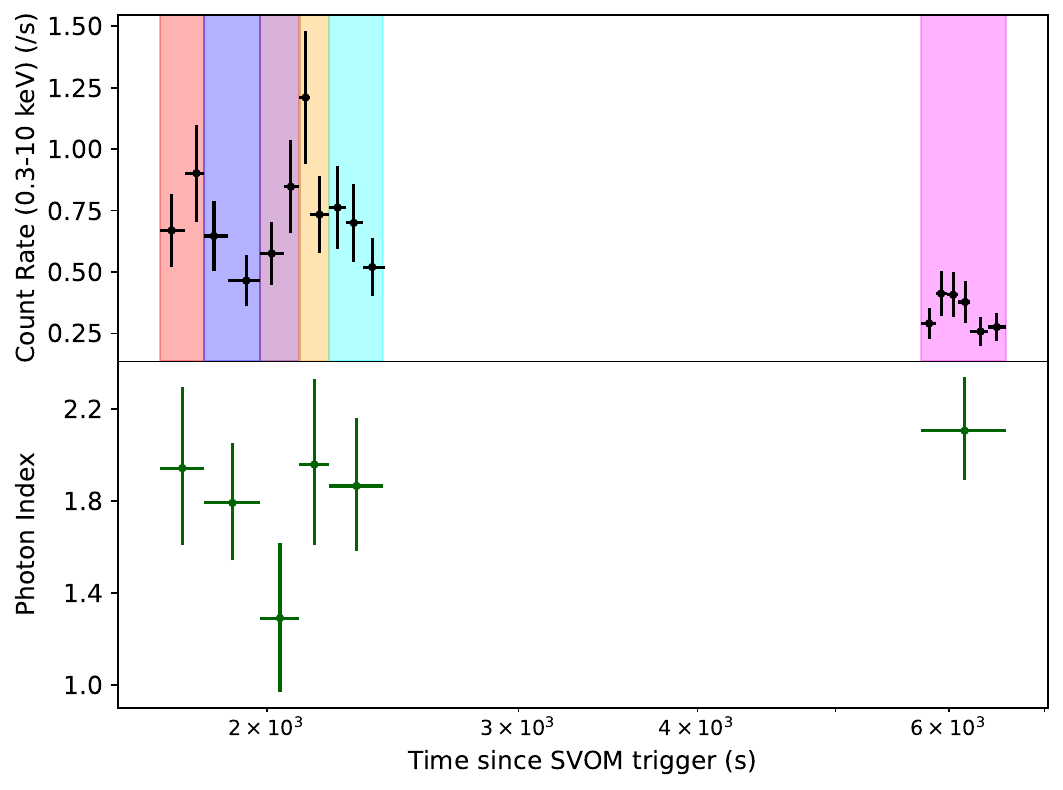}
    \caption{
    Top: \xrt count rate lightcurve of the three flaring episodes, where the highlighted sections indicate the six SNR = 5 spectral bins. The late-time \xrt data at \( \sim 3.35 \)~d does not meet the SNR threshold and so is not included in the spectral analysis. Bottom: evolution of the best-fit photon index with time.
    }
    \label{fig:xrtlc_gamma_snr5}
\end{figure}

\begin{figure}
    \centering
    \includegraphics[width=\columnwidth]{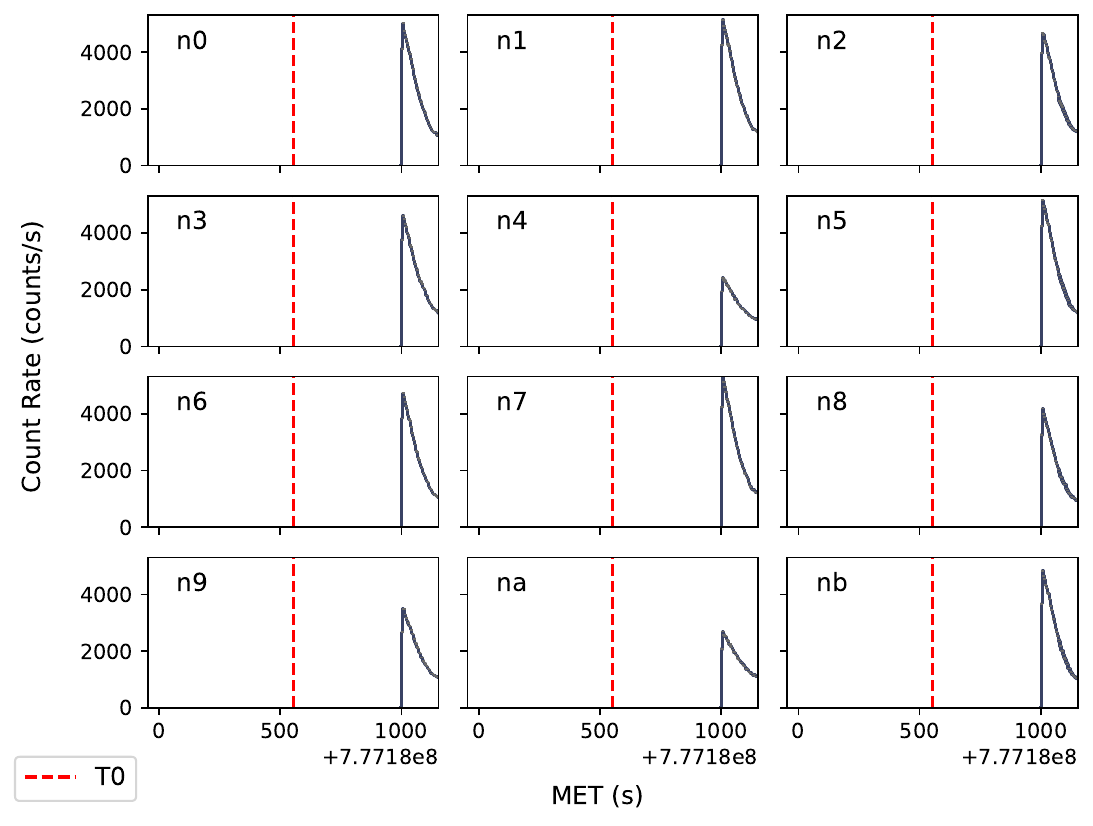}
    \caption{Light curves of all NaI detectors on \fermigbm between \( T_0-600 \)~s and \( T_0 + 600 \)~s (where \( T_0 \) is the \svom trigger time, indicated by the dashed red line). The lack of counts around \( T_0 \) is due to \fermi being over the South Atlantic Anomaly.
    }
    \label{Fig:NaIMulti}
\end{figure}

\begin{figure}
    \centering
    \includegraphics[width=\columnwidth]{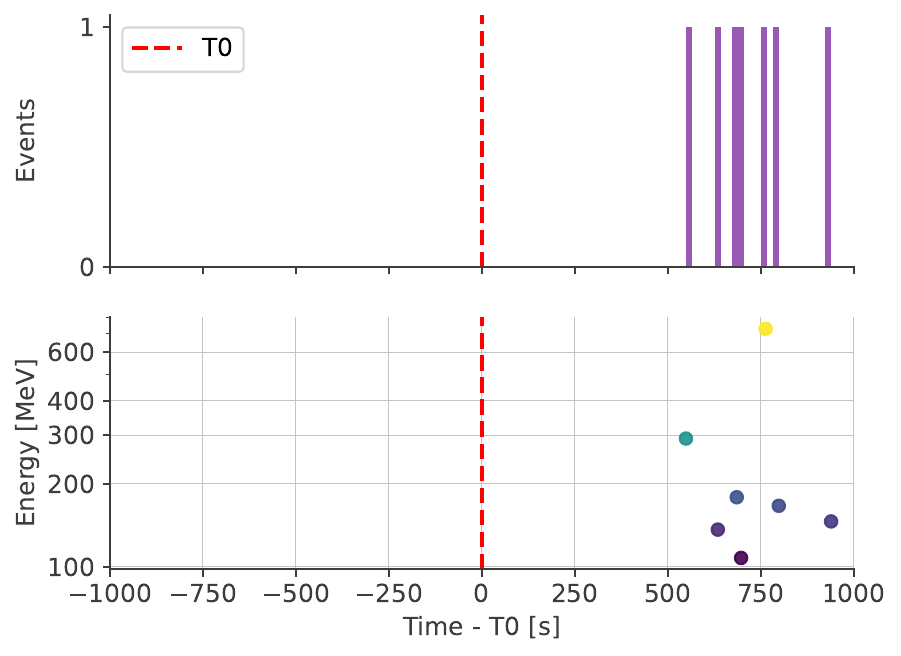}
    \caption{
    \fermilat event data between \( T_0-600\)~s and \( T_0 + 600 \)~s (where \( T_0 \) is the \svom trigger time, indicated by the red dashed line). Top: number of events detected by the \lat. Bottom: energy of detected events. The lack of counts around \( T_0 \) is due to \fermi being over the South Atlantic Anomaly.
    }
    \label{Fig:LAT}
\end{figure}

\begin{figure}
    \centering
    \includegraphics[width=\columnwidth]{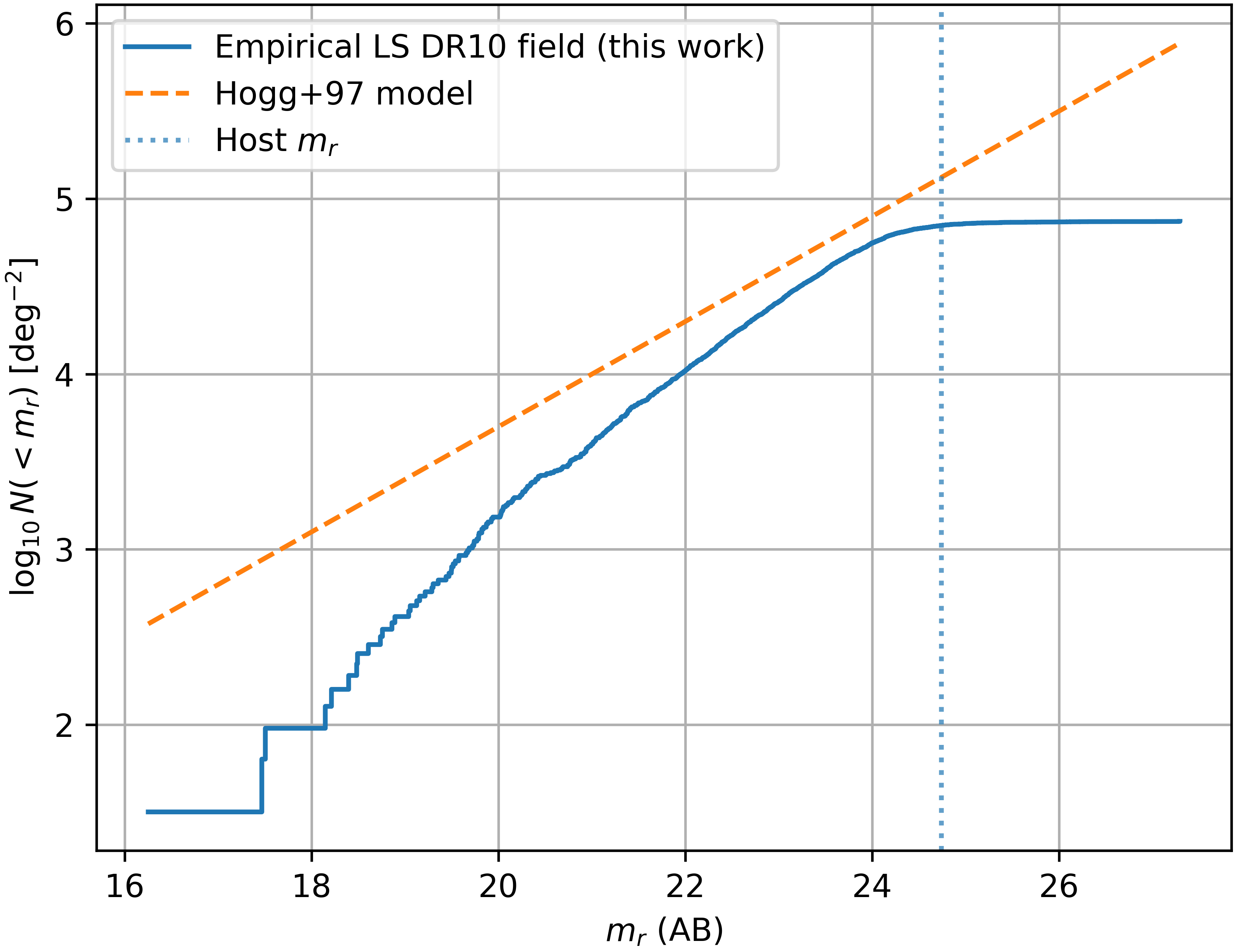}
    \caption{
    Cumulative galaxy counts in LS~DR10 around \grb{250818B}, measured within a \( 0.1^\circ \) cone and restricted to galaxy-like, unmasked sources with significant \( r \)-band flux (\texttt{type}~\( \in\{\mathrm{REX,EXP,DEV,SER}\} \), \texttt{allmask\_g = allmask\_r = allmask\_z = 0}, \texttt{flux\_r > 0}, \texttt{flux\_ivar\_r > 0}; blue step curve). The dashed line shows a simple \citealt{Hogg1997}) \( R \)-band number--counts model for comparison. The empirical counts follow a similar slope in \( \log_{10} N(<m_r) \) versus \( m_r \) but with a lower normalisation by a factor of \( \sim 2 \), likely due to our strict selection and the limited field size. The dotted vertical line marks the host-galaxy magnitude used to estimate \( \sigma(<m_r) \) for the chance-coincidence calculation.
    }
    \label{fig:lsdr10_counts}
\end{figure}

\begin{figure}
    \centering
    \includegraphics[width=\columnwidth]{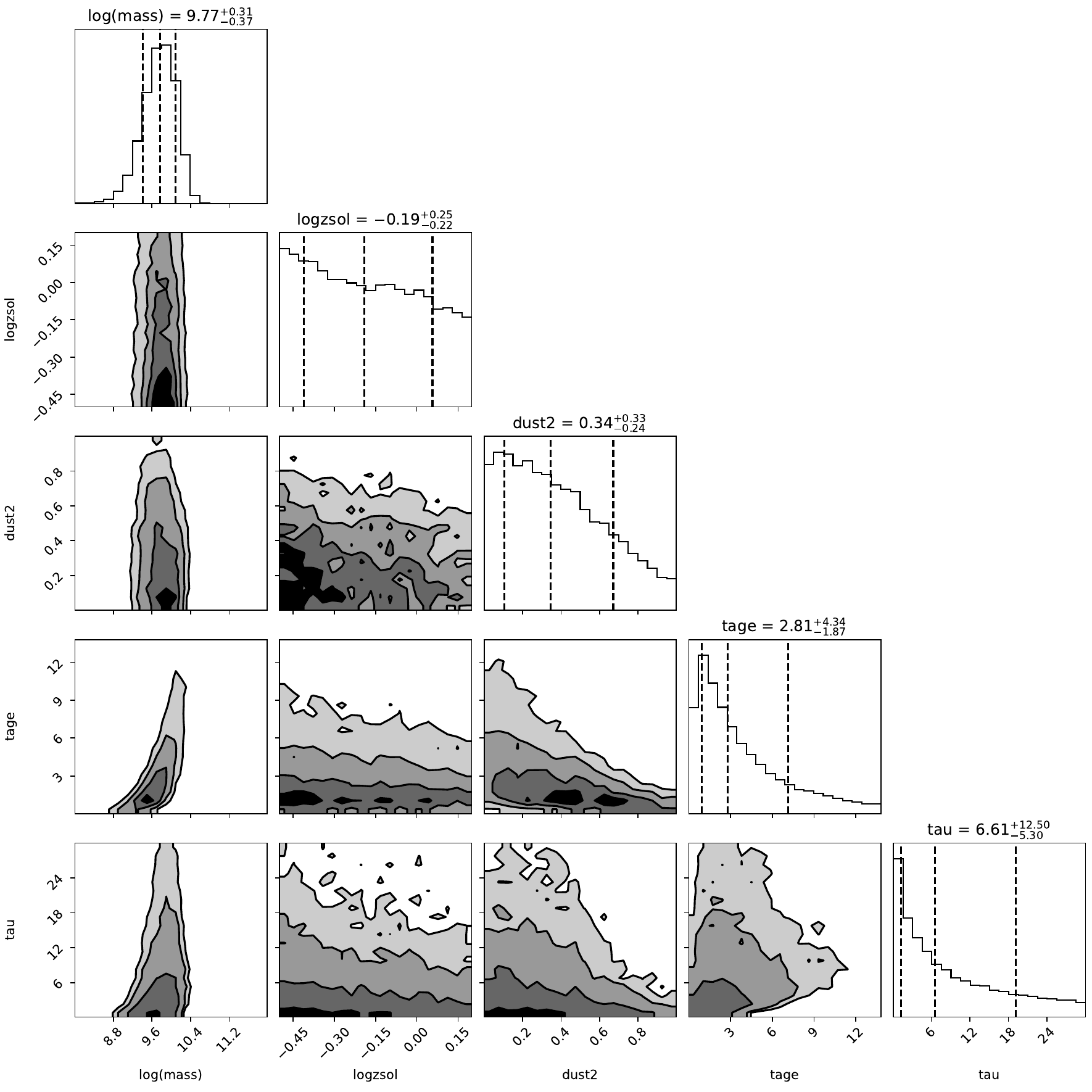}
    \caption{
    \textsc{Prospector} corner plot for the fitted parameters in the parametric delayed-\(\tau\) model (e.g. stellar mass, metallicity, attenuation, population age, and \(\tau\)). Shown for completeness; see Fig.~\ref{fig:host_corner_phys} for the physically interpreted subset.
    }
    \label{fig:host_corner_full}
\end{figure}

\begin{figure}
    \centering
    \includegraphics[width=\columnwidth]{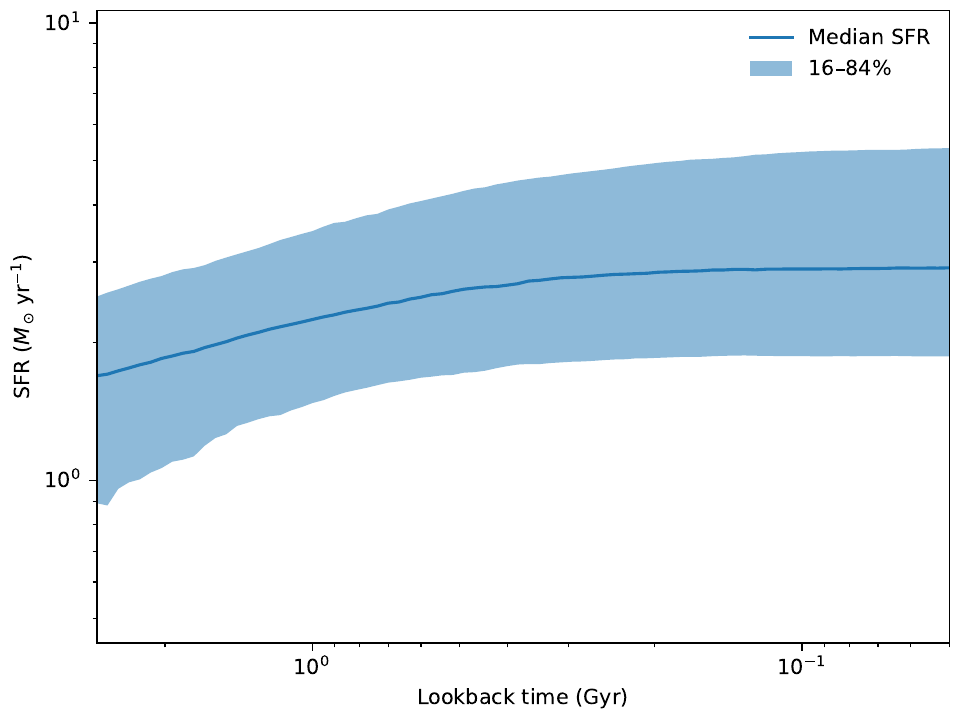}
    \caption{
    Star-formation history implied by the parametric delayed-\(\tau\) \textsc{Prospector} fit: posterior median SFR as a function of lookback time (solid line), with the 16--84\% credible interval (shaded). With the current photometric baseline, the recent SFR is weakly constrained; deeper blue/UV and NIR photometry would tighten the SFH constraints.
    }
    \label{fig:host_sfh}
\end{figure}

%%%%%%%%%%%%%%%%%%%%%%%%%%%%%%%%%%%%%%%%%%%%%%%%%%

\newpage
%\twocolumn
\section*{Affiliations}
$^{1}$School of Physics \& Astronomy, Monash University, Clayton VIC 3800, Australia\\
$^{2}$Astrophysics Research Institute, Liverpool John Moores University, 146 Brownlow Hill, Liverpool L3 5RF, UK\\
$^{3}$Department of Physics, University of Warwick, Gibbet Hill Road, Coventry CV4 7AL, UK\\
$^{4}$School of Physics and Astronomy, University of Birmingham, Edgbaston, Birmingham, B15 2TT, UK\\
$^{5}$Institute for Gravitational Wave Astronomy, University of Birmingham, Birmingham, B15 2TT, UK\\
$^{6}$Department of Astronomy, Cornell University, Ithaca, NY 14853, USA\\
$^{7}$School of Physics and Astronomy, University of Leicester, University Road, Leicester, LE1 7RH, UK\\
$^{8}$Department of Physics and Astronomy, Northwestern University, Evanston, IL 60208, USA\\
$^{9}$Center for Interdisciplinary Exploration and Research in Astrophysics (CIERA), Northwestern University, Evanston, IL 60208, USA\\
$^{10}$National Research University ‘Higher School of Economics’, Faculty of Physics, Myasnitskaya ul.~20, Moscow 101000, Russia\\
$^{11}$Space Research Institute of the Russian Academy of Sciences (IKI), Profsoyuznaya ul.~84/32, Moscow 117997, Russia\\
$^{12}$Radio Astronomy and Geodynamics Department of Crimean Astrophysical Observatory, 298688 Katsively, Crimea\\
$^{13}$Institute of Astronomy of the Russian Academy of Sciences, Moscow, Russia\\
$^{14}$Astrophysics Research Cluster, School of Mathematical and Physical Sciences, University of Sheffield, Sheffield, S3 7RH, UK\\
$^{15}$Research Software Engineering, University of Sheffield, Sheffield, S1 4DP, UK\\
$^{16}$Instituto de Astrof\'isica de Canarias, E-38205 La Laguna, Tenerife, Spain\\
$^{17}$Armagh Observatory \& Planetarium, College Hill, Armagh, BT61 9DG, Northern Ireland, UK\\
$^{18}$National Astronomical Research Institute of Thailand (NARIT), 260 Moo 4, T.~Donkaew, A.~Maerim, Chiangmai 50180, Thailand\\
$^{19}$Department of Physics \& Astronomy, University of Turku, Vesilinnantie 5, Turku, FI-20014, Finland\\
$^{20}$Jodrell Bank Centre for Astrophysics, Department of Physics and Astronomy, The University of Manchester, Manchester, M13 9PL, UK\\
$^{21}$Institute of Cosmology and Gravitation, University of Portsmouth, Portsmouth, PO1 3FX, UK\\
$^{22}$Departamento de Astrof\'isica, Univ. de La Laguna, E-38206 La Laguna, Tenerife, Spain\\
$^{23}$Department of Astrophysics/IMAPP, Radboud University, 6525 AJ Nijmegen, The Netherlands\\
$^{24}$Institute of Astronomy and Kavli Institute for Cosmology, University of Cambridge, Madingley Road, Cambridge CB3 0HA, UK\\
$^{25}$School of Physics, University College Cork, Cork, T12 K8AF, Ireland\\
$^{26}$School of Sciences, European University Cyprus, Diogenes Street, Engomi, 1516 Nicosia, Cyprus\\
$^{27}$Centre for Astrophysics Research, University of Hertfordshire, College Lane, Hatfield, AL10 9AB, UK\\
% Don't change these lines
%\bsp	% typesetting comment
\label{lastpage}
\end{document}